\documentclass[onecolumn,authoryear]{els-mrw} 

\usepackage{amsmath,amssymb,amsfonts,amsthm,makeidx,graphicx}
\usepackage{txfonts}
\usepackage{helvet}

\usepackage{aas_macros}
\usepackage{comment}
\usepackage{hyperref} 
\hypersetup{
    colorlinks=true,
    linkcolor=blue,
    urlcolor=blue
}

\begin{document}

\chapter{Weak Gravitational Lensing}\label{chap1}

\author[1]{Judit Prat}%
\author[2]{David Bacon}%

\address[1]{\orgname{Nordita}, \orgdiv{Stockholm University and KTH Royal Institute of Technology}, \orgaddress{Hannes Alfv\'ens v\"ag 12, SE-10691 Stockholm, Sweden }}
\address[2]{\orgname{Institute of Cosmology and Gravitation}, \orgdiv{University of Portsmouth}, \orgaddress{Portsmouth, PO1 3FX, UK}}

\articletag{Weak Lensing. \textit{This is a pre-print of a chapter for the Encyclopedia of Astrophysics (edited by I. Mandel, section editor C. Howlett) to be published by Elsevier as a Reference Module.}} 

\maketitle

\begin{abstract}[Abstract]
This chapter provides a comprehensive overview of weak gravitational lensing and its current applications in cosmology. We begin by introducing the fundamental concepts of gravitational lensing and derive the key equations for the deflection angle, lensing potential, convergence, and shear. We explore how weak lensing can be used as a cosmological probe, discussing cosmic shear, galaxy-galaxy lensing, and their combination with galaxy clustering in the 3$\times$2pt analysis. The chapter covers the theoretical framework for modeling lensing observables, shear estimation techniques, and major systematic effects such as intrinsic alignments and baryonic feedback. We review the current results of weak lensing cosmology from major surveys and outline prospects for future advancements in the field.
\end{abstract}

\section{Introduction}\label{chap1:intro}

Weak gravitational lensing has emerged as one of the most powerful tools in modern cosmology, providing a unique window into the distribution of matter in the Universe. Unlike many other cosmological probes, weak lensing is directly sensitive to both luminous and dark matter, making it an invaluable technique for studying the nature of dark matter and dark energy.
The phenomenon of gravitational lensing, predicted by Einstein's theory of General Relativity, occurs when light from distant galaxies is bent as it passes through the curved spacetime near massive objects. While strong lensing produces dramatic effects such as multiple images or giant arcs, weak lensing results in subtle distortions of galaxy shapes that are only detectable statistically over large populations of galaxies. This chapter aims to provide a comprehensive  overview of weak gravitational lensing, from its theoretical foundations to its practical applications in cosmology. We begin by developing the mathematical formalism of weak lensing, introducing key concepts such as the deflection angle, lensing potential, convergence, and shear. We then explore the challenges involved in measuring these effects from real astronomical data, including the complexities of galaxy shape measurement and the various systematic effects that must be accounted for.

We then discuss how measurements of cosmic shear -- the correlated distortion of galaxy shapes due to large-scale structure -- can be used to constrain cosmological parameters and test models of dark energy. We also introduce the powerful ``3$\times$2pt" analysis, which combines weak lensing with galaxy clustering measurements to break degeneracies and provide tighter constraints on cosmology.
Throughout the chapter, we highlight the current state of the art in weak lensing surveys, including results from major projects such as the Dark Energy Survey (DES), the Kilo-Degree Survey (KiDS), and the Hyper Suprime-Cam (HSC) survey. We also look ahead to the future of the field, discussing upcoming surveys and the challenges they will face.
By the end of this chapter, readers should have a solid understanding of the principles of weak lensing, its applications in cosmology, and the exciting prospects for future discoveries in this rapidly evolving field.

\section{Light propagation and the deflection angle} \label{cosmo:sec:deflection_angle}

From General Relativity we learn that the mass content in the universe shapes its metric, with massive bodies curving the space-time canvas around them. Gravitational lensing is caused by light traveling in such a curved space time, with a particular gravitational potential varying in space and time.  Light from distant galaxies is bent as it passes close to massive objects. In some cases, the bending of the light is so significant that multiple images of the galaxy are formed; this is referred to as strong gravitational lensing. In other cases, the light bending is subtle, and the images of galaxies are distorted, stretched and magnified in small amounts; this is referred to as weak gravitational lensing. In this section we will develop the weak lensing formalism, deriving the equations that describe the deflection of light rays in the presence of massive bodies.

Assuming the Friedmann-Lema\^itre-Robertson-Walker in the weak gravitational regime, the line element for a general metric that describes an expanding Universe with first-order perturbations to its homogeneity is given by (see derivation in e.g. \citealt{Hobson:2006se}):
\begin{align} \label{cosmo:eq: metric potential}
ds^2 = - \left( 1 + \frac{2 \Psi}{c^2 }\right)c^2 dt^2 + a^2(t) \left( 1 - \frac{2 \Phi}{c^2 } \right)dl^2 ,
\end{align}
where $dl^2 = d\chi^2+ S_k ^2 (\chi) d\Omega^2$. $\Omega$ is the solid angle, $\chi$ is the radial comoving distance, $S_k ^2 (\chi)=\chi$ for a flat Universe, $\sin \chi$ for a closed Universe with curvature $k=1$ and $\sinh \chi$ for an open Universe with  $k=-1$. The two potentials $\Psi$ and $\Phi$ describe weak gravitational fields with $\Psi$, $\Phi \ll c^2$. This condition is fulfilled for all masses $M$ and potentials $\Psi$, $\Phi \sim GM/R = (c^2/2)(R_s/R)$ whose extents $R$ are much larger than their Schwarzschild radius $R_s$, where $G$ is Newton's gravitational constant.  In General Relativity the two potentials are well approximated as equal for astrophysically realistic scenarios where dynamical shear is negligible, so that $\Psi \simeq \Phi$. Also, note that if there are no perturbations, the metric reduces to the standard FLRW metric. 

\begin{figure}[b]
\centering
\includegraphics[width=0.9\textwidth]%
	{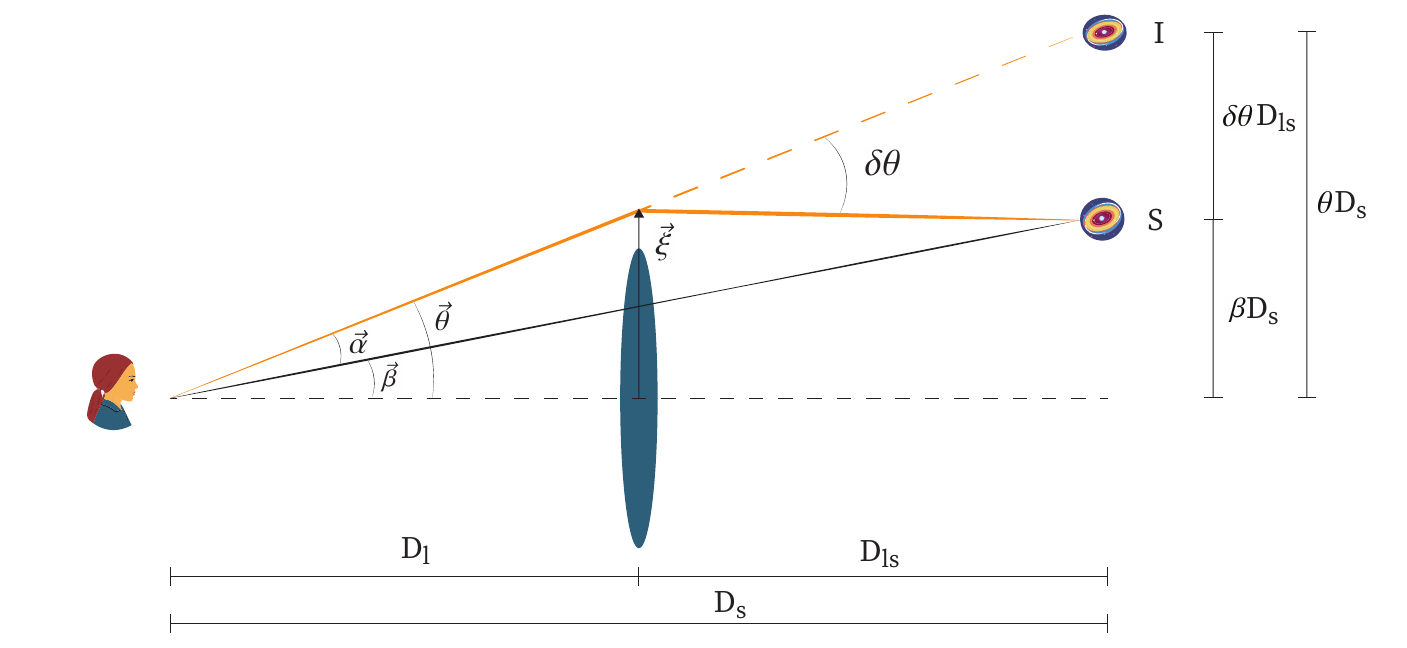}
	\caption{Sketch of a gravitational \textit{thin lens} system. The optical axis runs from the observer through the center of the lens. The angle between the source S and the optical axis is $\beta$, the angle between the image I and the optical axis is $\theta$. The light ray towards the image is bent by  $\delta \theta$, measured at the lens. The  deflection angle $\alpha$ is measured at the observer.}
	\label{cosmo:fig:lensing_diagram}
\end{figure}

\begin{figure}
	\centering
	\includegraphics[width=0.9\textwidth]%
	{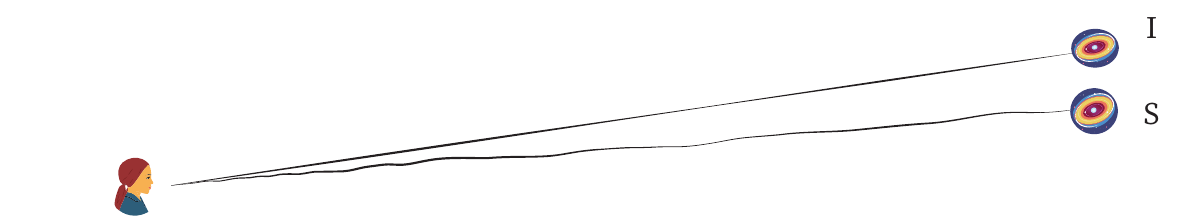}
	\caption{The Born approximation assumes that the perturbations to the light path due to  gravitational lensing by the structure along the line of sight are negligible and that we can compute the deflection angle integrating along the straight line back to the image position.}  
	\label{cosmo:fig:born_approx}
\end{figure}

Photons propagate on null geodesics, $ds^2=0$. That means that a measure of the time of light ray travel can be obtained from the metric equation (\ref{cosmo:eq: metric potential}). In General Relativity and for weak gravitational fields ($\Phi/c^2 \ll 1$), we obtain:
\begin{align}
t = \frac{1}{c} \int \left( 1 - \frac{2\Phi}{c^2}\right) dr,
\end{align}
where the integral is along the light path in physical coordinates $r = a \chi$. Then, we can make an analogy between the gravitational potential and a medium with variable refractive index $n=1- 2 \Phi/c^2$. After that, we can use Fermat's principle, which says that the light travels by the minimum-time path, $\delta t = 0$. Therefore, we obtain the Euler-Lagrange equations for the refractive index. Integrating these equations along the light path one can obtain the expression for the \textit{deflection angle} $\delta{\theta}$, which is the difference between the directions of the emitted and received light rays, illustrated in Fig.~\ref{cosmo:fig:lensing_diagram}: 

\begin{align} \label{cosmo:eq: deflection angle}
\delta{\theta} = - \frac{2}{c^2}  \int \nabla_\perp \Phi dr.
\end{align}
The gradient of the potential is taken perpendicular to the light path, with respect to physical coordinates. As it stands, this equation for the deflection angle is not very useful, as we would have to integrate over the actual light path, which is unknown. However, since $\Phi/c^2 \ll 1$, we expect the deflection angle to be small. Then, we can adopt the \textit{Born approximation} and integrate over the unperturbed light path, as illustrated in Fig.~\ref{cosmo:fig:born_approx}. Considering a point-like body of mass $M$ whose gravitational potential is $\Phi = -GM/r$, and using the Born approximation, the deflection angle reduces to:
\begin{align}
\delta{\theta} = \frac{4GM}{bc^2},
\end{align} 
where $b$ is the distance of closest approach to the lens, called the impact parameter. This result, assuming General Relativity, gives twice the classical prediction for the deflection angle using Newtonian dynamics. This factor of two comes from the fact that the perturbed Minkowski metric has equal perturbations in both its temporal and spatial components. In 1919, a team led by Franck Watson Dyson and Arthur Eddington, proved that General Relativity gave the correct factor by measuring the change in position of stars as their light passed near the sun during a solar eclipse.

We can now consider another particular case of a single lens, but extended in the transverse direction (perpendicular to the line of sight). This extension is still much simpler than a general gravitational potential but realistic enough in a wide number of cases. For instance even in the case of lensing by galaxy clusters, the physical size of the lens is generally much smaller than the distances between observer, lens and source. The deflection therefore arises along a very short section of the light path, and we can adopt the so-called \textit{thin lens approximation}, where the distribution of matter is assumed to be in the \textit{lens plane}.  Within this approximation, the lens matter distribution is fully described by its \textit{surface density},
\begin{align}
\Sigma(\vec{\xi}) = \int \rho(\vec{\xi}, z) \, dz, 
\end{align}
where $\vec{\xi}$ is a two-dimensional vector on the lens plane, also illustrated in Fig.~\ref{cosmo:fig:lensing_diagram},  $\rho$ is the three-dimensional density and  $z$ is the physical coordinate in the line of sight direction. As long as the thin lens approximation holds, the total deflection angle is obtained by summing the contribution of all the mass elements $\Sigma(\vec{\xi})d^2\vec{\xi}$:
\begin{align}
\delta \vec{\theta} \, (\vec{\xi})= \frac{4G}{c^2} \int \frac{(\vec{\xi}- \vec{\xi}')\, \Sigma(\vec{\xi'})}{|\vec{\xi}- \vec{\xi}'|^2} d^2\vec{\xi'}.
\end{align}

\begin{BoxTypeA}[chap1:box1]{Common approximations in weak lensing to simplify light propagation and lens geometry}

\subsection*{Thin Lens Approximation (see Fig.~\ref{cosmo:fig:lensing_diagram})}
\textit{Assumption:} The physical extent of the lens is much smaller than the distances $D_l$, $D_s$ and $D_{ls}$.
\textit{Application:} The mass distribution of the lens is projected onto a single plane perpendicular to the line of sight, described by the surface mass density.
\textit{Validity:} applicable to many astrophysical scenarios, including galaxy and cluster lensing by a single lens.
\subsection*{Born Approximation (see Fig.~\ref{cosmo:fig:born_approx})}
\textit{Assumption:} The deflection angle is small compared to the scale over which the gravitational potential varies significantly.
\textit{Application:} Integration is performed along the straight line to the image position instead of the actual (unknown) deflected path.
\textit{Validity:} Generally valid in weak lensing regimes where $\Phi/c^2 \ll 1$.
\end{BoxTypeA}

\subsection{The lens equation}\label{cosmo:subsec:lens_equation}

Figure~\ref{cosmo:fig:lensing_diagram} illustrates a thin lens system. Although not obvious from the figure, the angles have both an amplitude and a direction. The amplitude describes how the incoming ray is tilted with respect to the $z$-axis. The directions of the angles specify the locations in the plane perpendicular to the line of sight, the plane of the sky. Thus, the source $S$  position in this plane is given by $D_s \, \vec{\beta}$ and the image position $I$ is given by  $D_s \, \vec{\theta}$, where $D_s$ is the angular diameter distance to the source. Looking at the diagram and assuming the angles are small, we can relate these two positions to the deflection angle $\delta \vec{\theta}$ through the so-called \textit{lens equation}
\begin{align}
D_s \, \vec{\beta} = D_s \,  \vec{\theta} - D_{ls} \,\delta \vec{\theta},
\end{align}
where $D_{ls}$ is the angular diameter distance between the lens and the source.  As Fig.~\ref{cosmo:fig:lensing_diagram} shows, the lens equation is trivial to derive and only requires that the following Euclidean relation should exist between the angle enclosed by two lines and their separation: separation = angle $\times$ distance. It is not obvious that the same relation should also hold in curved spacetimes. However, angular diameter distances are defined exactly so that this relation holds. Dividing by $D_s$ and introducing the deflection angle from the point of view of the observer
\begin{align}
\vec{\alpha} ( \vec{\theta}) \equiv \frac{D_{ls}}{D_s} \delta \vec{\theta}
\end{align}
leads to the lens equation in its simplest form:
\begin{align}\label{cosmo:eq:lens_eq}
 \vec{\beta} = \vec{\theta} -  \vec{\alpha } (\vec{\theta}),
\end{align}
which actually hides quite a bit of complexity: the mapping from lens coordinates $\vec{\theta}$ to source coordinates $\vec{\beta}$ may be non-linear and have multiple solutions (only in the strong lensing regime) so that a given single point source at $\vec{\beta}$ has multiple images $\vec{\theta}$.

\begin{BoxTypeA}[chap1:box2]{Deflection angles: $\delta \vec{\theta}$ vs $\vec{\alpha }$}

\noindent Two distinct deflection angles are often confused and assigned different notations in the literature: $\delta \vec{\theta}$, the actual angular deflection measured at the lens plane, and $\vec{\alpha } $, the apparent deflection seen by the observer. They are related by $\vec{\alpha } ( \vec{\theta}) = \frac{D_{ls}}{D_s} \delta \vec{\theta}$, where $\vec{\alpha }$ is used in the lens equation $\vec{\beta} = \vec{\theta} -  \vec{\alpha } (\vec{\theta})$.
\end{BoxTypeA}
\subsection{Lensing potential}
\begin{figure}
	\centering
	\includegraphics[width=0.6\textwidth]{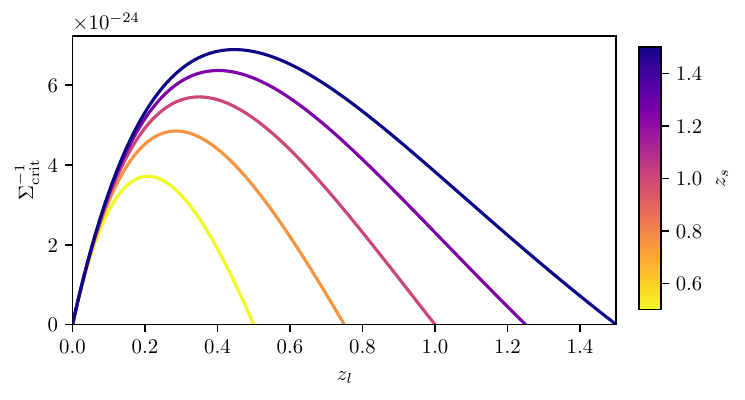}
	\caption{ Inverse critical surface density ($\Sigma^{-1}_\mathrm{crit}$) as a function of lens redshift ($z_l$) for various source redshifts ($z_s$). The plot shows how the lensing efficiency changes with lens redshift for different source redshifts, ranging from 0.5 to 1.5. Each line represents a different source redshift, with colors indicating the $z_s$ value according to the colorbar on the right. This figure illustrates the dependence of  lensing strength on the relative distances between the observer, lens, and source. A rule of thumb is that the lensing efficiency peaks at around half the redshift between the observer and the source. }
	\label{cosmo:fig:inv_sigma_crit}
\end{figure}

Still in the context of the thin lens approximation, an extended distribution of matter is characterized by its effective lensing potential, obtained by projecting the three-dimensional Newtonian potential along the line of sight or $z-$axis ($z$ is still the physical coordinate, not the redshift here):
\begin{align}\label{cosmo:eq:lensing_potential}
\psi (\vec{\theta})= \frac{2}{c^2  } \frac{D_{ls}}{D_l D_s} \int \Phi (D_l \vec{\theta}, z)\,  dz.
\end{align}
From the equation above we can learn in which cases the effect of gravitational lensing will be stronger. The contribution to $\Phi$ from inhomogeneities close to the source is suppressed by the $D_{ls}$ factor, and when the angular distance to the lens is similar to the angular distance between the lens and the source, the effect will be larger. Also, in the equation above we are again using the Born approximation, since we are integrating the potential along the line of sight between the apparent position $D_l\vec{\theta}$ and us, not along the path the light actually traveled, as illustrated in Fig.~\ref{cosmo:fig:born_approx}. More generally, if the potential is constant across the sky, there is no deflection. Lensing therefore emerges from changes in the projected gravitational potential across the sky. Mathematically, this is expressed by the gradient of the lensing potential, which yields the deflection angle $\vec{\alpha}$:
\begin{align}\label{cosmo:eq:gradient_potential}
  \vec{\alpha } (\vec{\theta}) = \vec{\nabla}_{\vec{\theta}}\,  \psi .
\end{align}
Moreover, the Laplacian of the potential is proportional to the surface-mass density at the lens plane position $D_l\vec{\theta}$ via the Poisson equation:
\begin{align}\label{cosmo:eq:laplacian_lensing_potencial}
\nabla^2_{\vec{\theta}}\,  \psi  = 2\frac{\Sigma(D_l\vec{\theta})}{\Sigma_\text{crit}} \equiv 2\kappa (\vec{\theta}) ,
\end{align}
where we have defined the dimensionless parameter $\kappa$, called \textit{convergence}, and the geometrical factor
\begin{align}
\Sigma_\text{crit} =\frac{c^2}{4\pi G}\frac{D_s}{D_l D_{ls}},
\end{align}
also called \textit{critical surface mass density}. We illustrate such a factor in Fig.~\ref{cosmo:fig:inv_sigma_crit}. 

\section{Linearised lens mapping: shear and magnification}\label{cosmo:sec:linearized_mapping}

Using Eq.~(\ref{cosmo:eq:gradient_potential}), the lens equation (\ref{cosmo:eq:lens_eq}) can be written in terms of the lensing potential in the following way:
\begin{align}
\vec{\beta} = \vec{\theta} - \vec{\nabla}\,  \psi .
\end{align}
If the extent of a source is much smaller than the scale of variation in the deflection angle, we can linearise the lens equation by defining the \textit{Jacobian matrix} $A$ and Taylor expanding the lens equation:
\begin{figure}
	\centering
	\includegraphics[width=0.75\textwidth]{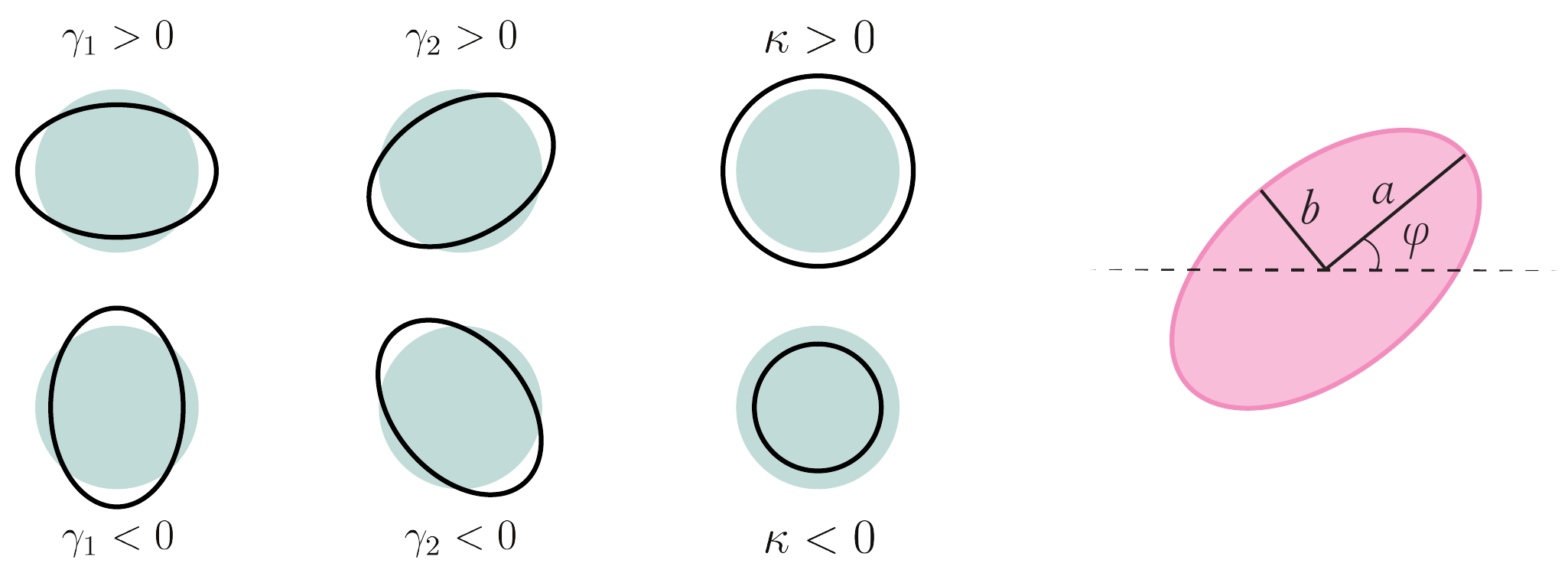}
	\caption{The left side of the image demonstrates the effects of the Jacobian matrix elements -- shear $\gamma = (\gamma_1, \gamma_2)$ and convergence $\kappa$ -- on an initially circular background object. The right side illustrates the key components that define an ellipse: the semi-major axis $a$, semi-minor axis $b$, and the position angle $\varphi$.}
	\label{cosmo:fig:shear_convergence}
\end{figure}
\begin{align}
\delta \vec{\beta} \approx A \delta \vec{\theta}
\end{align}
with $A$ having the components
\begin{align}\label{cosmo:eq:A_ij}
A_{ij} = \frac{\partial \beta_i}{\partial \theta_j} = \delta_{ij} - \frac{\partial \alpha_i}{\partial \theta_j} = \delta_{ij} - \partial_i \partial_j \psi, 
\end{align} 
where the partial derivatives are with respect to $\vec{\theta}$. This matrix describes the linear mapping between the lensed, $\vec{\theta}$, and the unlensed, $\vec{\beta}$, coordinates. In the absence of a lensing potential, the lens mapping is simply the identity. In the presence of a lens, the local properties of the lens  mapping are determined by the curvature of the lensing potential, expressed by the matrix of second derivatives of $\psi$. Moreover, for the physical interpretation of the Jacobian matrix, it is convenient to parametrize it in terms of the complex two-component shear field $\gamma\equiv \gamma_1 + i \gamma_2 = |\gamma| e^{2i\varphi}$ (see Fig.~\ref{cosmo:fig:shear_convergence} for an illustration of $\varphi$)  and the scalar convergence $\kappa$ as
\begin{align}
A=
\begin{pmatrix}
1-\kappa -\gamma_1 & -\gamma_2 \\
- \gamma_2 & 1 - \kappa + \gamma_1
\end{pmatrix}.
\end{align}

\begin{BoxTypeA}[wl:box_shear_as_spin]{The shear: a spin-2 quantity in weak lensing}

\noindent The shear (as well as the ellipticity or reduced shear) is a two-component quantity, typically represented as a complex number:
\begin{equation}
\gamma \equiv \gamma_1 + i \gamma_2 = |\gamma| e^{2i\varphi}
\end{equation}
The amplitude $|\gamma|$ quantifies the degree of distortion and the angle $\varphi$ describes the orientation of the shear. The factor of 2 in $e^{2i\varphi}$ indicates its spin-2 nature: shear transforms back to its original state after a 180° rotation. This complex number representation facilitates mathematical operations in weak lensing analyses while preserving the properties of shear as a spin-2 field. 
\end{BoxTypeA}

\noindent This means that the shear and the convergence can be expressed as second derivatives of the lensing potential $\psi$:
\begin{align}\label{cosmo:eq:derivatives_shear_kappa}
\begin{split}
\kappa &= \frac{1}{2} (\partial_1 \partial_1 + \partial_2 \partial_2) \psi = \frac{1}{2} \nabla^2 \psi  \\
\gamma_1 &= \frac{1}{2} (\partial_1 \partial_1 - \partial_2 \partial_2) \psi  \\
\gamma_2 &= \partial_1 \partial_2 \psi,
\end{split}
\end{align} 
as we had already anticipated for the convergence in Eq.~(\ref{cosmo:eq:laplacian_lensing_potencial}). We can also rewrite $A$ as
\begin{align}
A = (1-\kappa)
\begin{pmatrix}
1& 0 \\
0 & 1
\end{pmatrix} - 
|\gamma|
\begin{pmatrix}
\cos(2\varphi)& \sin(2\varphi) \\
\sin(2\varphi) & -\cos(2\varphi)
\end{pmatrix},
\end{align}
where we can see that the $(1-\kappa)$ term only affects the size and not the shape of the observed image. Thus, the convergence quantifies the isotropic change in the size of the source image; while the shear quantifies an anisotropic stretching, i.e. a change in the shape of the image, turning a circle into an ellipse as illustrated in Fig.~\ref{cosmo:fig:shear_convergence}. The Jacobian matrix actually tells us the inverse of what we typically want to know from weak gravitational lensing; it tells us how to go from lensed coordinates to source coordinates, while we may well wish to go from source coordinates to lensed coordinates. To obtain this other mapping, if A has a non-zero determinant, we can invert it:
\begin{align}
\delta   \vec{\theta} \approx A^{-1} \delta\vec{\beta}.
\end{align}
The Jacobi determinant is
\begin{align}
\det A = (1-\kappa)^2 - \vert \gamma \vert ^2 \approx 1 - 2\kappa
\end{align}
with  $\vert \gamma \vert ^2 = \gamma_1^2 + \gamma_2^2$, where the last approximation is only valid in the weak  lensing regime with $\gamma,\kappa \ll 1$. Thus, we can assume that in the weak lensing regime the linear lens mapping is invertible and that the inverse of the Jacobian matrix is
\begin{align}
A^{-1} = \frac{1}{\det A} \begin{pmatrix}
1-\kappa +\gamma_1 & \gamma_2 \\
 \gamma_2 & 1 - \kappa - \gamma_1
\end{pmatrix}
\end{align}
The overall factor in this expression indicates that the solid angle spanned by the image is changed compared to the solid angle covered by the source by the \textit{magnification} factor $\mu$:
\begin{align}\label{eq:magnification_kappa}
\mu = \frac{1}{\det A} = \frac{1}{(1-\kappa)^2 - \vert \gamma \vert ^2 } \approx 1+ 2\kappa,
\end{align}
where again the last approximation only holds in the weak lensing regime. Thus, in weak lensing, the magnification of an image is essentially determined by the convergence, not by the shear.

\subsection{Relation between shear and ellipticity} \label{sec:shear_ellipticity_relation}

\subsubsection{For circular objects}
The key to estimate shear and magnification is through measuring galaxy shapes, which are quantified by the \textit{ellipticity}. That is why it is very useful to relate the shape of the galaxy to the shear. We will start by estimating the distortions of an originally circular source. In this case, the circle will be deformed to become an ellipse whose semi-major and semi-minor axes, $a$ and $b$ respectively, illustrated in Fig.~\ref{cosmo:fig:shear_convergence}, are proportional to the eigenvalues $\lambda_{\pm} $, also respectively, of the inverse Jacobi matrix $A^{-1}$: 
\begin{align}\label{eq:ellipticity_circular}
\lambda_{\pm} \equiv \frac{1-\kappa \pm\vert \gamma \vert }{\det A} 
\end{align}
We can define the image \textit{ellipticity} $\epsilon$ of an originally circular source as:
\begin{align}\label{cosmo:eq:ellipticity_cicularsource}
 \epsilon  =\frac{a-b}{a+b} e^{2i \varphi} = \frac{\lambda_+ - \lambda_-}{\lambda_+ + \lambda_-}  e^{2i \varphi}= \frac{\gamma }{1-\kappa}.
\end{align}
The equation above shows that the ellipticity is determined by the \textit{reduced shear} quantity, which involves both the shear and the convergence:
\begin{align}\label{cosmo:eq:reduced_shear}
g \equiv \frac{\gamma}{ 1 - \kappa}.
\end{align}
Then the reduced shear is the relevant quantity that influences galaxy shapes. If $\kappa\ll 1$, which is generally the case for weak lensing, then the reduced shear is a good approximation to the shear.

\subsubsection{For arbitrary image shapes}

In the expressions above we have been considering an originally circular source to understand the effect of weak lensing in the shapes of galaxies. However, most sources are not initially circular, but have some intrinsic ellipticity $\epsilon_{\mathrm{int}}$ to start with. Moreover, real galaxies are  not perfectly elliptical; their isophotes (contours of constant brightness) are generally not true ellipses. That is why  we actually need a more complex definition of ellipticity than the one from Eq.~(\ref{eq:ellipticity_circular}). Fortunately, the ellipticity can also be quantified in terms of moments of the surface brightness. The surface brightness $I$ is defined as the flux of energy per unit time per unit area per solid angle, so has units of Energy Time$^{-1}$ Length$^{-2}$ steradian$^{-1}$ and is conserved in gravitational lensing processes. 

\begin{BoxTypeA}{Conservation of surface brightness}

\noindent Gravitational light deflection conserves surface brightness, following from Liouville’s theorem, since there are no emission or absorption processes, i.e. the photons are conserved.  It is also interesting to note that the surface brightness of a given object remains constant no matter how far the object is from us. That is because while the flux (energy per time per area) goes down as the object gets further, the physical size subtended by the same angle gets larger. The two effects scale by  the distance squared, and thus they cancel each other. 
\end{BoxTypeA}

\noindent If we define the quadrupole moments $Q_{ij}$ of the surface brightness as 
\begin{equation}
Q_{ij} \equiv \int d^2 \theta \,  I(\vec{\theta}) \theta_i \theta_j, \qquad i,j = 1,2.
\end{equation}
then we can define the two components of the ellipticity as:
\begin{equation}\label{cosmo:eq:ellipticity_measurement}
\epsilon_1 \equiv \frac{Q_{11} - Q_{22}	}{2 N_Q}, \qquad \epsilon_2 \equiv \frac{Q_{12}}{N_Q}, \qquad N_Q \equiv \frac{1}{2} \text{tr} Q + \sqrt{\det Q} = \frac{Q_{11} + Q_{22}}{2} + \sqrt{Q_{11} Q_{22} - Q^2_{12}}
\end{equation}
The Trace (tr) of $Q$ describes the size of the image, while the traceless part of $Q_{ij}$ contains the ellipticity information. Given this definition of ellipticity, the  relation between shear and ellipticity is given by\footnote{Note that $\epsilon = \frac{1 + g\epsilon_{\mathrm{int}}^*}{g^* + \epsilon^*_{\mathrm{int}}}$ if $|g| > 1$.} 
\begin{align}\label{cosmo:eq:ellipticity_shear}
\epsilon = \frac{\epsilon_{\mathrm{int}}+ g}{1+ g^* \epsilon_{\mathrm{int}}} \qquad \mathrm{if} \  |g|\leq 1  
\end{align}
where the asterisk denotes complex conjugation. When $\epsilon_{\mathrm{int}}=0$, it reduces to Eq.~(\ref{cosmo:eq:ellipticity_cicularsource}). Also, note that this expression depends on the convention used to define the ellipticities. There are two common conventions used in weak lensing, which lead to different relations between the ellipticity and the reduced shear; see the highlighted box below for further details on this.

\begin{BoxTypeA}[wl:box_ellipticity]{Alternative convention to define ellipticity}

\noindent There are multiple conventions for defining ellipticity in weak lensing literature. An alternative popular definition is 
\begin{equation}\label{cosmo:eq:ellipticity_measurement2}
\epsilon_1 \equiv \frac{Q_{11} - Q_{22}	}{Q_{11} + Q_{22}}, \qquad \epsilon_2 \equiv \frac{2 Q_{12}}{Q_{11} + Q_{22}} .
\end{equation}
Under this definition, 
\begin{equation}
    \epsilon = \frac{\epsilon_{\mathrm{int}} + 2g + g^2 \epsilon_{\mathrm{int}}^*  }{1+ |g|^2 + 2 \text{Re} \left[g \epsilon^*_{\mathrm{int}} \right] }
\end{equation}
replacing equation Eq.~(\ref{cosmo:eq:ellipticity_shear}) \citep{Dodelson2017}. Thus, the choice of definition affects the relationship between ellipticity and reduced shear, so it is crucial to be consistent and clear about which convention is being used in each analysis. Also note that under this convention, a circular source would have an image ellipticity
\begin{align}\label{cosmo:eq:ellipticity_cicularsource2}
 \epsilon  =\frac{a^2-b^2}{a^2+b^2} e^{2i \varphi} 
\end{align}
and then this implies
\begin{align}\label{cosmo:eq:reduced_shear2}
\epsilon = \frac{2\gamma}{ 1 - \kappa} \frac{1}{1-[\gamma_1^2 + \gamma_2^2]/(1-\kappa)^2},
\end{align}
replacing equation Eq.~(\ref{cosmo:eq:ellipticity_cicularsource}). The second factor on the right is almost always negligible, but note that it also only depends on the reduced shear. More importantly,  in this convention the reduced shear and the ellipticity are related by a factor of 2, while they are equivalent in the convention used in the main text. 

\end{BoxTypeA}

\subsection{Tangential and cross components of the shear}\label{cosmo:sec: tangential and cross}

\begin{figure}
	\centering
	\includegraphics[width=1.\textwidth]{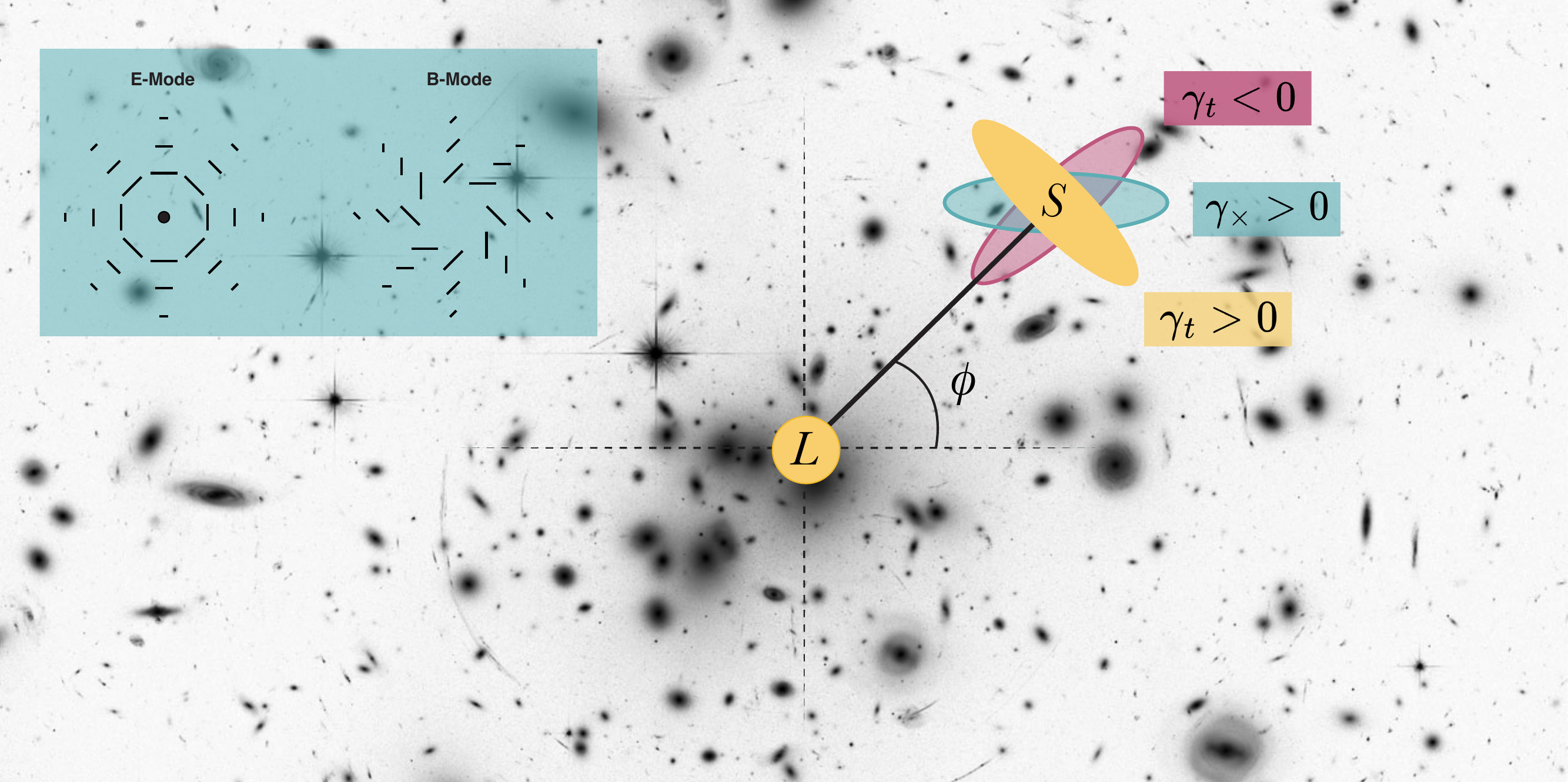}
	\caption{The massive foreground cluster (Abell 1689) causes the images of the background galaxies to be distorted, forming arcs, due to
strong gravitational lensing. The arcs are tangentially aligned, and so their ellipticity is oriented tangent to the direction of the foreground
mass, in this case the cluster, represented by the lens point $L$. Image taken with the Advanced Camera for Surveys on board the Hubble Space Telescope. Credits:
NASA, ESA, the Hubble Heritage Team (STScI/AURA), J. Blakeslee (NRC Herzberg Astrophysics Program, Dominion Astrophysical Observatory), and H. Ford (JHU).}
	\label{cosmo:fig: Abell 1689 cluster}
\end{figure}

The shear components $\gamma_1$ and $\gamma_2$ are defined relative to a reference Cartesian coordinate frame. However, because of the way background galaxies are distorted by a foreground mass, it is useful to consider the shear components in a rotated reference frame. For example, in the case of a \textit{spherical distribution} of matter, the shear at any point will be oriented tangentially to the direction toward the center of symmetry as can be seen in a real image in Fig.~\ref{cosmo:fig: Abell 1689 cluster}, where a massive cluster is bending the light of galaxies behind it. For a given lens-source pair of galaxies we define the \textit{tangential} and \textit{cross components} of the shear:
\begin{align}\label{cosmo:eq:gammat_euler}
\gamma_t = -\text{Re} \left[ \gamma e ^{-2i\phi} \right]  , \  \gamma_\times= -\text{Im} \left[ \gamma e ^{-2i\phi} \right],
\end{align}
where $\phi$ is the position angle of the source galaxy with respect to the horizontal axis of the Cartesian coordinate system, centered at the lens galaxy, as represented in Fig.~\ref{cosmo:fig: Abell 1689 cluster}. This can be expanded to yield
\begin{align}
\begin{split}
\gamma_t &= -\gamma_1 \cos(2\phi) - \gamma_2 \sin(2\phi)\\
\gamma_\times &= \gamma_1 \sin(2\phi) - \gamma_2 \cos(2\phi).
\end{split}
\end{align}
The tangential component  captures all the gravitational lensing signal produced by a spherically symmetric distribution of mass, while the cross-component of the shear $\gamma_\times$  vanishes if the mass distribution is spherically symmetric. Both components are represented in the lower left corner of Fig.~\ref{cosmo:fig: Abell 1689 cluster}.  There we can see that the cross-component has a curl pattern, something that cannot be produced by a scalar field such as the convergence $\kappa$. That is why, making an analogy with electromagnetism, where the electric field is the gradient of a scalar field and the magnetic field is the curl of a vector field, this cross pattern is sometimes called the B-mode, and the tangential component is usually called the E-mode. $\gamma_\times$ is often used as a null test to check the measurement is free of systematics for systems which are expected to satisfy spherical symmetry.

To better understand the physical interpretation of the tangential shear measurement, we can start from the  deflection angle for a spherically symmetric distribution, which is equal to \citep{Dodelson2017}
\begin{align}
\vec{\alpha } (\vec{\theta}) = \frac{\vec{\theta}}{\theta^2} \frac{M(\theta)}{\pi D_l^2 \Sigma_{\mathrm{crit}}},
\end{align}
where $M(\theta)$ is the mass enclosed within a cylinder of angular radius $\theta$. Obtaining the Cartesian shear components $\gamma_1$ and $\gamma_2$ by deriving the deflection angle with respect to the image position $\vec{\theta}$, as given in Eq.~(\ref{cosmo:eq:A_ij}), we can then rotate them to the tangential projection \citep{Dodelson2017}:
\begin{align}
\gamma_t (\theta)= - \frac{\theta}{2\pi D^2_L \Sigma_{\mathrm{crit}}}\frac{\partial}{\partial \theta} \left[ \frac{M(\theta)}{\theta^2} \right],
\end{align} 
which is the tangential shear in an annulus with radius $\theta$ produced by a spherical mass distribution. Performing the derivative we obtain
\begin{align}
\gamma_t (\theta)=  \bar{\kappa} (\leq\theta) -\kappa(\theta), 
\end{align}
where $\kappa$ is the surface density divided by the critical surface mass density $\Sigma_{\mathrm{crit}}$, as given in Eq.~(\ref{cosmo:eq:laplacian_lensing_potencial}) and $\bar{\kappa}$ is the average of the convergence within the angular radius $\theta$. If we multiply the expression above by the critical surface mass density we obtain the definition for the \textit{surface mass excess} $\Delta\Sigma$:
\begin{align}\label{cosmo:eq:delta_sigma}
\Delta\Sigma \equiv \gamma_t(\theta) \Sigma_{\mathrm{crit}} = \bar{\Sigma} (<\theta) - \Sigma(\theta),
\end{align}
which depends only on the lens properties, while $\gamma_t$ also depends on the geometry of the lens-source system through the $\Sigma_{\mathrm{crit}}$ factor. 

When both the lens and the source objects are galaxies, the \textit{mean tangential shear} is also usually referred to as \textit{galaxy-galaxy lensing} or \textit{galaxy-shear} correlations.

\subsection{Convergence reconstruction}

From the shear estimates, we can obtain the convergence, which offers several advantages in weak lensing analyses. The convergence is a scalar field and provides the most direct estimate of mass: as shown in Sec.~\ref{sec:convergence_project_matter}, the convergence is simply a weighted integral along the line of sight of the density field $\delta (\boldsymbol{x})$. This direct relationship to the underlying mass distribution is why convergence maps are often referred to as \textit{mass maps}.

\begin{figure}
	\centering
	\includegraphics[width=0.6\textwidth]{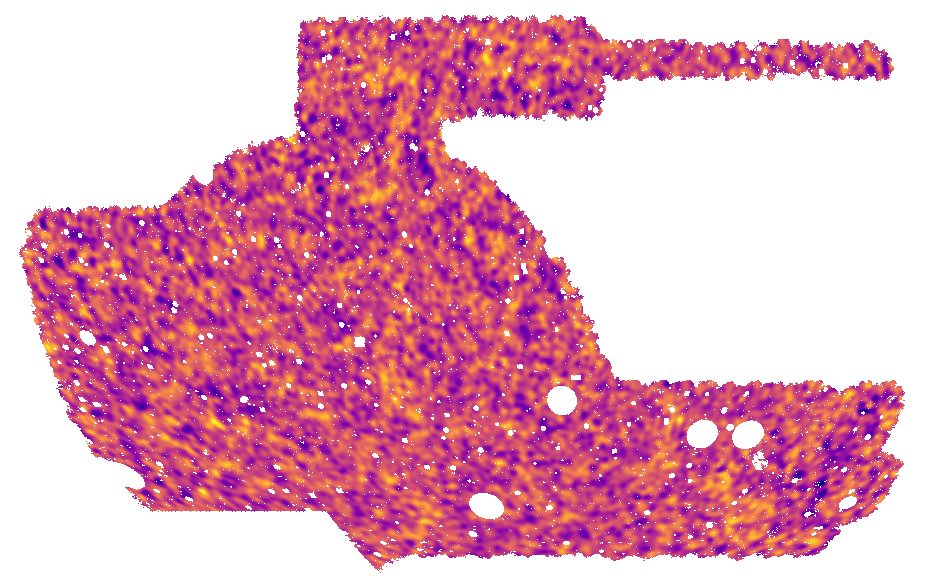}
	\caption{Smoothed convergence map obtained with the Kaiser-Squires algorithm using data from the first three years of observations from the Dark Energy Survey, constructed from shapes of over 100 million galaxies \citep{Jeffrey2021} placed around redshift of $z\simeq 1$. The  map covers $\sim 4500$ sq. deg.$^2$, which corresponds to roughly one eighth of the sky. Dark areas  represent under dense regions, and bright overdense, with the convergence $\kappa$ covering the range $\kappa\pm 0.025$.  }
	\label{cosmo:fig:kappa_map}
\end{figure}

We can \textit{reconstruct} the convergence from the shear by inverting the second derivatives from Eq.~(\ref{cosmo:eq:derivatives_shear_kappa}), reconstructing the lensing potential  $\psi$ from the shear, and then extracting $\kappa$ by differentiating $\psi$ again. This process, known as Kaiser-Squires  (KS; \citealt{KaiserSquires}) \textit{reconstruction}, is more conveniently performed in Fourier space. 
We can obtain harmonic coefficients $\hat{\psi}_{\ell m}$, $\hat{\kappa}_{\ell m}$ and $\hat{\gamma}_{\ell m}$ for $\psi$, $\kappa$ and $\gamma$ respectively as:
\begin{equation}
\gamma = \sum_{\ell m} \hat{\gamma}_{\ell m} \, {}_2Y_{\ell m} 
\end{equation}
with
\begin{equation}
\hat{\gamma}_{\ell m} = \int d\Omega \, \gamma(\theta,\varphi) \, {}_2Y_{\ell m}^*(\theta,\varphi).
\end{equation}
and $_sY_{lm}(\theta, \phi)$ denoting the spin-weight $s$ spherical harmonic basis functions. We can decompose the harmonic coefficients into real and imaginary parts: 
\begin{align}  \label{eq:EB_modes_kappa}
    \hat{\kappa}_{\ell m} &= \hat{\kappa}_{E,\ell m} + i\hat{\kappa}_{B,\ell m} \\
    \hat{\gamma}_{\ell m} &= \hat{\gamma}_{E,\ell m} + i\hat{\gamma}_{B,\ell m}.
\end{align}
Then, in harmonic space, the equations from (\ref{cosmo:eq:derivatives_shear_kappa}) become:
\begin{align}
\hat{\kappa}_{\ell m}  = - \frac{1}{2} \ell (\ell+1) \hat{\psi}_{\ell m}, \quad \hat{\gamma}_{lm} = \frac{1}{2}\sqrt{(\ell-1)\ell(\ell+1)(\ell+2)}\hat{\psi}_{\ell m}.
\end{align}
Thus, the shear and convergence are related by:
\begin{align}
\label{eq:mass_map_operator}
\hat{\gamma}_{lm} = -\sqrt{\frac{(\ell-1)(\ell+2)}{\ell(\ell+1)}} \hat{\kappa}_{\ell m}.
\end{align}
Thus, to perform the reconstruction, this relation is inverted. The equations presented here describe the reconstruction on a spherical surface, which is necessary for analyzing state-of-the-art data (see, for example, \citealt{Jeffrey2021}). It is worth noting that in some cases,  the flat sky approximation is still used instead of the spherical approach. We show an example of a curved-sky convergence map in Fig.~\ref{cosmo:fig:kappa_map}.

\subsection{The signal-to-noise}\label{cosmo:sec:shear_signal_noise}

Taking the weak lensing limit of Eq.~(\ref{cosmo:eq:ellipticity_shear}), the intrinsic source ellipticity $\epsilon_{\mathrm{int}}$ and the ellipticity caused by gravitational lensing can be added:
\begin{align}
\epsilon \approx \epsilon_{\mathrm{int}} + \gamma. 
\end{align}
If the intrinsic ellipticities of the galaxies are randomly aligned\footnote{This is not true in the presence of intrinsic alignments, which are one of the major systematic effects in weak lensing. Intrinsic alignments arise due to tidal interactions between galaxies. We discuss how we model such an effect in the context of cosmological analyses in Sec.~\ref{sec:IA}.}, the mean of the observed ellipticity over a sample of galaxies would yield the (tangential) shear, 
$\left< \epsilon\right> \approx \gamma_t$, since $\left< \epsilon_{\mathrm{int}} \right> =0 $. This is the signal. The variance of the intrinsic ellipticities is defined as the \textit{shape noise}: 
\begin{align} \label{eq:sigma_e}
\sigma^2_\epsilon \equiv \left< \epsilon^2 \right>
\end{align}
Typical values of $\sigma_\epsilon$ are of the order of 0.3. Then, for a given lens galaxy and one source galaxy, we would get a signal-to-noise of $\gamma_t / \sqrt{\sigma^2_\epsilon}$. In weak lensing, typical values of the shear will usually not be larger than $\gamma_t \sim 0.01$. Thus, we cannot estimate the shear for a single galaxy.  For a sample of $N_s$ source galaxies separated by a similar distance to the lens galaxy (and thus sharing the same signal), the signal-to-noise scales as $\sqrt{N_s}$:
\begin{align}
 \frac{S}{N} = \sqrt{\frac{N_s}{\sigma^2_\epsilon}} \gamma_{t}.
\end{align}
Thus weak lensing relies on averaging the shear for large numbers of source galaxies to beat down the shape noise.

\section{Shear estimation and calibration}

The accurate measurement and calibration of galaxy shapes constitute critical steps in weak lensing analyses. This section summarizes the main challenges, methodologies, and calibration techniques employed in shear estimation.

\subsection{The Point Spread Function}

A primary challenge in shear estimation are the distortions produced by the optical system, since they significantly influence observed galaxy shapes.  For telescopes in space, these distortions are due to aberrations and diffraction. When the telescope is on the ground, the light is further distorted due to the atmosphere. We encapsulate all these effects in the  \textit{point spread function} (PSF), which characterizes the response of the optical system to a point-like source. The observed image is the true image convolved with the point spread function. The observed surface brightness is distorted by the PSF  $\mathcal{P}(\vec{\theta})$:
\begin{equation}
    I^\mathrm{obs} (\vec{\theta})= \int d^2 \theta'  \mathcal{P}(\vec{\theta}) I(\vec{\theta} - \vec{\theta}')
\end{equation}
To obtain an unbiased shear estimate, it is imperative to accurately model and deconvolve the PSF from galaxy images. To learn the model, we can look at point sources such as stars, for which we know what the true image should look like.

\subsection{Ellipticity estimation}

Given the relationship between ellipticity and shear that we described in Sec.~\ref{sec:shear_ellipticity_relation}, the initial step in shear estimation involves the careful measurement of galaxy ellipticities. This process must account for the distortions introduced by the PSF while extracting the true galaxy shape information.
Model-fitting methods employ parametric models fitted to galaxy images, estimating ellipticities through moments of surface brightness. These methods are designed to disentangle the intrinsic galaxy shape from the PSF-induced distortions. For instance, the \textsc{ngmix} algorithm \citep{Sheldon2014} fits Gaussian mixture models simultaneously in the $riz$ photometric imaging bands, utilizing the \textsc{lensfit} algorithm \citep{Miller2007} for ellipticity estimation.
However, even with sophisticated fitting methods, raw ellipticity measurements remain subject to various sources of error, including PSF effects, noise, and potential biases in the fitting procedure. Consequently, rigorous calibration procedures are essential to translate these raw measurements into reliable estimates of the gravitational shear.
\subsection{Calibration methods}

Two primary approaches (which are often combined) are employed for the calibration of shear measurements:

\subsubsection{Calibration with image simulations}

This method utilizes simulations with known input shears to evaluate the accuracy of shear recovery. The quantitative metric is expressed as:
\begin{equation}\label{eq:shear_calibration}
\frac{\gamma_i^{\text{measured}}}{1-\kappa} = m_i \frac{\gamma_i^{\text{true}}}{1-\kappa} + c_i,
\end{equation}
where $m_i$ and $c_i$ represent  multiplicative and additive shear biases, respectively.

\subsubsection{Self-calibrating methods: Metacalibration}\label{sec:metacal}

\textsc{Metacalibration} \citep{Huff2017, Sheldon2017} implements a self-calibration of the mean shear using the data via a \textit{response} factor. This method applies a small known shear distortion to the image and calculates the response of a shear estimator to that applied shear\footnote{This description is simplified. In practice, the PSF must be deconvolved initially and subsequently reconvolved. For a more comprehensive description of this process, refer to \citet{Gatti2021}.}. The key idea of \textsc{Metacalibration} is that in the weak lensing regime the ellipticity $\boldsymbol{\epsilon}$ can be Taylor-expanded around zero shear:
\begin{equation}
\boldsymbol{\epsilon}(\boldsymbol{\gamma}) = \left. \boldsymbol{\epsilon}\right|_{\gamma=0} + \left.  \frac{\partial \boldsymbol{\epsilon}}{\partial \boldsymbol{\gamma}}\right|_{\gamma=0} \boldsymbol{\gamma} + ... \equiv  \left. \boldsymbol{\epsilon}\right|_{\gamma=0} +  \boldsymbol{R}_\gamma \boldsymbol{\gamma} + ... \, ,
\end{equation}
where $R_\gamma$ defines the shear response. Averaging over galaxies and inverting yields:
\begin{equation}
\left< \boldsymbol{\gamma} \right> \approx \left< \boldsymbol{R}_\gamma\right>^{-1} \left< \boldsymbol{\epsilon} \right>. 
\end{equation}
Then the shear response is measured for each galaxy by artificially shearing the images:
\begin{equation}
 R_{\gamma, i, j} =  \frac{\epsilon_i^+ -\epsilon_i^-}{\Delta \gamma_j}
\end{equation}
where $\epsilon_i^+$ and $\epsilon_i^-$ denote the ellipticity measurements on component $i$ made on an image sheared by $\pm\gamma_j$, with $\Delta \gamma_j = 2\gamma_j$. 

Also, note that while self-calibrating methods reduce dependence on image simulations, they still require additional calibration. We usually do so by measuring a residual multiplicative shear bias with image simulations, which is typically  an order of magnitude smaller compared to non-self-calibrating methods.

\paragraph{Selection response} In addition to the shear response, \textsc{Metacalibration} also accounts for selection effects through a selection response. This is crucial because any cuts made on the catalog (e.g. based on signal-to-noise ratio or size) can introduce biases that depend on shear. The selection response is calculated by measuring how these cuts change when applied to the artificially sheared images. The total response is then the sum of the shear and selection responses. This approach significantly reduces shear-dependent selection biases, which are particularly important for tomographic weak lensing analyses.

\paragraph{Blending} In crowded fields, blending of galaxy images can affect the measured shapes and the efficacy of the calibration process. The impact of blending on \textsc{Metacalibration} can be partially mitigated by employing advanced deblending algorithms or by calibrating the residual bias using image simulations that include realistic blending scenarios. Moreover, blending also impacts the redshift estimation, which can be accounted for jointly with the shear uncertainty, e.g. as in \citet{MacCrann2022}.

\paragraph*{\textsc{Metadetection}} An extension of \textsc{Metacalibration}, known as \textsc{Metadetection} \citep{Sheldon2023, Yamamoto2025}, addresses the issue of shear-dependent detection biases. While \textsc{Metacalibration} applies shears to measured objects, \textsc{Metadetection} applies shears to the original images before object detection. This allows for the calibration of biases introduced by the object detection process itself, which can be significant, especially for faint objects near the detection threshold.

\section{Weak lensing as a cosmological probe}\label{cosmo:sec:cosmological_weak_lensing}
In this section we want to relate weak lensing observables to cosmological parameters, in order to understand how we can extract cosmological information from shear measurements.

\subsection{Generalization of the lensing potential}
First it is useful to generalize the lensing potential from Eq.~(\ref{cosmo:eq:lensing_potential}) to extended lenses in redshift, i.e. beyond the thin-lens approximation, since this approximation is not appropriate to describe the lensing by the large-scale structures in the Universe. To achieve this, we only need to move the distance factors inside the integral. It will also be useful to convert the angular diameter distances we have been using so far to comoving distances. Making these changes and assuming a flat universe, the lensing potential now reads as:
\begin{align}\label{cosmo:eq:generalized_lensing_potential}
\psi (\vec{\theta})= \frac{2}{c^2  }\int_0^{\chi_s}  \frac{\chi_s - \chi_l }{\chi_s \chi_l}  \Phi' (\chi_l \vec{\theta}, \chi_l)\,  d \chi_l,
\end{align}
where we have replaced the factor of angular diameter distances by:
\begin{align}
\frac{D_{ls}}{D_l D_s} \longrightarrow (1+z_l)\frac{\chi_s- \chi_l}{\chi_s \chi_l}.
\end{align}
Note that angular diameter distances are not additive since $D_{ls}$ is defined as
\begin{align}
D_{ls} \equiv \frac{\chi_s-\chi_l}{(1+z_s)}  \qquad \text{(Flat universe)}
\end{align}
in a flat universe. To be consistent with the change, we also replaced the differential $dz$ from Eq.~(\ref{cosmo:eq:lensing_potential}), which was in proper distance, to comoving scales: $dz = 1/(1+z_l) d\chi$. Also, notice that the 3D Newtonian potential $\Phi$ is now a function of comoving distance and therefore has a different form than in Eq.~(\ref{cosmo:eq:lensing_potential}), and is labeled as $\Phi'$. Finally, even though here we are assuming a flat geometry for simplicity, the general reasoning of the rest of the section is still valid with non-zero curvature, replacing the above distance changes with general curvature ones and propagating them in every equation below. 

\subsection{Convergence as the projected matter density} \label{sec:convergence_project_matter}

Starting from the 2-D Laplacian of the lensing potential (\ref{cosmo:eq:derivatives_shear_kappa}) and the usual 3-D Poisson equation in comoving coordinates, 
\begin{align}
\kappa = \frac{\nabla^2 \psi }{2}  , \qquad \nabla^2 \Phi = \frac{4\pi G}{c^2}\bar{\rho}_m\,  a^2 \, \delta
\end{align}
together with the mean matter density
\begin{align}
\bar{\rho}_m = \frac{3H_0^2}{8\pi G} \Omega_m a^{-3} 
\end{align}
and the generalized lensing potential (\ref{cosmo:eq:generalized_lensing_potential}), we can relate the convergence $\kappa$ to the density contrast $\delta$ to obtain:
\begin{align}\label{cosmo:eq:convergence_projection}
\kappa (\vec{\theta}, \chi)= \frac{3H_0^2 \Omega_m}{2c^2} 
\int_0^{\chi} d\chi'  \frac{\chi'(\chi- \chi')}{\chi} \frac{\delta(\chi' \vec{\theta}, \chi')}{a(\chi')}.
\end{align}
$H_0$ is the expansion rate today and $\Omega_m$ is the total matter density today. 
Then we can think of the convergence $\kappa$ as the 2D projected analogue of the matter overdensity $\delta$,  weighted by the lensing efficiency factor. For a distribution of sources in comoving distance $n_\chi (\chi) $, the convergence becomes:
\begin{align}\label{cosmo:eq:convergece_projected_sources}
\kappa(\vec{\theta}) = \int_0^{\chi_{h}}  d\chi \, n_{\chi} \, (\chi) \, \kappa(\vec{\theta}, \chi) = \frac{3H_0^2 \Omega_m}{2c^2}  \int_0^{\chi_{h}} d\chi \,  g(\chi) \, \chi\, \frac{\delta( \chi\vec{\theta}, \chi)}{a(\chi)},
\end{align}
where $\chi_{h}$ is the comoving distance to the horizon and where we have defined $g(\chi)$ as the source-redshift weighted lens efficiency factor:
\begin{equation}\label{cosmo:eq: lens efficiency factor}
 g(\chi) = \int_\chi^{\chi_\text{h}} d \chi'  \frac{n_s\,(z) }{\bar{n}_s} \frac{dz}{d\chi'}\frac{\chi'- \chi}{\chi'},
\end{equation}
which indicates the lensing strength of the combined source distribution at a distance $\chi$. $n_s(z)$ is the redshift distribution of the source galaxies, $\bar{n}_s$ is the mean number density of the source galaxies.

\subsection{Angular correlation function and angular power spectrum}
\label{sec:angular_correlation}
We cannot predict the exact positions of overdensities and underdensities in the Universe, but only the statistical properties of the density field. Therefore we cannot predict the lensing effects produced by the density field along one particular line of sight. Moreover, the mean shear and the mean convergence vanish: $\left<\kappa\right> =\left<\gamma_1\right>  = \left<\gamma_2\right> =0$.  That is why we  use two-point correlation functions  to capture the signal, since we expect the shear and convergence to be correlated for regions that are close in the sky. These two-point statistics capture the Gaussian information in the matter field, which is sensitive to cosmological parameters that describe the history and content of the Universe. If the weak lensing field was Gaussian, the two-point correlation function, or its Fourier transform of the power spectrum, would capture all the information. This is not the case due to the matter fluctuations being non-Gaussian in the late-time Universe at small scales, but they are still very useful tools since they are the easiest to model analytically. They capture the degree to which lensing quantities such as the lensing potential, the deflection angle, the convergence and the shear are correlated with each other, constituting the lensing \textit{auto-correlations}, or how they correlate with the density field, comprising the so-called \textit{cross-correlations}.

\begin{BoxTypeA}[chap1:box3]{Gaussian vs non-Gaussian fields in Cosmology}

\noindent For Gaussian fields, two-point correlation functions (or power spectra) contain all the statistical information. For non-Gaussian fields, higher-order statistics (e.g., three-point functions, bispectra)  are needed to capture all the information. The weak lensing field is non-Gaussian at small scales due to non-linear structure formation. This transition from Gaussian to non-Gaussian fields as we move from the early to the late Universe highlights the need for more sophisticated statistical tools in weak lensing analyses to extract all available cosmological information as opposed to e.g. the CMB. 

Until now though the standard approach in weak lensing analyses has been to focus on two-point statistics, effectively extracting only Gaussian information. These analyses have provided headline results in cosmology, constraining key parameters and contributing to our understanding of the Universe's composition and evolution. However, this decade marks a turning point as the field begins to seriously explore methods to effectively and robustly extract information beyond two-point statistics. This shift is driven by the recognition that significant cosmological information is locked in the non-Gaussian features of the weak lensing field. It is currently an open research question to understand how much more information there actually is.

One of the leading techniques in this new frontier is simulation-based inference, see e.g. \citet{Jeffrey2021} for a comprehensive explanation and application of the method to weak lensing fields. This approach is particularly useful for higher order statistics because most of the time we cannot analytically model statistics beyond the two-point function (note three-point functions can be modeled but are often too expensive to compute to become practical). Currently analyses that aim to capture non-Gaussian information can be categorized into two main groups:
\begin{itemize}
    \item  Using a summary statistic that encapsulates higher-order information; e.g see \citet{Gatti2024} for such an application to data of second and third moments of the convergence field, wavelet phase harmonics and the scattering transform;
    \item At the map level, working directly with the full weak lensing field, such as  \citet{Jeffrey2024}, which applied convolutional neural networks to compress the convergence maps.
\end{itemize}

\end{BoxTypeA}

\begin{figure}
	\centering
	\includegraphics[width=1.\textwidth]{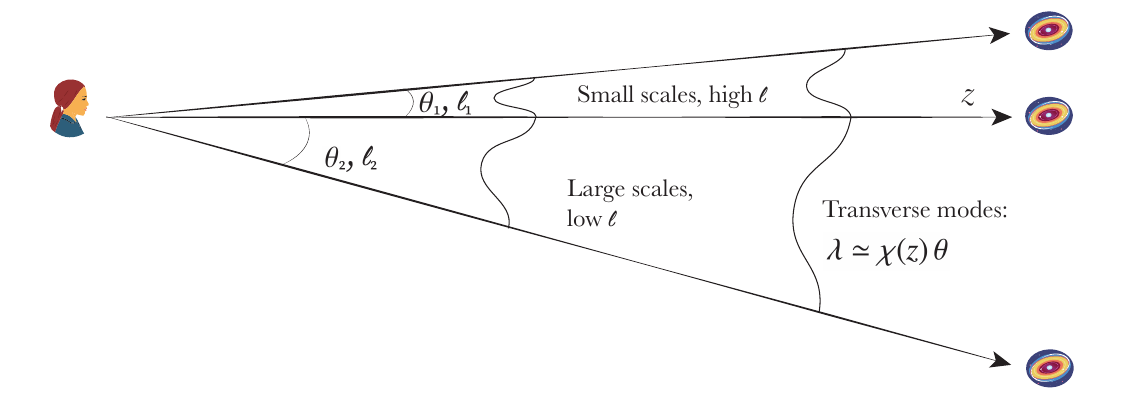}
	\caption{Sketch of the correlation of two quantities under the Limber approximation observed from two galaxies separated by a small angular separation $\theta_1$, which corresponds to high multipoles $\ell$. The Limber approximation assumes that only the modes transverse to the line of sight (the ones illustrated in the graphic that have a perturbation wavelength of $\lambda \simeq \chi(z) \, \theta $) contribute to the integral of the projection of a 3D power spectrum to its corresponding 2D quantity. This is usually a good approximation for small scales, while it can break down at the largest scales.  }
	\label{cosmo:fig:limber}
\end{figure}

Here we are interested in modeling \textit{angular two-point correlation functions}, which are the ones commonly used in photometric surveys, where the redshift is not known with enough accuracy to measure 3D correlation functions. Angular two-point correlation functions are a 2D projection of the 3D version, integrating over all galaxies in a certain redshift range, i.e. correlating 2D quantities instead of 3D ones. They are defined as:
\begin{align}
\xi_{\alpha \beta} (\theta) \equiv \left< \alpha(\vec{\theta'}) \, \beta(\vec{\theta'} + \vec{\theta}) \right>,
\end{align}
where $\alpha$ and  $\beta$ are the two quantities being correlated at the angular positions $\vec{\theta}'$ and $\vec{\theta}'+ \vec{\theta}$. If $\alpha = \beta$, and this is a lensing quantity, then $\xi$ is a lensing autocorrelation function, often referred to as  \textit{cosmic shear}. If one of the two quantities is the projected density field of foreground galaxies, $\xi$ is a cross-correlation between lensing and the density field, such as galaxy-shear correlations. The angular two-point correlation function only depends on the absolute value of $\theta$, but not on its orientation due to the assumed isotropy of the Universe. The corresponding Fourier-transform of the angular correlation function, called the \textit{angular power spectrum}, is more convenient in some situations and is defined as
\begin{align}
\mathcal{C}(\ell) = \int d^2\theta\,  \xi(\theta) \, \mathrm{e}^{-i \vec{\ell}\cdot \vec{\theta}},
\end{align}
where $\vec{\ell}$ is the two-dimensional wave vector conjugate to the angular separation $\vec{\theta}$, with $\vec{\theta} = 2\pi/\vec{\ell}$. However, to reduce computing time, quite often the lensing power spectra are calculated using the \textit{Limber approximation}, which assumes that only the modes transverse to the line-of-sight contribute (see Fig.~\ref{cosmo:fig:limber}, ignoring the modes parallel to the line of sight. Under this assumption, if the quantities $\alpha(\vec{\theta})$ and $\beta(\vec{\theta})$, defined in two dimensions, are a projection of the quantities $a(\vec{r})$ and $b(\vec{r})$, defined in three dimensions, with a window function $W(\chi)$ as in 
\begin{align}
\alpha (\vec{\theta}) = \int_0^{\chi_{h}} d\chi \, W_a(\chi)\,  a(\chi \vec{\theta}, \chi), \qquad \beta (\vec{\theta}) = \int_0^{\chi_{h}} d\chi \, W_b(\chi)\,  b(\chi \vec{\theta}, \chi),
\end{align}
then the angular cross-power spectrum of $\alpha$ and $\beta$ is given by \citep{Bartelmann2001}:
\begin{align}\label{cosmo:eq:limber}
\mathcal{C}_{\alpha \beta}(\ell) = \int_0^{\chi_{h}} d\chi \frac{W_\alpha(\chi)W_\beta(\chi)}{\chi ^2} P_{ab} \left(\frac{\ell}{\chi} \right),
\end{align}
where $P_{ab}(k)$ is the 3D cross-power spectrum of $a$ and $b$, taken at the wave number $k = \ell/\chi$. Limber's approximation holds if $a, b$ vary on length scales much smaller than the typical length scale of the window functions $W_a, W_b$, which is usually the case at most scales except for the largest ones (lowest $\ell$ multipoles). Thus, the Limber approximation allows us to compute the statistics of any projected quantity as an integral over the statistics of the 3D quantity.

\begin{BoxTypeA}{Correspondence between scales: 3D vs projected, real vs harmonic}

\noindent Under the Limber approximation a given $k$ scale in 3D harmonic space corresponds to a projected 2D scale:
\begin{equation} \label{eq:k_ell}
\ell = k \, \chi(z),
\end{equation}
where $\chi(z)$ is the comoving distance from us to the lens plane. Then, under the Limber approximation only modes transverse to the line of sight contribute, and hence $\lambda \simeq \chi(z) \, \theta$, where $\lambda$ is the wavelength of the perturbation, $\chi(z)$ is the distance from us to the perturbation (which in the case of lensing is the distance to the lens plane) and $\theta$ is the angle between the two quantities we are correlating. Then  since $\lambda = 2\pi/k$ and combining with Eq.~(\ref{eq:k_ell}):
\begin{equation} \label{eq:theta_k}
\theta \simeq 2 \pi /\ell
\end{equation}
Note that  this correspondence is not sharp, and typically a given angle $\theta$ has contributions from many scales $k$ around these values. Different two point estimators  can have different correspondences between real and harmonic scales depending on their kernel, and we find the most clear example in the cosmic shear two point estimators $\xi_+$ vs $\xi_-$, which are defined later in Eq.~(\ref{eq:xi}): $\xi_+$ includes larger physical scales than $\xi_{-}$ for the same angular separation $\theta$.

\subsection*{Limber approximation: subtleties and variations}

Actually, there are many small variations that can be made on the Limber approximation, see Table 1 from \citet{Kilbinger2017} for a comprehensive list.  Here we will summarize the most used variations.

The standard Limber approximation uses $k = \ell/\chi$, while an extended version of this approximation uses $k = (\ell+1/2)/\chi$, which changes the conversion from real to harmonic scales of Eq.~(\ref{eq:k_ell}). Moreover, as we will see later, this conversion appears both in the  power-spectrum argument and as a prefactor when converting from harmonic space correlation functions to real space two-point functions. This extended version is actually a worse approximation than standard Limber, since the approximated prefactor converges only with $\mathcal{O}(\ell)$. A hybrid version is typically more standard, with $\ell$ in the prefactor, but  $k = (\ell+ 1/2)$ in the integral. 

\end{BoxTypeA}

\subsubsection{The lensing  power spectra}

Using Limber's approximation described above, we are ready to compute the convergence power spectrum and express it in terms of the 3D matter power spectrum $P_{\delta\delta}$. This is useful because there are many efforts aimed at computing the 3D matter power spectrum, using either linear theory or non-linear fitting formulae obtained from $N$-body simulations. Equation~(\ref{cosmo:eq:convergece_projected_sources}) shows that the convergence is a projection of the density contrast $\delta$ with the following window function: 
\begin{equation} \label{eq:lensing_window}
q_s(\chi) = \frac{3H^2_0 \Omega_m }{2c^2}  \frac{\chi}{a(\chi)} g(\chi),
\end{equation}
where $a$ is the scale factor of the source and $g(\chi)$ is the lensing efficiency kernel defined in Eq.~(\ref{cosmo:eq: lens efficiency factor}). Then, using Eq.~(\ref{cosmo:eq:limber}), we obtain the \textit{convergence power spectrum} as a projection of the 3D matter power spectrum:
\begin{equation}\label{eq:Ckappakappa}
C_{\kappa \kappa} (\ell)= \int_0^{\chi_h} d\chi \frac{ q^2_s(\chi)}{\chi^2} P_{\delta \delta} \left(\frac{\ell}{\chi}\right),
\end{equation}
Moreover, a simple derivation shows that the shear and convergence have identical power spectra. To see this, we transform the defining equations for $\kappa$ and $\gamma$ from Eq.~(\ref{cosmo:eq:derivatives_shear_kappa}) into Fourier space:
\begin{align}
\tilde{\kappa}(\vec{\ell}) = -\frac{\ell^2}{2} \tilde{\psi}(\vec{\ell}), \qquad 
\tilde{\gamma}_1 (\vec{\ell}) = \frac{-(\ell_1^2 - \ell_2^2)}{2} \tilde{\psi}(\vec{\ell}), \qquad
\tilde{\gamma}_2 (\vec{\ell}) = -\ell_1 \ell_2 \tilde{\psi}(\vec{\ell})
\end{align}
Then we can compute
\begin{align} 4 |\tilde{\gamma}|^2 = \left[   (\ell_1^2-\ell_2^2)^2 + 4 \ell_1^2 \ell_2^2\right] |\tilde{\psi}|^2 = (\ell_1^2 + \ell_2^2)^2|\tilde{\psi}|^2 = 4|\tilde{\kappa}|^2,
\end{align}
which shows that the convergence and the shear power spectra are the same, $\mathcal{C}_{\kappa\kappa}  = \mathcal{C}_{\gamma\gamma}$, in the weak lensing regime.

\subsubsection{Covariance of  the lensing power spectra}

The covariance matrix for the  convergence (or shear) power spectrum can be expressed as \citep{Dodelson2017}:
\begin{equation}
\text{Cov}[C_{\kappa\kappa}(\ell), C_{\kappa\kappa}(\ell')] = \frac{2}{(2\ell+1)f_{\text{sky}}\Delta\ell} \left[C_{\kappa\kappa}(\ell) + \frac{\sigma_{\epsilon}^2}{n_\mathrm{eff}}\right]^2 \delta_{\ell\ell'},
\end{equation}
where $f_{\text{sky}}$ is the fraction of the sky covered by the survey, $\Delta\ell$ is the width of the multipole bin (just one if we are estimating the covariance for a single multipole), $\sigma_{\epsilon}$ is the intrinsic ellipticity dispersion of source galaxies, defined in Eq.~(\ref{eq:sigma_e}),  $n_\mathrm{eff}$ is the effective number density of source galaxies (see highlighted text  box below) and $\delta_{\ell\ell'}$ is the Kronecker delta function. The first term in the square brackets represents the \textit{cosmic variance}, which is a noise term proportional to the signal; while the second term accounts for the \textit{shape noise}, discussed in Sec.~\ref{cosmo:sec:shear_signal_noise}. The cosmic variance  comes from the fact that for each $\ell$ we use a finite number of modes to estimate the power spectra, i.e. there are only a certain number of galaxies that will be separated by the distance corresponding to a given $\ell$. That is also why  $f_{\text{sky}}$ enters the equation. The larger the survey area,  the more modes that we can find in a given $\ell$, and thus the smaller the errorbars. Moreover, we cannot probe the modes that are beyond the area of a survey. Thus, the cosmic variance term is larger at large scale separations (smaller multipoles $\ell$), while the shape noise term becomes more significant at smaller angular scales (higher multipoles), and for surveys with lower galaxy number densities. It is also worth noting that the equation above only holds for Gaussian fields, which is a good approximation at large-scales. At small scales, nonlinearities introduce non-Gaussianities, and additional terms in both the data vector modeling and its corresponding covariance need to be considered.

\begin{BoxTypeA}{Estimating $n_\mathrm{eff}$ and  $\sigma^2_\epsilon$ in practice}\label{box:ns}

\noindent Most shear catalogs have a weight factor per galaxy. The weights are commonly used to increase the overall signal-to-noise of the measurement, by applying typically an inverse variance weighting scheme.  Moreover, under the \textsc{Metacalibration} or \textsc{Metadetection} scheme described in Sec.~\ref{sec:metacal} we also have a response factor associated with each source galaxy. The presence of  weights and responses complicates the computation of the number density of sources and the variance of the ellipticity components in practice. Below we give the state-of-the-art equations that are used in the most recent weak lensing analyses. From the definition in \citet{Heymans2012}, the effective number density of the sources is:
\begin{equation}
n_\mathrm{eff} = \frac{1}{A} \frac{\left(\sum_i w_i \right)^2 }{\sum_i w^2_i }
\end{equation}
where $A$ is the  effective area of the survey footprint and $w_i$ is the weight for the galaxy $i$. Also from \citet{Heymans2012} a common expression for the effective variance of the combined ellipticity is:
\begin{equation}
\sigma_{\epsilon}^2 = \frac{1}{2} \left[ \frac{\Sigma w_i^2(e_{i,1}-\langle e_1 \rangle)^2}{(\Sigma w_i R)^2} + \frac{\Sigma w_i^2(e_{i,2}-\langle e_2 \rangle)^2}{(\Sigma w_i R)^2}\right] \frac{(\Sigma w_i)^2}{\Sigma w_i^2} 
\end{equation}
where the mean shear is subtracted first. Also, sometimes the variance per ellipticity component is given, so one needs to be careful to input the proper value to the covariance estimation to avoid this common mistake. Note also that for harmonic space estimators or other kinds of map-based analyses, $\sigma^2_{\epsilon}$  needs to be estimated per pixel, instead of per galaxy, which leads to a different estimator in practice. 

An additional complication is that $\sigma^2_\epsilon$ and $n_\mathrm{eff}$ can vary as a function of angular scale, since different pairs of galaxies are used in each angular bin. This often needs to be tested to ensure it is negligible for a given analysis.

\end{BoxTypeA}

\subsubsection{Lensing real space two-point function}

For cosmic shear the real-space two-point correlation function is related to the angular power spectrum from Eq.~(\ref{eq:Cgammagamma}) via a Hankel transform \citep{Bartelmann2001}:
    \begin{equation} 
    \xi_{\pm}(\theta) = \int \frac{d\ell \ell}{2\pi} C_{\kappa \kappa} J_{0/4}(\ell \theta),
    \label{eq:xi}
    \end{equation}
under the flat-sky and the standard Limber approximation. Here the $J_{0/4}$ represent the Bessel functions of the first kind, where $J_{0}$ and $J_{4}$ are used for $\xi_{+}$ and $\xi_{-}$ respectively. We can also compute the real space correlation functions using a spherical transformation, also sometimes referred to as  \textit{full sky} or \textit{curved sky} transformation. In this case we need to  decompose $\kappa$ into E- and B-mode components as in Eq.(\ref{eq:EB_modes_kappa}) and then \citep{Secco2022}:
\begin{equation}
\xi_{\pm}(\theta) = \sum_{\ell} \frac{2\ell + 1}{2\pi\ell^2 (\ell + 1)^2} 
[G_{\ell,2}^+(\cos\theta) \pm G_{\ell,2}^-(\cos\theta)]
\times [C_{EE}(\ell) \pm C_{BB}(\ell)],
\end{equation}
where the functions $G_\ell^\pm(x)$ are computed from Legendre polynomials $P_\ell(x)$ and averaged over angular bins (see \citealt{Krause2021} Eqs. 19 and 20). This is more accurate but slower, so this choice will be determined by the area and precision of a given survey and analysis.

Note that besides $\xi_\pm$ there are other real space quantities that are often used for cosmic shear analyses. COSEBIs (Complete Orthogonal Sets of E/B-mode Integrals, \citealt{Schneider2010}) are among the most popular alternatives, which have the advantage of cleanly separating E and B modes while being less sensitive to small-scale systematics. 

\begin{figure}
	\centering
	\includegraphics[width=1.\textwidth]{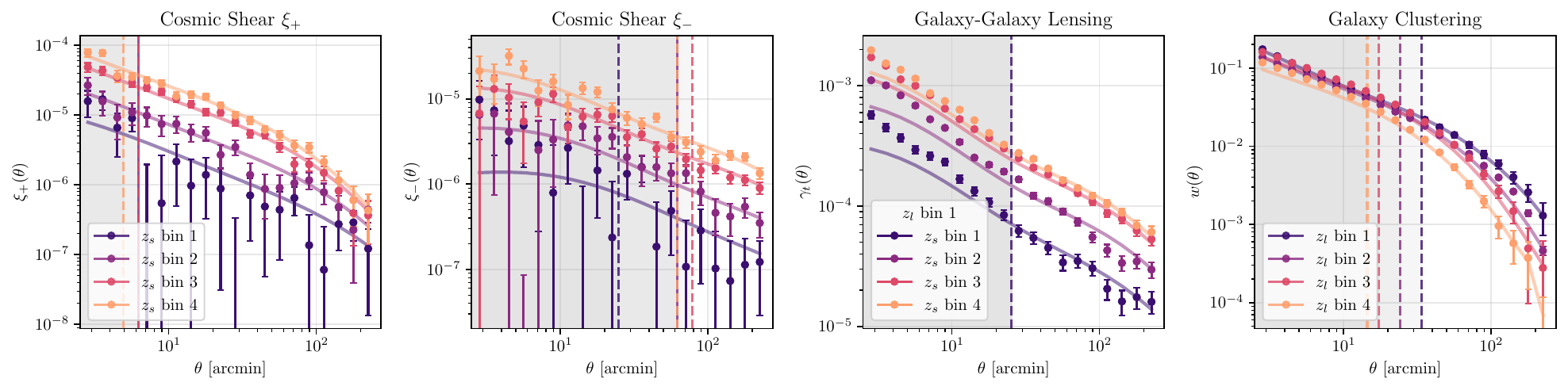}
	\caption{Measured 3$\times$2pt datavector and best-fit model in real space for the Dark Energy Survey Y3 dataset. Cosmic shear correlation functions $\xi_{+} (\theta)$ and $\xi_{-}(\theta)$ on the left for the autocorrelations (\citealt{Amon2022, Secco2022}), the mean tangential shear for galaxy-galaxy lensing in the middle right for the first lens bin \citep{Prat2022} and angular clustering on the right panel \citep{Monroy2022}. The shaded regions represent the scale cuts applied in the analysis to mitigate non-linearities in the matter power spectrum and the galaxy-matter connection, as well as baryonic effects.  }
	\label{cosmo:fig:desy3-dv}
\end{figure}

\subsubsection{Measurements of the two-point correlation functions: real space estimators}

In this section we summarize how we estimate in practice such correlation functions. $\xi_{\pm}$ is defined in terms of the tangential ($t$) and cross ($\times$) components of the ellipticity $\hat{\epsilon}$ defined along the line that connects each pair of galaxies $a,b$:
\begin{equation}
\xi_{\pm}(\theta)=
\frac{\underset{ab}{\sum} w_a w_b \left(\hat{\epsilon}_{t,a}\,\hat{\epsilon}_{t,b}
\pm
\hat{\epsilon}_{\times,a}\,\hat{\epsilon}_{\times,b}\right)}
{\underset{ab}{\sum}w_{a}w_{b},
},\label{eq:xipm_estimator}
\end{equation} 
\noindent
with inverse variance weighting $w$ and  where the sums run over pairs of galaxies $a,b$, for which the angular separation falls within the range $|\boldsymbol{\theta}-\delta \vec{\theta}|$ and $|\boldsymbol{\theta}+\delta \vec{\theta}|$. The ellipticities that enter Eq.~\eqref{eq:xipm_estimator} are typically corrected for residual mean shear, such that $\hat{\epsilon}_k\equiv \epsilon_k - \langle \epsilon_k \rangle$ for components $k\in (1,2)$ and a given sample of galaxies. In Fig.~\ref{cosmo:fig:desy3-dv} we show state-of-the art cosmic shear measurements.

\subsection{Weak lensing surveys}

\begin{table}[t]
\caption{Characteristics of major current and forthcoming weak lensing surveys. }
\label{tab:surveys}
\begin{tabular*}{\textwidth}{@{\extracolsep{\fill}}lllclc}
\toprule
 & Survey & Area (deg$^2$) & Source density [arcmin$^{-2}$] & Redshift range & Observing Status \\
\midrule
Stage III & DES Y3 & 4,143 & 5.59 & 0.2 - 1.3 & Completed \\
 & KiDS-1000 & 777 & 6.17 & 0.1 - 1.2 & Completed \\
 & HSC Y3  & 416 & 15  & 0.3 - 1.5 & Ongoing \\
\midrule
Stage IV & Euclid & 14,000 & $\sim$27 & 0 - 2+ & Ongoing \\
 & LSST & 18,000 & $\sim$27 & 0 - 3+ & Future (2025+) \\
 & Roman & 2,000 &  $\sim$50 & 0 - 3+ & Future (2027+) \\
\bottomrule
\end{tabular*}
\end{table}

\begin{figure}
    \centering
    \includegraphics[width=0.6\textwidth]{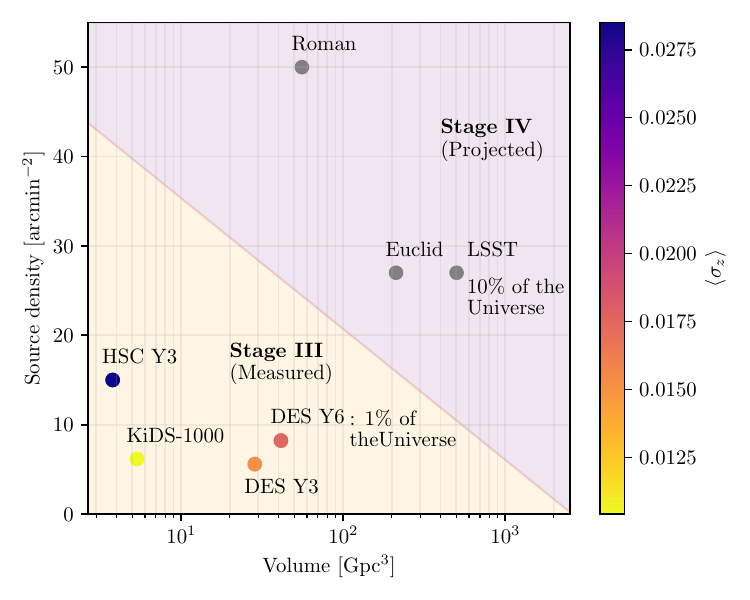}
    \caption{Source density for current and upcoming weak lensing surveys, shown against the volume they probe. The background shading separates Stage III from Stage IV  surveys. The color scale indicates the average photometric redshift uncertainty as measured in Stage III surveys. DES Y6 probes 1\% of the Universe's volume out to $z=15$, while LSST will reach 10\%.}
    \label{cosmo:fig:volume_density}
\end{figure}

\begin{figure}
    \centering
    \includegraphics[width=\textwidth]{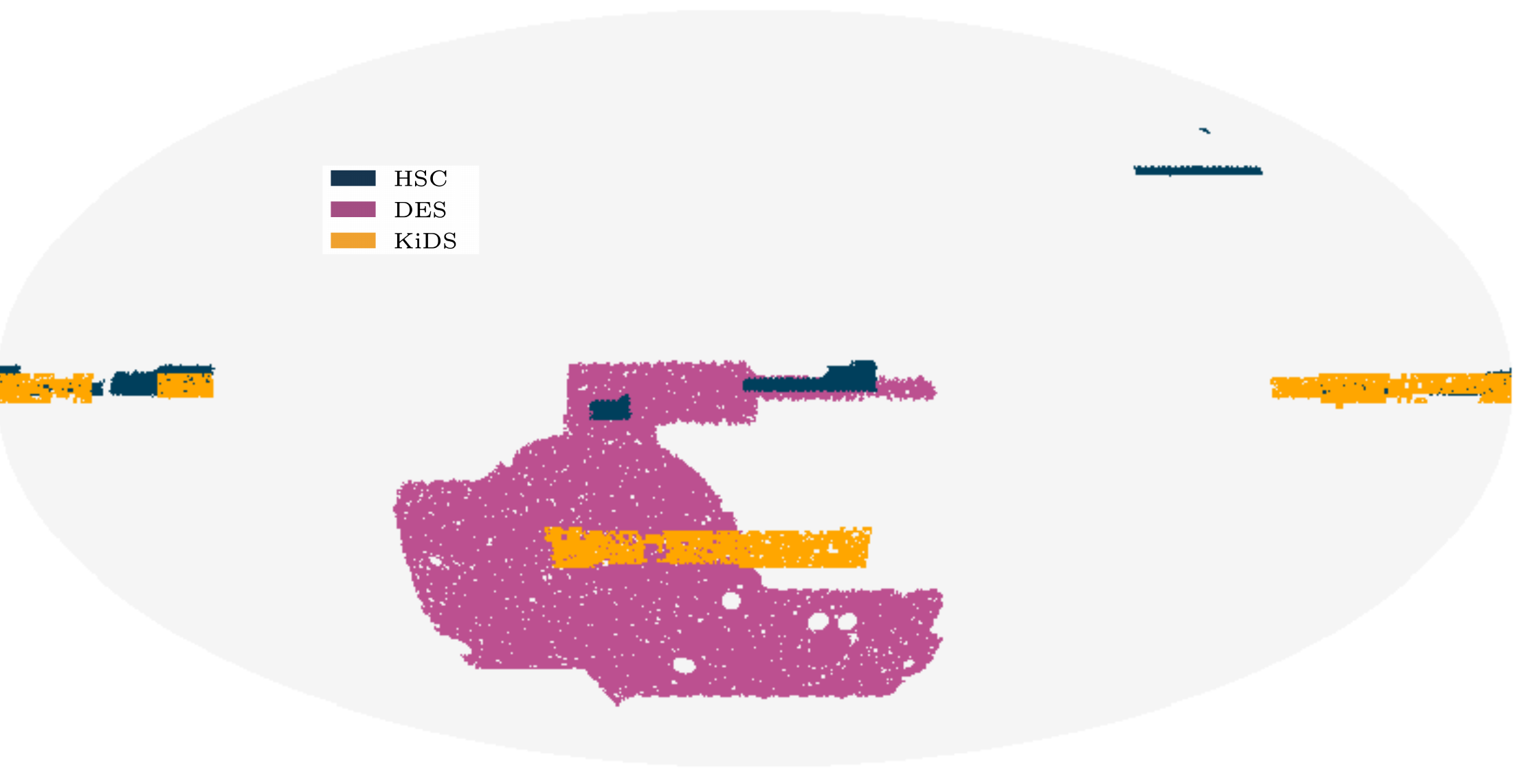}
    \caption{Footprints of the main Stage III Weak lensing surveys, for HSC Y3, KiDS-100 and DES Y3 releases.}
    \label{cosmo:fig:footprints}
\end{figure}

To provide context for the current state and future direction of weak lensing research, it is helpful to overview the available and upcoming surveys. The Dark Energy Survey \citep[DES,][]{Flaugher2005}, Hyper Suprime-Cam survey \citep[HSC,][]{Aihara2018}, and the Kilo-Degree Survey \citep[KiDS,][]{deJong2013} are classified as Stage-III\footnote{The Stage-III and Stage-IV classification was introduced in the Dark Energy Task Force report \citep{Albrecht2006}. Stage-III refers to dark energy experiments that began in the 2010s, while Stage-IV denotes those starting in the 2020s.} weak lensing surveys. These are all wide-field photometric surveys, meaning they capture images of large areas of the sky through multiple color filters. This approach allows for efficient observation of millions of galaxies and estimation of their redshifts based on their observed colors. This process, known as photometric redshift estimation, is sophisticated and demands careful calibration of uncertainties. State-of-the-art techniques for estimating photometric redshifts involve complex algorithms and extensive validation (see, e.g., \citealt{Myles2021} for a comprehensive overview of how this process was done for the DES analysis of the first three years of observations, DES Y3).

We are now entering the era of Stage-IV galaxy surveys. The ESA satellite Euclid \citep{Euclid2024} has already launched, while the Rubin Observatory Legacy Survey of Space and Time \citep[LSST,][]{Ivezic2019} and NASA's Nancy Grace Roman Space Telescope \citep{Akeson2019} are set to begin operations in the coming years. These next-generation surveys will also employ photometric techniques, but with enhanced capabilities in terms of sky coverage, depth, and resolution.

\begin{figure}
    \centering
    \includegraphics[width=0.7\textwidth]{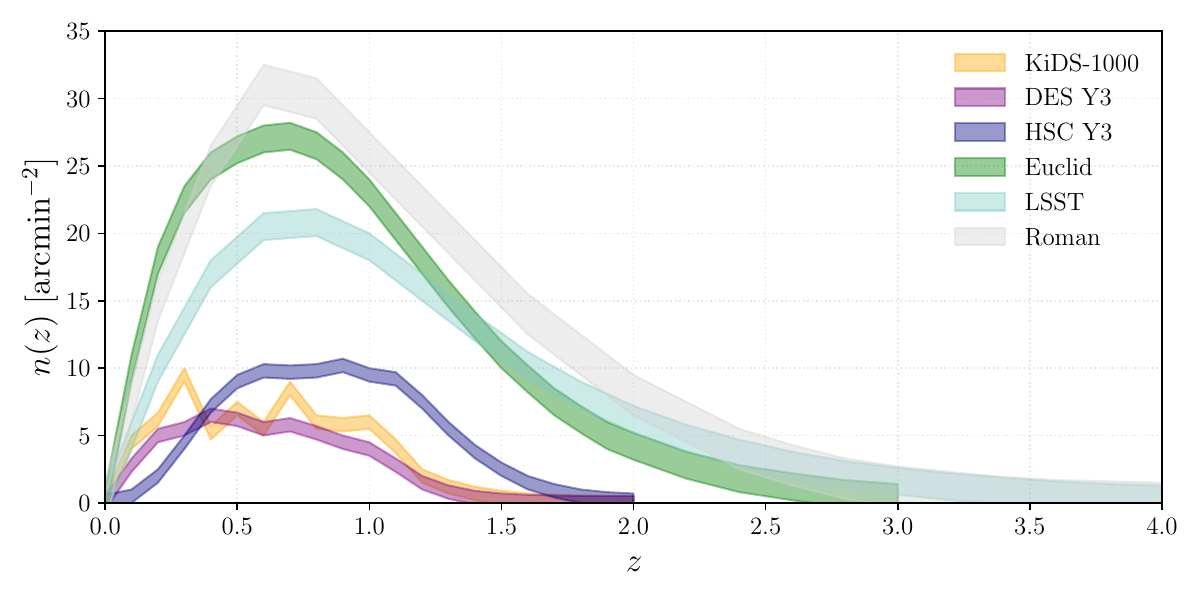}
    \caption{Redshift distribution $n(z)$ of galaxies for major current (as measured) and future weak lensing surveys (forecasted). Distributions are normalised to the mean number density of sources used in the lensing analyses. This comparison illustrates the significant leap in galaxy sampling and redshift coverage anticipated from next-generation surveys, which will enable more precise cosmological measurements and enhanced sensitivity to the growth of structure over cosmic time.}
    \label{cosmo:fig:nzs}
\end{figure}

Table~\ref{tab:surveys} and Fig.~\ref{cosmo:fig:volume_density} summarize the main specifications of each survey. Additionally, we illustrate the survey footprints in Fig.~\ref{cosmo:fig:footprints} and the redshift distributions in Fig.~\ref{cosmo:fig:nzs}. The data for these figures and specifications are sourced as follows: DES Y3 shape catalog from \citet{Gatti2021, Myles2021}, DES Y6  from \citet{Yamamoto2025} and Yin et al. (in prep), KiDS-1000 catalog from \citet{Giblin2021, Asgari2021}, HSC-Y3 catalog from \citet{Dalal2023}, Euclid forecasts from the recent \citet{Euclid2024}, LSST forecasts from \citet{DESC2018}, and Roman forecasts from \citet{Eifler2021}.

\subsection{What does cosmic shear tell us about the Universe?}\label{y1key:subsec:params}

Measurements of such two-point correlation functions are then typically used to estimate cosmological parameters in the framework of a given cosmological model. The most popular models include  flat $\Lambda$CDM, the current Standard Cosmological model and flat $w$CDM, a straightforward extension to it that allows for evolving dark energy. As a summary, $\Lambda$CDM contains three free energy densities in units of the critical density that evolve with time: the cold dark matter, baryon, and massive neutrino energy densities\footnote{The radiation component is usually neglected because its density is much smaller compared to the others at the present time given the observed Cosmic Microwave Background temperature.}, $\Omega_{\mathrm{CDM}}, \Omega_b,$ and $\Omega_\nu$, while keeping the dark energy component constant with time as a cosmological constant. We will also refer to the total matter density $\Omega_m$, which is the sum of  $\Omega_{\mathrm{CDM}}, \Omega_b,$ and $\Omega_\nu$. Within the $w$CDM model, the dark energy equation of state parameter $w$  is taken as an additional free parameter instead of being fixed at $w=-1$ as in $\Lambda$CDM. $w$CDM thus contains seven cosmological parameters, while $\Lambda$CDM has six. Other typical extensions include  models in which $w$ is allowed to vary in time or the assumption of flatness is released, where $\Omega_m + \Omega_\Lambda$ can differ from unity.  

Cosmic shear is sensitive to the amplitude of the power spectrum in the late Universe.  In the early Universe, we use $A_s$ to describe the amplitude of the primordial power spectrum. For the late times, it is common to replace $A_s$ with the RMS amplitude of mass fluctuations on an 8 $h^{-1}$ Mpc scale in linear theory, $\sigma_8$, since this is somewhat closer to what we can measure from the late-time Universe. $\sigma_8$ is  defined as:
\begin{equation}\label{cosmo:eq:sigma_8}
\sigma_8^2 = \frac{1}{2\pi^2} \,\int\, k^2 P(k)\, \left[ \frac{3j_1(kR)}{kR}\right]^2 \,dk,
\end{equation}
where $j_1(kR)$ is the spherical Bessel function of the first kind of order one. In general we can write the following expression to quantify fluctuations for the smoothed density field in a given scale $R$:
\begin{equation}
\sigma_R^2 \equiv \left< \delta_R^2(x)\right> = \frac{1}{2\pi^2}\, \int\, k^2 P(k) \,|\tilde{W}(kR)|^2 dk.
\end{equation}
where $W(\vec{x}')$ is the window function that smooths the density fluctuations on a given sphere in the following way:
\begin{equation}
\delta(\vec{x}) = \int d^3 \vec{x}'\  W(\vec{x}') \delta(\vec{x} + \vec{x}').
\end{equation}
The most common window function is a spherical top-hat of radius $R$: $W_R(r) = 3/(4\pi R^3)$ for $r<R$ and $W_R(r)=0$ for $r>R$.  The Fourier transform of this particular window function is $\tilde{W}_R(kR) = 3j_1(kR)/(kR)$, which is the factor that appears in the $\sigma_8$ definition. Then the case $R = 8$ Mpc/h, using the Fourier transform of the top-hat as the window function is the one that defines the famous cosmological parameter $\sigma_8$. The reason for choosing this particular scale is that the relative fluctuations of the galaxy number density in the local Universe are of order unity if one considers spheres of radius $R = 8$ Mpc/h, which is the typical scale of massive galaxy clusters. 

\begin{figure}
	\centering
	\includegraphics[width=0.8\textwidth]{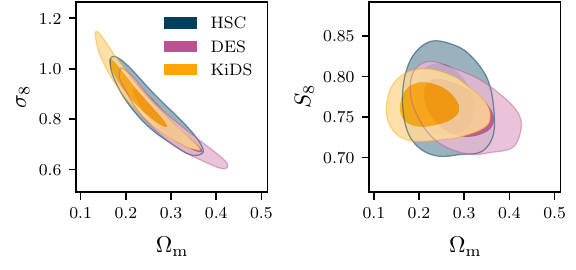}
	\caption{This figure presents cosmological constraints from cosmic shear measurements for the three primary Weak Lensing Stage III surveys assuming the $\Lambda$CDM model. The contours show results from real-space two-point correlation analyses of cosmic shear data. The Hyper Suprime-Cam (HSC) results are based on their first three years of observations (HSC Y3), similar to the Dark Energy Survey (DES) which also uses their Year 3 data (DES Y3). The Kilo-Degree Survey (KiDS) results are derived from the KiDS-1000 data release. These results represent the current state-of-the-art in weak lensing cosmology.}
	\label{cosmo:fig:cosmo_cosmic_shear}
\end{figure}

However, for analyses  from cosmic shear two-point correlation functions, $\sigma_8$ and $\Omega_m$ are usually very correlated. Therefore, when reporting constraints for cosmic shear it is useful to define another parameter that removes this correlation and it  much better constrained:
\begin{equation}
S_8\equiv \sigma_8 \left( \frac{\Omega_m}{0.3}\right)^{0.5}, \label{y1key:eq:s8}
\end{equation}
where hereafter $\Omega_m$ is defined as the total matter density today.
In Fig.~\ref{cosmo:fig:cosmo_cosmic_shear} we show the latest results on $\Omega_m$, $\sigma_8$ and $S_8$ for the three main Stage III weak lensing surveys assuming the $\Lambda$CDM model. Notice the correlation between $\sigma_8$ and $\Omega_m$ on the left panel, and the decorrelated parameter $S_8$ on the right one.

\subsection{Bayesian Inference: How do we actually obtain constraints on cosmological parameters? }
From a statistical point of view, the weak lensing and clustering observations are a random sample from an unknown population, and the goal of a statistical analysis is to infer the population that is most likely to have generated the sample, specifically the probability distribution corresponding to the population. For instance, in cosmology, and specifically here, the two-point correlation functions measured from the data are one sample of all possible two-point correlation functions that could be measured from all realizations of the density field. This information is encapsulated in the  \textit{likelihood} $\mathcal{L}$ function, which is the probability that a given experiment would get the data $\vec{D}$ it did given a theory, parametrized by the parameters $\vec{p}$: 
\begin{equation}
\mathcal{L}(\vec{p})\equiv  \mathcal{P}(\vec{D}\, |\,  \vec{p}\, ).
\end{equation}
Assuming the two-point correlation functions, which constitute our observations, are multivariate random variables, we can write the likelihood function proportional to a multivariate Gaussian with covariance $C$ between the measurements:
\begin{equation} \label{y1key:eq:likelihood}
 \mathcal{L}(\vec{p}) \propto e^{-\frac{1}{2}\sum_{ij} \left[D_i-T_i(\vec p)\right]\, C^{-1}{}_{ij} \left[D_j-T_j(\vec p)\right] },
\end{equation}
where $\vec p$ includes both  cosmological and nuisance parameters\footnote{In addition to the cosmological parameters,  models typically include on the order of $\sim$20 nuisance parameters. See Section~\ref{sec:3x2pt} for more information on such parameters.} and $T_i(\vec p)$ are the theoretical predictions for the two-point functions.

Eventually our objective is to infer cosmological parameters from a set of measurements, in this case the two-point correlation functions. Thus, we are interested in the probability that the theoretical parameters $\vec{p}$ take some value given a set of measurements $\vec{D}$, i.e. we would like to have $\mathcal{P}( \vec{p}\, |\,  \vec{D} )$. Conveniently, we can use \textit{Bayes' theorem} to convert the likelihood function  into the probability we are actually interested in:
\begin{equation}
\mathcal{P}(\vec{p}\, |\,  \vec{D} ) = \frac{\mathcal{P}(\vec{D}\, |\,  \vec{p}\, )\,  \mathcal{P}(\vec{p})}{\mathcal{P}(\vec{D})} \propto  \mathcal{L}(\vec{p}) \mathcal{P}(\vec{p}),
\end{equation}
where $\mathcal{P}(\vec{p})$ is called the $\textit{prior}$, which encloses the previous information we might have on the parameters $\vec{p}$. The quantity on the left side of the equation is called the \textit{posterior} distribution, which is what we want to obtain, and the denominator is called the \textit{evidence}, which is not relevant for parameter inference, but becomes important when carrying out model comparison. Then, we can determine the parameters $\vec p$ of the theory along with their errors by sampling the likelihood in the many-dimensional parameter space using a Markov Chain Monte Carlo  (MCMC) and then multiplying the likelihood by the prior $\mathcal{P}(\vec p)$. In particular, it is common to sample the logarithm of the likelihood function instead of the likelihood itself, since otherwise it is easy to run into numerical issues:
\begin{equation} 
\ln\mathcal{L}(\vec p) = -\frac{1}{2}\sum_{ij} \left[D_i-T_i(\vec p)\right] C^{-1}{}_{ij} \left[D_j-T_j(\vec p)\right] .\label{y1key:eq:likeli}
\end{equation}

\begin{BoxTypeA}{Gaussian likelihood}\label{box:gaussian_likelihood}

\noindent Both in Eq.~(\ref{y1key:eq:likelihood}) and in (\ref{y1key:eq:likeli}) we are assuming $D_i$ are drawn from a Gaussian distribution. This is a very common assumption and it is a generally a good approximation for large scales \citep{Sellentin2018}.  However, in upcoming observations when the statistical uncertainties become much smaller, or when using higher order statistics or small scales, this assumption might not be good enough and should be tested. One approach to do this is validation  using the so-called \textit{simulation-based inference} framework, see e.g. Fig. 9 from \citet{Gatti2024sims}. 
\end{BoxTypeA}

\subsection{The 3$\times$2pt combination} \label{sec:3x2pt}

\begin{figure}
	\centering
	\includegraphics[width=0.7\textwidth]{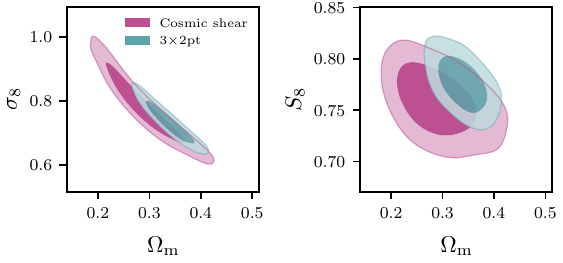}
	\caption{This figure compares the cosmological constraints from   cosmic shear with the 3$\times$2pt combination  for the analysis with the first three years of observations of the Dark Energy Survey (DES Y3) assuming the $\Lambda$CDM model \citep{desy3-3x2}.}
	\label{cosmo:fig:cosmo_3x2}
\end{figure}

 Recent analyses of galaxy surveys have  shown that it is even more effective to combine galaxy clustering and weak lensing in a multi-probe approach to jointly infer cosmology.  In particular, a common approach is to combine three two-point functions of the galaxy density field $\delta_{g}$ and the weak lensing shear field $\gamma$: galaxy clustering, galaxy-galaxy lensing  and cosmic shear. In these measurements, we usually refer to the galaxy sample used for $\delta_{g}$ as the {\it lens galaxies}, and the sample used for weak lensing as the \textit{source galaxies}.  
This combination has been recognized for more than a decade to contain a tremendous amount of complementary information, and to be remarkably resilient to the presence of nuisance parameters that describe systematic errors and non-cosmological information \citep{Hu:2003pt,Bernstein2009,Joachimi:2009ez,Nicola:2016eua}. 

At the present time, the 3$\times$2pt analysis has become a standard tool in modern cosmology and the main  weak lensing surveys have all presented results using this combination. For the most recent analyses, the results can be found for DES in \citet{desy3-3x2}, for KiDS in \citet{kids1000}, and for HSC in \citet{hscy3-3x2}\footnote{Note that each of the collaborations still need to release one more round of analyses, which will constitute the  results using the final data sets of each of the surveys.}. We show cosmological results from cosmic shear two-point correlation functions in comparison with 3$\times$2pt in Fig.~\ref{cosmo:fig:cosmo_3x2} for the DES Y3 analysis. One important distinction is that 3$\times$2pt is able to break the degeneracy between $\sigma_8$ and $\Omega_m$ and can place constraints on $\Omega_m$, while cosmic shear is not very sensitive to $\Omega_m$ alone.

These surveys, while all employing the 3$\times$2pt analysis, differ in their specific setups, particularly in their choice of lens samples. DES used two distinct lens samples: a magnitude-limited sample called \textsc{MagLim} and a red sequence-based sample named \textsc{redMaGiC}. HSC  also employed a color-selected sample of luminous red galaxies (LRGs) as their lens population.  In contrast, KiDS  utilized external spectroscopic data from the Baryon Oscillation Spectroscopic Survey (BOSS) and the 2-degree Field Lensing Survey (2dFLenS) for their lens sample. These choices significantly affect the resulting constraining power and need to be held in mind when considering the $3\times$2pt  results of each survey, which we show in Fig.~\ref{fig:3x2t_vs_planck}. 

This section describes the ingredients needed to perform a tomographic\footnote{Typically we do not only use one sample of galaxies to extract cosmological information from weak lensing measurements, but we instead slice the samples into redshift bins; this is usually referred to as tomography.} $3\times2$pt analysis, including a discussion of the major systematic effects that must be accounted for in current analyses to produce the latest results.  Some equations might be redundant with the previous cosmic shear section, but we put it all together for clarity here. 

The Limber approximation is used, see highlighted text in Sec.~\ref{sec:angular_correlation}; and assuming a flat Universe cosmology, the three two-point correlation functions involved in a $3\times2$pt analysis in harmonic space are:
\begin{equation}\label{eq:Cgammagamma}
C^{ij}_{\gamma \gamma} (\ell)= \int d\chi \frac{ q_s^i(\chi)\,  q_s^j(\chi)}{\chi^2} P_{\delta \delta} \left(k = \frac{\ell}{\chi},z(\chi)\right),
\end{equation}
\begin{equation}\label{eq:C_gkappa}
    C_{\delta_{g}\gamma}^{ij}(\ell) = \int d\chi \frac{N_l^i(\chi)\, q_s^j(\chi)}{\chi^2}P_{\rm g \delta}\left(k = \frac{\ell}{\chi},z(\chi)\right)\,.
\end{equation}
\begin{equation}\label{eq:C_gg}
    C_{\delta_{g}\delta_{g}}^{ij}(\ell) = \int d\chi \frac{N_l^i(\chi)\, N_l^j(\chi)}{\chi^2}P_{\rm gg}\left(k = \frac{\ell}{\chi},z(\chi)\right)\,.
\end{equation}
The first expression is the one we were already familiar with from the previous cosmic shear sections; the second one is the galaxy-galaxy lensing or galaxy-shear correlation, and the third one is the galaxy clustering harmonic space correlation function. $i$, $j$ correspond to the tomographic bins, which for cosmic shear involves two source redshift bins, for galaxy clustering two lens bins, and for galaxy-galaxy lensing a lens bin $i$ and a source bin $j$.  $q_s(\chi)$ is the window function of the given source population of galaxies, as defined in Eq.~(\ref{eq:lensing_window}), 
and $N_l(\chi)$ is the lens window function:
\begin{equation}
N_l^i(\chi) = \frac{n^i_l\,(z)}{\bar{n}^i_l}\frac{dz}{d\chi}, 
\end{equation}
where $n^i_l$ is the lens redshift distribution and $\bar{n}^i_l$ is the mean number density of the lens galaxies. 

To infer cosmological information from these two-point measurements, we model these functions in terms of the \textit{cosmological parameters}, but we also need to add some \textit{nuisance parameters} to account for astrophysical and observational uncertainties. Below we describe in detail the nuisance parameters that are most commonly considered in such analyses, that model  uncertainties in photometric redshifts, shear calibration, the galaxy bias, and the contribution of intrinsic alignment to the shear spectra. 

\subsubsection{Galaxy bias model}

Galaxies are biased tracers of the underlying dark matter field.
The galaxy bias model captures the statistical relation between galaxies and dark matter. For this relation, the simplest model that can be assumed is the  linear biasing model ($\delta_g = b \:\delta_m$, where $b$ is modeled as a constant for each redshift bin), so the galaxy power spectrum and the galaxy-matter power spectrum relate to the matter power spectrum by different factors of the galaxy bias:
\begin{align}
\label{eq:linearbias}
 P_{\rm gg} &= b^2  P_{\delta \delta}, \\
 P_{\rm g\delta} &= b \,  P_{\delta \delta}.
\end{align} 
The linear biasing model, while widely used for its simplicity, has several important limitations. In reality, galaxy bias is scale-dependent, especially on smaller scales where non-linear clustering becomes significant. Secondly, this model neglects higher-order clustering effects and assembly bias, when halo properties depend on properties beyond just their mass, such as formation history or environment, which can be particularly important for certain galaxy populations. Furthermore, it does not account for stochastic bias, where there is not a deterministic relationship between galaxy and matter densities.  See \citet{Wechsler2018} for a comprehensive review on the galaxy-halo connection and a description of some of the limitations and extended models to smaller scales.  However, currently due to a lack of precise enough knowledge in the galaxy bias relation at small scales, analyses are often restricted to larger scales where the linear bias approximation remains valid, discarding valuable information from smaller scales. One possible approach is to define a scale cut that ensures the systematic bias in cosmological parameter estimation remains below an acceptable level (e.g., $0.3\sigma_{2D}$ for the $2D$ $\Omega_m - S_8$ parameter space).

\subsubsection{Intrinsic alignment model} \label{sec:IA}

Intrinsic alignments (IAs) are a crucial systematic effect in weak lensing analyses that arise from correlations between the intrinsic shapes of galaxies. These alignments can mimic the signal from gravitational lensing, potentially leading to biases in cosmological parameter estimation if not properly accounted for. For a comprehensive guide on this important and complex effect, readers are directed to \citet{Lamman2024}. In this section, we will introduce the concept of IAs and describe how they are typically modeled in cosmic shear and galaxy-galaxy lensing analyses.

The impact of IAs on the two-point correlation functions manifests as additional terms in the observed power spectra. For cosmic shear, the observed angular power spectrum $C^{ij}_{GG} (\ell)$ between tomographic bins $i$ and $j$ can be expressed as:
\begin{equation}
C^{ij}{GG} (\ell)= C^{ij}_{\gamma \gamma} (\ell) + C^{ij}_{II} (\ell) + C^{ij}_{\gamma I} (\ell) + C^{ji}_{\gamma I } (\ell) \ ,
\end{equation}
Here, $C^{ij}_{\gamma \gamma} (\ell)$ represents the pure cosmic shear signal, while the additional terms account for different IA contributions. $C^{ij}_{II} (\ell)$ describes the correlation between intrinsic shapes of physically close galaxies (II term), and $C^{ij}_{\gamma I} (\ell)$ represents the correlation between the intrinsic shape of a foreground galaxy and the shear experienced by a background galaxy (GI term).
Similarly, for galaxy-galaxy lensing, the observed angular power spectrum $C_{\delta_{g}G}^{ij}(\ell)$ includes an IA contribution:
\begin{equation}
C_{\delta_{g}G}^{ij}(\ell)  =  C_{\delta_{g} \gamma}^{ij}(\ell)  +  C_{\delta_{g}I}^{ij}(\ell),
\end{equation}
where $G$ represents the total observed shape of galaxies, encompassing both the cosmological shear $\gamma$ and the intrinsic shape $I$.
To model these IA contributions, we need to project the relevant three-dimensional power spectra onto the sky. The projections for the II, GI, and galaxy-IA terms are given by:
\begin{equation}
C^{ij}_{II} (\ell) = \int d\chi \frac{ N_s^i(\chi) \,  N_s^j(\chi)}{\chi^2} P_{\rm II} \left(k = \frac{\ell}{\chi},z(\chi)\right)
\end{equation}
\begin{equation}
C^{ij}_{\gamma I} (\ell) = \int d\chi \frac{ q_s^i(\chi) \,   N_s^j(\chi)}{\chi^2} P_{\rm \gamma
I} \left(k = \frac{\ell}{\chi},z(\chi)\right)
\end{equation}
\begin{equation}
C^{ij}_{\delta{g} I} (\ell) = \int d\chi \frac{ N_l^i(\chi) \,  N_s^j(\chi)}{\chi^2} P_{\rm \gamma  I} \left(k = \frac{\ell}{\chi},z(\chi)\right).
\end{equation}
These equations introduce the source window function $N_s^i(\chi)$, analogous to the lens one:
\begin{equation}
N_s^i(\chi) = \frac{n^i_s(z)}{\bar{n}^i_s}\frac{dz}{d\chi}.
\end{equation}
To model the IA power spectra $P_{\rm II}$ and $P_{\rm \gamma I}$, one popular choice is to pick the nonlinear alignment (NLA) model. This model assumes that galaxy shapes align with the local tidal field, resulting in IA power spectra that are proportional to the matter power spectrum:
\begin{equation}
P_{\rm I I} (k,z) = A(z)^2 P_{mm} (k,z) , ,
\end{equation}
\begin{equation}
P_{\rm \gamma I} (k,z) = A(z) P_{mm} (k,z).
\end{equation}
The NLA model, first proposed by \citet{Bridle2007}, introduces a flexible parameterization of the IA amplitude:
\begin{equation}
A(z) = -A_{IA} \bar{C}1 \frac{3H_0^2\Omega_m}{8\pi G} D^{-1} \left( \frac{1+z}{1+z_0}\right)^{\eta{IA}}.
\end{equation}
This equation includes two free parameters: $A_{IA}$, a dimensionless amplitude governing the overall strength of the IA signal, and $\eta_{IA}$, which controls its redshift evolution. $D(z)$ is the linear growth factor, and $G$ is the gravitational constant. The normalization constant $\bar{C}1 = 5 \times 10^{-14} M\odot^{-1} h^{-2} Mpc^{3}$ is typically fixed based on measurements from the SuperCOSMOS Sky Survey \citep{Brown2002}. A typical value for the pivot redshift $z_0$ is 0.62.
The NLA model, despite its simplicity, has proven remarkably effective in capturing the main features of IA contamination in weak lensing surveys. By including these two additional free parameters ($A_{IA}$ and $\eta_{IA}$) in cosmological analyses, we can marginalize over the uncertainty in the IA signal, reducing potential biases in cosmological parameter estimation. More complex IA models exist, such as the TATT (Tidal Alignment and Tidal Torquing, see e.g. \citealt{Blazek_2019}), which was used as the fiducial in the DES Y3 cosmic shear analysis. It is currently an open question in the community how much complexity intrinsic alignment models warrant in practise for cosmological analyses. 

The combination of galaxy-galaxy lensing and cosmic shear measurements provides complementary information on intrinsic alignments, due to their different sensitivity to this effect, and thus provides much more tight constrains on the IA parameters than cosmic shear alone. This has also been explored using galaxy lensing ratios, via the so called shear-ratio technique \citep{Sanchez2022}.

\subsubsection{Redshift and shear marginalization}

When obtaining cosmological constraints, we need to marginalize over observational systematics such as redshift and shear calibration uncertainties. The simplest parametrization of the redshift uncertainties is to marginalize over a shift in the mean redshift $\Delta z^i$ for both the lens and source input redshift distributions $n_{\mathrm{input}}^i$:
\begin{equation}
n^i (z) = n_{\mathrm{input}}^i (z - \Delta z^i).
\end{equation}
However, this approach may be insufficient, particularly for the lens sample, where the width of the distribution can significantly impact cosmological inferences and it is also often marginalized over.  Furthermore, to account for more complex changes in the shape of the redshift distribution, we can employ methods that provide samples of the entire $n(z)$ function and then  we can then marginalize over these samples during the parameter inference process. 

For the shear, a multiplicative shear bias $m$ per each source bin is typically used, which modifies the shear and galaxy-shear angular power spectra in the following way:
\begin{equation}
     C_{\delta_{g}G}^{ij}(\ell) = (1+m^j)  C^{ij}_{\delta_{g}G, {\mathrm{input}}} (\ell)
\end{equation}
\begin{equation}
    C^{ij}_{GG} (\ell)= (1+m^i)(1+m^j) \ C^{ij}_{GG, {\mathrm{input}}} (\ell)
\end{equation}

\subsubsection{Lens magnification}

Lens magnification is the effect of magnification produced on the lens galaxy sample by the structure that is between the lens galaxies and the observer. In this section we describe how lens magnification affects  galaxy-galaxy lensing and galaxy clustering. 

In the weak gravitational lensing picture, besides having shape distortions described by the shear, the solid angle spanned by the image is changed compared to the solid angle covered by the source by the so-called \textit{magnification factor} $\mu$. This change in solid angle can alter the number density of a given sample via two different mechanisms: (1) The number density decreases by a factor $\mu$ due to the sky being locally stretched by the same factor; and (2) since the area increases but the surface brightness is conserved, the flux of individual galaxies rises, and some galaxies that would otherwise not have been detected pass the relevant flux threshold for a particular sample. These are two competing effects and the dominant one depends on the specifics of the galaxy sample. Then, to understand how lens magnification affects two-point  measurements, it is useful to express the observed density contrast for the lens sample as the sum of the intrinsic galaxy density contrast and the ``artificial'' one produced by lens magnification:
\begin{equation}
    \delta_g^\text{obs} =  \delta_g^\text{int} + \delta_g^\text{mag} .
\end{equation}
Then we can make the assumption that the change in number density produced by magnification is  proportional to the convergence. In that case, we can write
\begin{equation}\label{eq:definition_C}
\delta_g^\text{mag} (\theta) = C \kappa_l (\theta) \, , 
\end{equation}
where $\kappa_l$ is the convergence field at the lens redshift and $C$ is just a proportionality factor. At this point we can separate the area effect and the flux effect on the number density change: $C_\mathrm{total} = C_\mathrm{area} + C_\mathrm{flux}$, since it can be shown that $C_\mathrm{area}= - 2$ \citep*{Elvin-Poole2023} while $C_\mathrm{flux}$ will depend on the sample. That is why this proportionality factor is also sometimes written  as  $C_\mathrm{total} = 2 (\alpha -1)$, where $\alpha$ is a property of the sample and is equivalent to $C_\mathrm{flux}/2$.  The $\alpha$ parameters have to be carefully measured and calibrated for a given sample using realistic $N$-body simulations and realistic image simulations.

Lens magnification becomes relevant because the change in number density produced in the lens sample is correlated with the large scale structure that is between the lens galaxies and the observer. That means that for a given sample of lens galaxies, some lines-of-sight with, for instance, more matter between the lens galaxies and us could be over-sampled if $\alpha >1$, or down-sampled if $\alpha<1$, and the tangential shear measurement would be biased, as seen in the following equation:
\begin{equation}
\left< \delta_g^\text{obs} \gamma \right> = \left< \delta_g^\text{int} \gamma \right> + 2 (\alpha -1)\left< \kappa_l\gamma \right>
\end{equation}
The first term is just the usual galaxy-galaxy lensing signal, and the additional lens magnification term is modeled in the following way before performing the projection to real space:
\begin{equation}
   \left< \kappa_l \gamma \right> \equiv C_{mm}^{ij}(\ell) = \int d\chi \frac{q_l^i(\chi)\, q_s^j(\chi)}{\chi^2}P_{mm}\left(k = \frac{\ell}{\chi},z(\chi)\right)\,,
\end{equation}
where the lensing window function $q_s$ is defined in Eq.~(\ref{eq:lensing_window}), and the analogous window function for the lens sample is given by $q_l$. The $i$ index represents the lens tomographic bin and $j$ the source one.

\subsubsection{Baryonic feedback}

Baryonic feedback effects significantly influence the small-scale matter power spectrum, presenting a leading systematic uncertainty in cosmic shear surveys. Active Galactic Nuclei (AGN) feedback processes suppress power at scales around $k \sim 10h/\text{Mpc}$, while at smaller scales, power is enhanced due to more efficient cooling and star formation. These effects on the matter power spectrum are complex and can be modeled using various approaches, including empirical halo models \citep{Mead2021} and \textit{baryonification models} \citep{Arico2021}. However, the uncertainty in the models is still large.

To mitigate potential biases, some analyses adopt a gravity-only power spectrum and limit measurements to larger angular scales. The minimum scale at which baryonic effects may have significant impact is determined by contaminating synthetic cosmic shear data vectors with baryonic effects measured from hydrodynamic simulations e.g., OWLS-AGN (OverWhelmingly Large Simulations project, \citealt{OWLS, vanDaalen11}). Then, in a similar fashion to the galaxy bias case, a minimum scale cut can be chosen in a way that the parameters of interest are not affected significantly by this effect. Alternatively, baryonic feedback can be modelled in the power spectrum, adding some uncertainty to be marginalized over, see e.g. \citet{Mead2021}. 

\subsubsection{Real space projection}

Finally,  each real-space two-point correlation function is related to the total angular power spectrum for galaxy-galaxy lensing and galaxy clustering via a Hankel transform
    \begin{equation}
    \gamma_{t}^{ij}(\theta) = \int \frac{d\ell \ell}{2\pi} C_{\delta_{g}G}^{ij} (\ell) J_{2}(\ell \theta),
    \end{equation}
    \begin{equation}
    w^{ij}(\theta) = \int \frac{d \ell \ell}{2\pi} C_{\delta_{g}\delta_{g}}^{ij}(\ell) J_{0}(\ell \theta),
    \label{eq:wtheta}
    \end{equation}
under the flat-sky approximation. The cosmic shear expression can be found in Eq.~\ref{eq:xi}, but now replacing the shear angular power spectra by the observed one $ C^{ij}_{GG} (\ell)$ before performing the projection. If instead we want to use the curved sky projection:
\begin{align}
\gamma_t^{ij}(\theta) &= \sum_\ell \frac{2\ell+1}{4\pi\ell(\ell+1)}P^2_\ell(\cos\theta)  C_{\delta_{g}G}^{ij} (\ell),
\end{align}
\begin{equation}
w^{ij}(\theta) = \sum_{\ell} \frac{2\ell + 1}{4\pi} P_{\ell}(\cos \theta)C_{\delta_{g}\delta_{g}}^{ij} (\ell) ,
\end{equation}
where $P_\ell^2$ are the associated Legendre polynomial and $P_{\ell}$ are the Legendre polynomials. Typically these need to be averaged over the same angular binning that we apply to the measurements, as specified in Eqs.~19 and 20. from \citet{Krause2021}. 

\subsubsection{Real space estimators for galaxy-galaxy lensing and galaxy clustering}

In this section we summarize how we estimate in practice such correlation functions. For cosmic shear, we already described how to estimate $\xi_{\pm}$ in Eq.~(\ref{eq:xipm_estimator}).  For galaxy-galaxy lensing, the mean tangential shear estimator is usually expressed as:
\begin{equation}
\hat{\gamma}^{ij}_t(\theta)  = \frac{\sum_{LS} w_{LS} \, \hat{\epsilon}_{t,LS}^{ij}(\theta)}{\sum_{LS} w_{LS}(\theta)} - \frac{\sum_{RS} w_{RS} \, \hat{\epsilon}_{t,RS}^{ij}(\theta)}{\sum_{RS} w_{RS}(\theta)},
\label{eq:gt}
\end{equation}
where $LS$ refers to lens-source galaxy pairs that are separated by a given angular scale that falls within the bin. $RS$ refers to random-source pairs, and this term removes the tangential shear around a sample of random points  to correct for mask effects. See e.g. \citet{Prat2022} for a comprehensive description of all the important effects affecting the measurement and modeling of the mean tangential shear.  Finally, for the angular 2-point correlation function, $w(\theta)$ the typical estimator is the  Landy-Szalay estimator which can be written as: 
\begin{equation}
 \hat{w}^{ij}(\theta) = \frac{D^i D^j - D^i R^j - D^j R^i + R^i R^j }{R^i R^j} \, ,
\label{eq:wtheta_estimator}
\end{equation}\\ 
where $DD$, $RR$ and $DR$ are the normalized galaxy-galaxy, random-random and galaxy-random pair counts within an angular bin, respectively. See e.g. \citet{Monroy2022} for an extended description of such estimator. This work is particularly useful also for understanding how to mitigate systematic biases in clustering measurements, which is crucial for extracting accurate cosmological information from large-scale structure data.

In Fig.~\ref{cosmo:fig:desy3-dv} we show state-of-the art galaxy-galaxy lensing and angular clustering measurements.

\subsection{Cosmological model testing}
\begin{figure}
	\centering
	\includegraphics[width=0.8\textwidth]{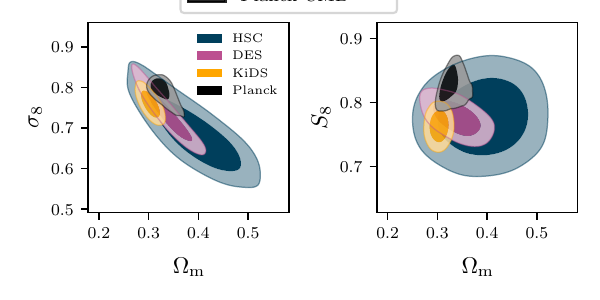}
	\caption{This figure presents cosmological constraints from the 3$\times$2pt combination for DES Y3 \citep{desy3-3x2}, KiDS-1000 \citep{kids1000} and HSC Y3 \citep{hscy3-3x2} in comparison with early Universe measurements from the \textit{Planck} \citep{Planck2018} satellite assuming the $\Lambda$CDM model. }
	\label{fig:3x2t_vs_planck}
\end{figure}

Weak lensing measurements can be compared  with early Universe results coming from the Cosmic Microwave Background. The CMB measures the state of the Universe when it was 380,000 years old, while galaxy surveys measure the matter distribution in the Universe roughly ten billion years later.  Then, comparing these results is one of the most  stringent tests we can perform of the current standard cosmological model $\Lambda$CDM.  In Fig.~\ref{fig:3x2t_vs_planck} we display  $3\times$2pt results from the three main Stage III surveys in comparison with results from the CMB, in particular from  the \textit{Planck} satellite.  There is a parameter difference of 1.5 (2) $\sigma$ assuming the $\Lambda$CDM for DES (KiDS), as reported in \citet{desy3-3x2, kids1000} respectively, considering the full parameter space. Future analyses including larger data sets will improve the precision of such tests and provide even more decisive tests of $\Lambda$CDM, together with exploring potential model extensions.

\section{Summary and Outlook}\label{chap1:summary}

Weak gravitational lensing has emerged as a powerful probe of cosmology, providing unique insights into the distribution of matter and the nature of dark energy. Current surveys have demonstrated the ability of weak lensing to provide competitive constraints on cosmological parameters, particularly $S_8$, while also highlighting the importance of careful systematic control.

Looking ahead, upcoming surveys such as LSST/Rubin, Euclid, and Roman promise to revolutionize the field. These surveys will observe billions of galaxies over unprecedented areas, dramatically improving statistical precision. However, this precision will place even greater demands on systematic control, driving innovations in shear measurement, photometric redshift estimation, and theoretical modeling. Key areas for future work include:
\begin{enumerate}
    \item Improved modeling of nonlinear structure formation, including baryonic effects.
    \item Better understanding and modeling of intrinsic alignments.
    \item Consolidated  techniques for extracting non-Gaussian information from weak lensing data.
\end{enumerate}

As the field progresses, weak lensing will play a crucial role in addressing fundamental questions in cosmology, potentially shedding light on the nature of dark energy, the properties of dark matter, and possible deviations from General Relativity on cosmological scales.

The computational challenges posed by the enormous datasets of future surveys will necessitate the development of innovative data processing and analysis techniques. Machine learning and artificial intelligence are likely to play an increasingly important role in areas such as galaxy shape measurement and photometric redshift estimation.

Furthermore, the combination of weak lensing with other cosmological probes will become increasingly important. Cross-correlations between weak lensing and other tracers of large-scale structure, such as the CMB, galaxies, and the Lyman-alpha forest, will provide complementary information and help break degeneracies between cosmological parameters. Moreover, weak gravitational lensing by galaxy clusters enables the accurate calibration of their mass scale and the use of cluster number counts as an independent cosmological probe also provides significant additional cosmological information. 

In conclusion, weak gravitational lensing stands at the forefront of observational cosmology. As we enter the era of precision cosmology, weak lensing will be instrumental in testing the $\Lambda$CDM model and exploring possible extensions or alternatives.

\begin{ack}[Acknowledgments]

We would like to thank Surhud More, Arthur Loureiro and Benjamin Giblin for providing the data products for HSC and KiDS needed to plot the survey footprints. We would also like to thank  Ami Choi for helping us collect the Roman forecasts. 

\end{ack}

\begin{glossary}[Glossary]

\term{Shear} The stretching of galaxy images due to gravitational lensing.

\term{Convergence} The dilation of galaxy images due to gravitational lensing.

\term{Tangential shear} The component of shear aligned tangentially to the line connecting the lens and source.

\term{Cosmic shear} The correlation of galaxy shapes due to weak lensing by large-scale structure.

\term{Galaxy-galaxy lensing} The correlation between foreground galaxy positions and background galaxy shapes.

\term{3$\times$2pt analysis} A combined analysis of galaxy clustering, cosmic shear, and galaxy-galaxy lensing.

\term{Intrinsic alignments} Correlations between the intrinsic shapes of galaxies that can mimic the weak lensing signal.

\term{Photometric redshift} An estimate of a galaxy's redshift based on its observed colors in different photometric bands.
\end{glossary}

\begin{glossary}[Nomenclature]
\begin{tabular}{@{}lp{34pc}@{}}
DES & Dark Energy Survey\\
KiDS & Kilo-Degree Survey\\
HSC & Hyper Suprime-Cam\\

\end{tabular}
\end{glossary}


\bibliographystyle{Harvard}
\renewcommand*{\bibfont}{\normalfont\footnotesize}
\bibliography{reference}

\begin{thebibliography*}{57}
\providecommand{\bibtype}[1]{}
\providecommand{\natexlab}[1]{#1}
{\catcode`\|=0\catcode`\#=12\catcode`\@=11\catcode`\\=12
|immediate|write|@auxout{\expandafter\ifx\csname
  natexlab\endcsname\relax\gdef\natexlab#1{#1}\fi}}
\renewcommand{\url}[1]{{\tt #1}}
\providecommand{\urlprefix}{URL }
\expandafter\ifx\csname urlstyle\endcsname\relax
  \providecommand{\doi}[1]{doi:\discretionary{}{}{}#1}\else
  \providecommand{\doi}{doi:\discretionary{}{}{}\begingroup
  \urlstyle{rm}\Url}\fi
\providecommand{\bibinfo}[2]{#2}
\providecommand{\eprint}[2][]{\url{#2}}

\bibtype{Article}%
\bibitem[{Abbott} et al.(2022)]{desy3-3x2}
\bibinfo{author}{{Abbott} TMC}, \bibinfo{author}{{Aguena} M},
  \bibinfo{author}{{Alarcon} A}, \bibinfo{author}{{Allam} S},
  \bibinfo{author}{{Alves} O}, \bibinfo{author}{{Amon} A},
  \bibinfo{author}{{Andrade-Oliveira} F}, \bibinfo{author}{{Annis} J},
  \bibinfo{author}{{Avila} S}, \bibinfo{author}{{Bacon} D},
  \bibinfo{author}{{Baxter} E}, \bibinfo{author}{{Bechtol} K},
  \bibinfo{author}{{Becker} MR}, \bibinfo{author}{{Bernstein} GM},
  \bibinfo{author}{{Bhargava} S}, \bibinfo{author}{{Birrer} S},
  \bibinfo{author}{{Blazek} J}, \bibinfo{author}{{Brandao-Souza} A},
  \bibinfo{author}{{Bridle} SL}, \bibinfo{author}{{Brooks} D},
  \bibinfo{author}{{Buckley-Geer} E}, \bibinfo{author}{{Burke} DL},
  \bibinfo{author}{{Camacho} H}, \bibinfo{author}{{Campos} A},
  \bibinfo{author}{{Carnero Rosell} A}, \bibinfo{author}{{Carrasco Kind} M},
  \bibinfo{author}{{Carretero} J}, \bibinfo{author}{{Castander} FJ},
  \bibinfo{author}{{Cawthon} R}, \bibinfo{author}{{Chang} C},
  \bibinfo{author}{{Chen} A}, \bibinfo{author}{{Chen} R},
  \bibinfo{author}{{Choi} A}, \bibinfo{author}{{Conselice} C},
  \bibinfo{author}{{Cordero} J}, \bibinfo{author}{{Costanzi} M},
  \bibinfo{author}{{Crocce} M}, \bibinfo{author}{{da Costa} LN},
  \bibinfo{author}{{da Silva Pereira} ME}, \bibinfo{author}{{Davis} C},
  \bibinfo{author}{{Davis} TM}, \bibinfo{author}{{De Vicente} J},
  \bibinfo{author}{{DeRose} J}, \bibinfo{author}{{Desai} S},
  \bibinfo{author}{{Di Valentino} E}, \bibinfo{author}{{Diehl} HT},
  \bibinfo{author}{{Dietrich} JP}, \bibinfo{author}{{Dodelson} S},
  \bibinfo{author}{{Doel} P}, \bibinfo{author}{{Doux} C},
  \bibinfo{author}{{Drlica-Wagner} A}, \bibinfo{author}{{Eckert} K},
  \bibinfo{author}{{Eifler} TF}, \bibinfo{author}{{Elsner} F},
  \bibinfo{author}{{Elvin-Poole} J}, \bibinfo{author}{{Everett} S},
  \bibinfo{author}{{Evrard} AE}, \bibinfo{author}{{Fang} X},
  \bibinfo{author}{{Farahi} A}, \bibinfo{author}{{Fernandez} E},
  \bibinfo{author}{{Ferrero} I}, \bibinfo{author}{{Fert{\'e}} A},
  \bibinfo{author}{{Fosalba} P}, \bibinfo{author}{{Friedrich} O},
  \bibinfo{author}{{Frieman} J}, \bibinfo{author}{{Garc{\'\i}a-Bellido} J},
  \bibinfo{author}{{Gatti} M}, \bibinfo{author}{{Gaztanaga} E},
  \bibinfo{author}{{Gerdes} DW}, \bibinfo{author}{{Giannantonio} T},
  \bibinfo{author}{{Giannini} G}, \bibinfo{author}{{Gruen} D},
  \bibinfo{author}{{Gruendl} RA}, \bibinfo{author}{{Gschwend} J},
  \bibinfo{author}{{Gutierrez} G}, \bibinfo{author}{{Harrison} I},
  \bibinfo{author}{{Hartley} WG}, \bibinfo{author}{{Herner} K},
  \bibinfo{author}{{Hinton} SR}, \bibinfo{author}{{Hollowood} DL},
  \bibinfo{author}{{Honscheid} K}, \bibinfo{author}{{Hoyle} B},
  \bibinfo{author}{{Huff} EM}, \bibinfo{author}{{Huterer} D},
  \bibinfo{author}{{Jain} B}, \bibinfo{author}{{James} DJ},
  \bibinfo{author}{{Jarvis} M}, \bibinfo{author}{{Jeffrey} N},
  \bibinfo{author}{{Jeltema} T}, \bibinfo{author}{{Kovacs} A},
  \bibinfo{author}{{Krause} E}, \bibinfo{author}{{Kron} R},
  \bibinfo{author}{{Kuehn} K}, \bibinfo{author}{{Kuropatkin} N},
  \bibinfo{author}{{Lahav} O}, \bibinfo{author}{{Leget} PF},
  \bibinfo{author}{{Lemos} P}, \bibinfo{author}{{Liddle} AR},
  \bibinfo{author}{{Lidman} C}, \bibinfo{author}{{Lima} M},
  \bibinfo{author}{{Lin} H}, \bibinfo{author}{{MacCrann} N},
  \bibinfo{author}{{Maia} MAG}, \bibinfo{author}{{Marshall} JL},
  \bibinfo{author}{{Martini} P}, \bibinfo{author}{{McCullough} J},
  \bibinfo{author}{{Melchior} P}, \bibinfo{author}{{Mena-Fern{\'a}ndez} J},
  \bibinfo{author}{{Menanteau} F}, \bibinfo{author}{{Miquel} R},
  \bibinfo{author}{{Mohr} JJ}, \bibinfo{author}{{Morgan} R},
  \bibinfo{author}{{Muir} J}, \bibinfo{author}{{Myles} J},
  \bibinfo{author}{{Nadathur} S}, \bibinfo{author}{{Navarro-Alsina} A},
  \bibinfo{author}{{Nichol} RC}, \bibinfo{author}{{Ogando} RLC},
  \bibinfo{author}{{Omori} Y}, \bibinfo{author}{{Palmese} A},
  \bibinfo{author}{{Pandey} S}, \bibinfo{author}{{Park} Y},
  \bibinfo{author}{{Paz-Chinch{\'o}n} F}, \bibinfo{author}{{Petravick} D},
  \bibinfo{author}{{Pieres} A}, \bibinfo{author}{{Plazas Malag{\'o}n} AA},
  \bibinfo{author}{{Porredon} A}, \bibinfo{author}{{Prat} J},
  \bibinfo{author}{{Raveri} M}, \bibinfo{author}{{Rodriguez-Monroy} M},
  \bibinfo{author}{{Rollins} RP}, \bibinfo{author}{{Romer} AK},
  \bibinfo{author}{{Roodman} A}, \bibinfo{author}{{Rosenfeld} R},
  \bibinfo{author}{{Ross} AJ}, \bibinfo{author}{{Rykoff} ES},
  \bibinfo{author}{{Samuroff} S}, \bibinfo{author}{{S{\'a}nchez} C},
  \bibinfo{author}{{Sanchez} E}, \bibinfo{author}{{Sanchez} J},
  \bibinfo{author}{{Sanchez Cid} D}, \bibinfo{author}{{Scarpine} V},
  \bibinfo{author}{{Schubnell} M}, \bibinfo{author}{{Scolnic} D},
  \bibinfo{author}{{Secco} LF}, \bibinfo{author}{{Serrano} S},
  \bibinfo{author}{{Sevilla-Noarbe} I}, \bibinfo{author}{{Sheldon} E},
  \bibinfo{author}{{Shin} T}, \bibinfo{author}{{Smith} M},
  \bibinfo{author}{{Soares-Santos} M}, \bibinfo{author}{{Suchyta} E},
  \bibinfo{author}{{Swanson} MEC}, \bibinfo{author}{{Tabbutt} M},
  \bibinfo{author}{{Tarle} G}, \bibinfo{author}{{Thomas} D},
  \bibinfo{author}{{To} C}, \bibinfo{author}{{Troja} A},
  \bibinfo{author}{{Troxel} MA}, \bibinfo{author}{{Tucker} DL},
  \bibinfo{author}{{Tutusaus} I}, \bibinfo{author}{{Varga} TN},
  \bibinfo{author}{{Walker} AR}, \bibinfo{author}{{Weaverdyck} N},
  \bibinfo{author}{{Wechsler} R}, \bibinfo{author}{{Weller} J},
  \bibinfo{author}{{Yanny} B}, \bibinfo{author}{{Yin} B},
  \bibinfo{author}{{Zhang} Y}, \bibinfo{author}{{Zuntz} J} and
  \bibinfo{author}{{DES Collaboration}} (\bibinfo{year}{2022}),
  \bibinfo{month}{Jan.}
\bibinfo{title}{{Dark Energy Survey Year 3 results: Cosmological constraints
  from galaxy clustering and weak lensing}}.
\bibinfo{journal}{{\em \prd}} \bibinfo{volume}{105} (\bibinfo{number}{2}),
  \bibinfo{eid}{023520}. \bibinfo{doi}{\doi{10.1103/PhysRevD.105.023520}}.
\eprint{2105.13549}.

\bibtype{Article}%
\bibitem[{Aihara} et al.(2018)]{Aihara2018}
\bibinfo{author}{{Aihara} H}, \bibinfo{author}{{Arimoto} N},
  \bibinfo{author}{{Armstrong} R}, \bibinfo{author}{{Arnouts} S},
  \bibinfo{author}{{Bahcall} NA}, \bibinfo{author}{{Bickerton} S},
  \bibinfo{author}{{Bosch} J}, \bibinfo{author}{{Bundy} K},
  \bibinfo{author}{{Capak} PL}, \bibinfo{author}{{Chan} JHH},
  \bibinfo{author}{{Chiba} M}, \bibinfo{author}{{Coupon} J},
  \bibinfo{author}{{Egami} E}, \bibinfo{author}{{Enoki} M},
  \bibinfo{author}{{Finet} F}, \bibinfo{author}{{Fujimori} H},
  \bibinfo{author}{{Fujimoto} S}, \bibinfo{author}{{Furusawa} H},
  \bibinfo{author}{{Furusawa} J}, \bibinfo{author}{{Goto} T},
  \bibinfo{author}{{Goulding} A}, \bibinfo{author}{{Greco} JP},
  \bibinfo{author}{{Greene} JE}, \bibinfo{author}{{Gunn} JE},
  \bibinfo{author}{{Hamana} T}, \bibinfo{author}{{Harikane} Y},
  \bibinfo{author}{{Hashimoto} Y}, \bibinfo{author}{{Hattori} T},
  \bibinfo{author}{{Hayashi} M}, \bibinfo{author}{{Hayashi} Y},
  \bibinfo{author}{{He{\l}miniak} KG}, \bibinfo{author}{{Higuchi} R},
  \bibinfo{author}{{Hikage} C}, \bibinfo{author}{{Ho} PTP},
  \bibinfo{author}{{Hsieh} BC}, \bibinfo{author}{{Huang} K},
  \bibinfo{author}{{Huang} S}, \bibinfo{author}{{Ikeda} H},
  \bibinfo{author}{{Imanishi} M}, \bibinfo{author}{{Inoue} AK},
  \bibinfo{author}{{Iwasawa} K}, \bibinfo{author}{{Iwata} I},
  \bibinfo{author}{{Jaelani} AT}, \bibinfo{author}{{Jian} HY},
  \bibinfo{author}{{Kamata} Y}, \bibinfo{author}{{Karoji} H},
  \bibinfo{author}{{Kashikawa} N}, \bibinfo{author}{{Katayama} N},
  \bibinfo{author}{{Kawanomoto} S}, \bibinfo{author}{{Kayo} I},
  \bibinfo{author}{{Koda} J}, \bibinfo{author}{{Koike} M},
  \bibinfo{author}{{Kojima} T}, \bibinfo{author}{{Komiyama} Y},
  \bibinfo{author}{{Konno} A}, \bibinfo{author}{{Koshida} S},
  \bibinfo{author}{{Koyama} Y}, \bibinfo{author}{{Kusakabe} H},
  \bibinfo{author}{{Leauthaud} A}, \bibinfo{author}{{Lee} CH},
  \bibinfo{author}{{Lin} L}, \bibinfo{author}{{Lin} YT},
  \bibinfo{author}{{Lupton} RH}, \bibinfo{author}{{Mand elbaum} R},
  \bibinfo{author}{{Matsuoka} Y}, \bibinfo{author}{{Medezinski} E},
  \bibinfo{author}{{Mineo} S}, \bibinfo{author}{{Miyama} S},
  \bibinfo{author}{{Miyatake} H}, \bibinfo{author}{{Miyazaki} S},
  \bibinfo{author}{{Momose} R}, \bibinfo{author}{{More} A},
  \bibinfo{author}{{More} S}, \bibinfo{author}{{Moritani} Y},
  \bibinfo{author}{{Moriya} TJ}, \bibinfo{author}{{Morokuma} T},
  \bibinfo{author}{{Mukae} S}, \bibinfo{author}{{Murata} R},
  \bibinfo{author}{{Murayama} H}, \bibinfo{author}{{Nagao} T},
  \bibinfo{author}{{Nakata} F}, \bibinfo{author}{{Niida} M},
  \bibinfo{author}{{Niikura} H}, \bibinfo{author}{{Nishizawa} AJ},
  \bibinfo{author}{{Obuchi} Y}, \bibinfo{author}{{Oguri} M},
  \bibinfo{author}{{Oishi} Y}, \bibinfo{author}{{Okabe} N},
  \bibinfo{author}{{Okamoto} S}, \bibinfo{author}{{Okura} Y},
  \bibinfo{author}{{Ono} Y}, \bibinfo{author}{{Onodera} M},
  \bibinfo{author}{{Onoue} M}, \bibinfo{author}{{Osato} K},
  \bibinfo{author}{{Ouchi} M}, \bibinfo{author}{{Price} PA},
  \bibinfo{author}{{Pyo} TS}, \bibinfo{author}{{Sako} M},
  \bibinfo{author}{{Sawicki} M}, \bibinfo{author}{{Shibuya} T},
  \bibinfo{author}{{Shimasaku} K}, \bibinfo{author}{{Shimono} A},
  \bibinfo{author}{{Shirasaki} M}, \bibinfo{author}{{Silverman} JD},
  \bibinfo{author}{{Simet} M}, \bibinfo{author}{{Speagle} J},
  \bibinfo{author}{{Spergel} DN}, \bibinfo{author}{{Strauss} MA},
  \bibinfo{author}{{Sugahara} Y}, \bibinfo{author}{{Sugiyama} N},
  \bibinfo{author}{{Suto} Y}, \bibinfo{author}{{Suyu} SH},
  \bibinfo{author}{{Suzuki} N}, \bibinfo{author}{{Tait} PJ},
  \bibinfo{author}{{Takada} M}, \bibinfo{author}{{Takata} T},
  \bibinfo{author}{{Tamura} N}, \bibinfo{author}{{Tanaka} MM},
  \bibinfo{author}{{Tanaka} M}, \bibinfo{author}{{Tanaka} M},
  \bibinfo{author}{{Tanaka} Y}, \bibinfo{author}{{Terai} T},
  \bibinfo{author}{{Terashima} Y}, \bibinfo{author}{{Toba} Y},
  \bibinfo{author}{{Tominaga} N}, \bibinfo{author}{{Toshikawa} J},
  \bibinfo{author}{{Turner} EL}, \bibinfo{author}{{Uchida} T},
  \bibinfo{author}{{Uchiyama} H}, \bibinfo{author}{{Umetsu} K},
  \bibinfo{author}{{Uraguchi} F}, \bibinfo{author}{{Urata} Y},
  \bibinfo{author}{{Usuda} T}, \bibinfo{author}{{Utsumi} Y},
  \bibinfo{author}{{Wang} SY}, \bibinfo{author}{{Wang} WH},
  \bibinfo{author}{{Wong} KC}, \bibinfo{author}{{Yabe} K},
  \bibinfo{author}{{Yamada} Y}, \bibinfo{author}{{Yamanoi} H},
  \bibinfo{author}{{Yasuda} N}, \bibinfo{author}{{Yeh} S},
  \bibinfo{author}{{Yonehara} A} and  \bibinfo{author}{{Yuma} S}
  (\bibinfo{year}{2018}), \bibinfo{month}{Jan.}
\bibinfo{title}{{The Hyper Suprime-Cam SSP Survey: Overview and survey
  design}}.
\bibinfo{journal}{{\em \pasj}} \bibinfo{volume}{70}, \bibinfo{eid}{S4}.
  \bibinfo{doi}{\doi{10.1093/pasj/psx066}}.
\eprint{1704.05858}.

\bibtype{Article}%
\bibitem[{Akeson} et al.(2019)]{Akeson2019}
\bibinfo{author}{{Akeson} R}, \bibinfo{author}{{Armus} L},
  \bibinfo{author}{{Bachelet} E}, \bibinfo{author}{{Bailey} V},
  \bibinfo{author}{{Bartusek} L}, \bibinfo{author}{{Bellini} A},
  \bibinfo{author}{{Benford} D}, \bibinfo{author}{{Bennett} D},
  \bibinfo{author}{{Bhattacharya} A}, \bibinfo{author}{{Bohlin} R},
  \bibinfo{author}{{Boyer} M}, \bibinfo{author}{{Bozza} V},
  \bibinfo{author}{{Bryden} G}, \bibinfo{author}{{Calchi Novati} S},
  \bibinfo{author}{{Carpenter} K}, \bibinfo{author}{{Casertano} S},
  \bibinfo{author}{{Choi} A}, \bibinfo{author}{{Content} D},
  \bibinfo{author}{{Dayal} P}, \bibinfo{author}{{Dressler} A},
  \bibinfo{author}{{Dor{\'e}} O}, \bibinfo{author}{{Fall} SM},
  \bibinfo{author}{{Fan} X}, \bibinfo{author}{{Fang} X},
  \bibinfo{author}{{Filippenko} A}, \bibinfo{author}{{Finkelstein} S},
  \bibinfo{author}{{Foley} R}, \bibinfo{author}{{Furlanetto} S},
  \bibinfo{author}{{Kalirai} J}, \bibinfo{author}{{Gaudi} BS},
  \bibinfo{author}{{Gilbert} K}, \bibinfo{author}{{Girard} J},
  \bibinfo{author}{{Grady} K}, \bibinfo{author}{{Greene} J},
  \bibinfo{author}{{Guhathakurta} P}, \bibinfo{author}{{Heinrich} C},
  \bibinfo{author}{{Hemmati} S}, \bibinfo{author}{{Hendel} D},
  \bibinfo{author}{{Henderson} C}, \bibinfo{author}{{Henning} T},
  \bibinfo{author}{{Hirata} C}, \bibinfo{author}{{Ho} S},
  \bibinfo{author}{{Huff} E}, \bibinfo{author}{{Hutter} A},
  \bibinfo{author}{{Jansen} R}, \bibinfo{author}{{Jha} S},
  \bibinfo{author}{{Johnson} S}, \bibinfo{author}{{Jones} D},
  \bibinfo{author}{{Kasdin} J}, \bibinfo{author}{{Kelly} P},
  \bibinfo{author}{{Kirshner} R}, \bibinfo{author}{{Koekemoer} A},
  \bibinfo{author}{{Kruk} J}, \bibinfo{author}{{Lewis} N},
  \bibinfo{author}{{Macintosh} B}, \bibinfo{author}{{Madau} P},
  \bibinfo{author}{{Malhotra} S}, \bibinfo{author}{{Mand el} K},
  \bibinfo{author}{{Massara} E}, \bibinfo{author}{{Masters} D},
  \bibinfo{author}{{McEnery} J}, \bibinfo{author}{{McQuinn} K},
  \bibinfo{author}{{Melchior} P}, \bibinfo{author}{{Melton} M},
  \bibinfo{author}{{Mennesson} B}, \bibinfo{author}{{Peeples} M},
  \bibinfo{author}{{Penny} M}, \bibinfo{author}{{Perlmutter} S},
  \bibinfo{author}{{Pisani} A}, \bibinfo{author}{{Plazas} A},
  \bibinfo{author}{{Poleski} R}, \bibinfo{author}{{Postman} M},
  \bibinfo{author}{{Ranc} C}, \bibinfo{author}{{Rauscher} B},
  \bibinfo{author}{{Rest} A}, \bibinfo{author}{{Roberge} A},
  \bibinfo{author}{{Robertson} B}, \bibinfo{author}{{Rodney} S},
  \bibinfo{author}{{Rhoads} J}, \bibinfo{author}{{Rhodes} J},
  \bibinfo{author}{{Ryan} Russell J}, \bibinfo{author}{{Sahu} K},
  \bibinfo{author}{{Sand} D}, \bibinfo{author}{{Scolnic} D},
  \bibinfo{author}{{Seth} A}, \bibinfo{author}{{Shvartzvald} Y},
  \bibinfo{author}{{Siellez} K}, \bibinfo{author}{{Smith} A},
  \bibinfo{author}{{Spergel} D}, \bibinfo{author}{{Stassun} K},
  \bibinfo{author}{{Street} R}, \bibinfo{author}{{Strolger} LG},
  \bibinfo{author}{{Szalay} A}, \bibinfo{author}{{Trauger} J},
  \bibinfo{author}{{Troxel} MA}, \bibinfo{author}{{Turnbull} M},
  \bibinfo{author}{{van der Marel} R}, \bibinfo{author}{{von der Linden} A},
  \bibinfo{author}{{Wang} Y}, \bibinfo{author}{{Weinberg} D},
  \bibinfo{author}{{Williams} B}, \bibinfo{author}{{Windhorst} R},
  \bibinfo{author}{{Wollack} E}, \bibinfo{author}{{Wu} HY},
  \bibinfo{author}{{Yee} J} and  \bibinfo{author}{{Zimmerman} N}
  (\bibinfo{year}{2019}), \bibinfo{month}{Feb.}
\bibinfo{title}{{The Wide Field Infrared Survey Telescope: 100 Hubbles for the
  2020s}}.
\bibinfo{journal}{{\em arXiv e-prints}} ,
  \bibinfo{eid}{arXiv:1902.05569}\eprint{1902.05569}.

\bibtype{Article}%
\bibitem[{Albrecht} et al.(2006)]{Albrecht2006}
\bibinfo{author}{{Albrecht} A}, \bibinfo{author}{{Bernstein} G},
  \bibinfo{author}{{Cahn} R}, \bibinfo{author}{{Freedman} WL},
  \bibinfo{author}{{Hewitt} J}, \bibinfo{author}{{Hu} W},
  \bibinfo{author}{{Huth} J}, \bibinfo{author}{{Kamionkowski} M},
  \bibinfo{author}{{Kolb} EW}, \bibinfo{author}{{Knox} L},
  \bibinfo{author}{{Mather} JC}, \bibinfo{author}{{Staggs} S} and
  \bibinfo{author}{{Suntzeff} NB} (\bibinfo{year}{2006}), \bibinfo{month}{Sep.}
\bibinfo{title}{{Report of the Dark Energy Task Force}}.
\bibinfo{journal}{{\em arXiv e-prints}} ,
  \bibinfo{eid}{astro-ph/0609591}\eprint{astro-ph/0609591}.

\bibtype{Article}%
\bibitem[{Amon} et al.(2022)]{Amon2022}
\bibinfo{author}{{Amon} A}, \bibinfo{author}{{Gruen} D},
  \bibinfo{author}{{Troxel} MA}, \bibinfo{author}{{MacCrann} N},
  \bibinfo{author}{{Dodelson} S}, \bibinfo{author}{{Choi} A},
  \bibinfo{author}{{Doux} C}, \bibinfo{author}{{Secco} LF},
  \bibinfo{author}{{Samuroff} S}, \bibinfo{author}{{Krause} E},
  \bibinfo{author}{{Cordero} J}, \bibinfo{author}{{Myles} J},
  \bibinfo{author}{{DeRose} J}, \bibinfo{author}{{Wechsler} RH},
  \bibinfo{author}{{Gatti} M}, \bibinfo{author}{{Navarro-Alsina} A},
  \bibinfo{author}{{Bernstein} GM}, \bibinfo{author}{{Jain} B},
  \bibinfo{author}{{Blazek} J}, \bibinfo{author}{{Alarcon} A},
  \bibinfo{author}{{Fert{\'e}} A}, \bibinfo{author}{{Lemos} P},
  \bibinfo{author}{{Raveri} M}, \bibinfo{author}{{Campos} A},
  \bibinfo{author}{{Prat} J}, \bibinfo{author}{{S{\'a}nchez} C},
  \bibinfo{author}{{Jarvis} M}, \bibinfo{author}{{Alves} O},
  \bibinfo{author}{{Andrade-Oliveira} F}, \bibinfo{author}{{Baxter} E},
  \bibinfo{author}{{Bechtol} K}, \bibinfo{author}{{Becker} MR},
  \bibinfo{author}{{Bridle} SL}, \bibinfo{author}{{Camacho} H},
  \bibinfo{author}{{Carnero Rosell} A}, \bibinfo{author}{{Carrasco Kind} M},
  \bibinfo{author}{{Cawthon} R}, \bibinfo{author}{{Chang} C},
  \bibinfo{author}{{Chen} R}, \bibinfo{author}{{Chintalapati} P},
  \bibinfo{author}{{Crocce} M}, \bibinfo{author}{{Davis} C},
  \bibinfo{author}{{Diehl} HT}, \bibinfo{author}{{Drlica-Wagner} A},
  \bibinfo{author}{{Eckert} K}, \bibinfo{author}{{Eifler} TF},
  \bibinfo{author}{{Elvin-Poole} J}, \bibinfo{author}{{Everett} S},
  \bibinfo{author}{{Fang} X}, \bibinfo{author}{{Fosalba} P},
  \bibinfo{author}{{Friedrich} O}, \bibinfo{author}{{Gaztanaga} E},
  \bibinfo{author}{{Giannini} G}, \bibinfo{author}{{Gruendl} RA},
  \bibinfo{author}{{Harrison} I}, \bibinfo{author}{{Hartley} WG},
  \bibinfo{author}{{Herner} K}, \bibinfo{author}{{Huang} H},
  \bibinfo{author}{{Huff} EM}, \bibinfo{author}{{Huterer} D},
  \bibinfo{author}{{Kuropatkin} N}, \bibinfo{author}{{Leget} P},
  \bibinfo{author}{{Liddle} AR}, \bibinfo{author}{{McCullough} J},
  \bibinfo{author}{{Muir} J}, \bibinfo{author}{{Pandey} S},
  \bibinfo{author}{{Park} Y}, \bibinfo{author}{{Porredon} A},
  \bibinfo{author}{{Refregier} A}, \bibinfo{author}{{Rollins} RP},
  \bibinfo{author}{{Roodman} A}, \bibinfo{author}{{Rosenfeld} R},
  \bibinfo{author}{{Ross} AJ}, \bibinfo{author}{{Rykoff} ES},
  \bibinfo{author}{{Sanchez} J}, \bibinfo{author}{{Sevilla-Noarbe} I},
  \bibinfo{author}{{Sheldon} E}, \bibinfo{author}{{Shin} T},
  \bibinfo{author}{{Troja} A}, \bibinfo{author}{{Tutusaus} I},
  \bibinfo{author}{{Tutusaus} I}, \bibinfo{author}{{Varga} TN},
  \bibinfo{author}{{Weaverdyck} N}, \bibinfo{author}{{Yanny} B},
  \bibinfo{author}{{Yin} B}, \bibinfo{author}{{Zhang} Y},
  \bibinfo{author}{{Zuntz} J}, \bibinfo{author}{{Aguena} M},
  \bibinfo{author}{{Allam} S}, \bibinfo{author}{{Annis} J},
  \bibinfo{author}{{Bacon} D}, \bibinfo{author}{{Bertin} E},
  \bibinfo{author}{{Bhargava} S}, \bibinfo{author}{{Brooks} D},
  \bibinfo{author}{{Buckley-Geer} E}, \bibinfo{author}{{Burke} DL},
  \bibinfo{author}{{Carretero} J}, \bibinfo{author}{{Costanzi} M},
  \bibinfo{author}{{da Costa} LN}, \bibinfo{author}{{Pereira} MES},
  \bibinfo{author}{{De Vicente} J}, \bibinfo{author}{{Desai} S},
  \bibinfo{author}{{Dietrich} JP}, \bibinfo{author}{{Doel} P},
  \bibinfo{author}{{Ferrero} I}, \bibinfo{author}{{Flaugher} B},
  \bibinfo{author}{{Frieman} J}, \bibinfo{author}{{Garc{\'\i}a-Bellido} J},
  \bibinfo{author}{{Gaztanaga} E}, \bibinfo{author}{{Gerdes} DW},
  \bibinfo{author}{{Giannantonio} T}, \bibinfo{author}{{Gschwend} J},
  \bibinfo{author}{{Gutierrez} G}, \bibinfo{author}{{Hinton} SR},
  \bibinfo{author}{{Hollowood} DL}, \bibinfo{author}{{Honscheid} K},
  \bibinfo{author}{{Hoyle} B}, \bibinfo{author}{{James} DJ},
  \bibinfo{author}{{Kron} R}, \bibinfo{author}{{Kuehn} K},
  \bibinfo{author}{{Lahav} O}, \bibinfo{author}{{Lima} M},
  \bibinfo{author}{{Lin} H}, \bibinfo{author}{{Maia} MAG},
  \bibinfo{author}{{Marshall} JL}, \bibinfo{author}{{Martini} P},
  \bibinfo{author}{{Melchior} P}, \bibinfo{author}{{Menanteau} F},
  \bibinfo{author}{{Miquel} R}, \bibinfo{author}{{Mohr} JJ},
  \bibinfo{author}{{Morgan} R}, \bibinfo{author}{{Ogando} RLC},
  \bibinfo{author}{{Palmese} A}, \bibinfo{author}{{Paz-Chinch{\'o}n} F},
  \bibinfo{author}{{Petravick} D}, \bibinfo{author}{{Pieres} A},
  \bibinfo{author}{{Romer} AK}, \bibinfo{author}{{Sanchez} E},
  \bibinfo{author}{{Scarpine} V}, \bibinfo{author}{{Schubnell} M},
  \bibinfo{author}{{Serrano} S}, \bibinfo{author}{{Smith} M},
  \bibinfo{author}{{Soares-Santos} M}, \bibinfo{author}{{Tarle} G},
  \bibinfo{author}{{Thomas} D}, \bibinfo{author}{{To} C},
  \bibinfo{author}{{Weller} J} and  \bibinfo{author}{{DES Collaboration}}
  (\bibinfo{year}{2022}), \bibinfo{month}{Jan.}
\bibinfo{title}{{Dark Energy Survey Year 3 results: Cosmology from cosmic shear
  and robustness to data calibration}}.
\bibinfo{journal}{{\em \prd}} \bibinfo{volume}{105} (\bibinfo{number}{2}),
  \bibinfo{eid}{023514}. \bibinfo{doi}{\doi{10.1103/PhysRevD.105.023514}}.
\eprint{2105.13543}.

\bibtype{Article}%
\bibitem[{Aric{\`o}} et al.(2021)]{Arico2021}
\bibinfo{author}{{Aric{\`o}} G}, \bibinfo{author}{{Angulo} RE},
  \bibinfo{author}{{Contreras} S}, \bibinfo{author}{{Ondaro-Mallea} L},
  \bibinfo{author}{{Pellejero-Iba{\~n}ez} M} and  \bibinfo{author}{{Zennaro} M}
  (\bibinfo{year}{2021}), \bibinfo{month}{Sep.}
\bibinfo{title}{{The BACCO simulation project: a baryonification emulator with
  neural networks}}.
\bibinfo{journal}{{\em \mnras}} \bibinfo{volume}{506} (\bibinfo{number}{3}):
  \bibinfo{pages}{4070--4082}. \bibinfo{doi}{\doi{10.1093/mnras/stab1911}}.
\eprint{2011.15018}.

\bibtype{Article}%
\bibitem[{Asgari} et al.(2021)]{Asgari2021}
\bibinfo{author}{{Asgari} M}, \bibinfo{author}{{Lin} CA},
  \bibinfo{author}{{Joachimi} B}, \bibinfo{author}{{Giblin} B},
  \bibinfo{author}{{Heymans} C}, \bibinfo{author}{{Hildebrandt} H},
  \bibinfo{author}{{Kannawadi} A}, \bibinfo{author}{{St{\"o}lzner} B},
  \bibinfo{author}{{Tr{\"o}ster} T}, \bibinfo{author}{{van den Busch} JL},
  \bibinfo{author}{{Wright} AH}, \bibinfo{author}{{Bilicki} M},
  \bibinfo{author}{{Blake} C}, \bibinfo{author}{{de Jong} J},
  \bibinfo{author}{{Dvornik} A}, \bibinfo{author}{{Erben} T},
  \bibinfo{author}{{Getman} F}, \bibinfo{author}{{Hoekstra} H},
  \bibinfo{author}{{K{\"o}hlinger} F}, \bibinfo{author}{{Kuijken} K},
  \bibinfo{author}{{Miller} L}, \bibinfo{author}{{Radovich} M},
  \bibinfo{author}{{Schneider} P}, \bibinfo{author}{{Shan} H} and
  \bibinfo{author}{{Valentijn} E} (\bibinfo{year}{2021}), \bibinfo{month}{Jan.}
\bibinfo{title}{{KiDS-1000 cosmology: Cosmic shear constraints and comparison
  between two point statistics}}.
\bibinfo{journal}{{\em \aap}} \bibinfo{volume}{645}, \bibinfo{eid}{A104}.
  \bibinfo{doi}{\doi{10.1051/0004-6361/202039070}}.
\eprint{2007.15633}.

\bibtype{Article}%
\bibitem[Bartelmann and Schneider(2001)]{Bartelmann2001}
\bibinfo{author}{Bartelmann M} and  \bibinfo{author}{Schneider P}
  (\bibinfo{year}{2001}), \bibinfo{month}{jan}.
\bibinfo{title}{{Weak gravitational lensing}}.
\bibinfo{journal}{{\em Phys. Rep.}} \bibinfo{volume}{340}
  (\bibinfo{number}{4-5}): \bibinfo{pages}{291--472}.
ISSN \bibinfo{issn}{03701573}.
  \bibinfo{doi}{\doi{10.1016/S0370-1573(00)00082-X}}.
\bibinfo{url}{\url{http://linkinghub.elsevier.com/retrieve/pii/S037015730000082X}}.

\bibtype{Article}%
\bibitem[Bernstein(2009)]{Bernstein2009}
\bibinfo{author}{Bernstein GM} (\bibinfo{year}{2009}), \bibinfo{month}{apr}.
\bibinfo{title}{{COMPREHENSIVE TWO-POINT ANALYSES OF WEAK GRAVITATIONAL LENSING
  SURVEYS}}.
\bibinfo{journal}{{\em \apj}} \bibinfo{volume}{695} (\bibinfo{number}{1}):
  \bibinfo{pages}{652--665}.
ISSN \bibinfo{issn}{0004-637X}.
  \bibinfo{doi}{\doi{10.1088/0004-637X/695/1/652}}.
\bibinfo{url}{\url{http://stacks.iop.org/0004-637X/695/i=1/a=652?key=crossref.f97c38b48449946e0fcb679780bb1133}}.

\bibtype{Article}%
\bibitem[{Blazek} et al.(2019)]{Blazek_2019}
\bibinfo{author}{{Blazek} JA}, \bibinfo{author}{{MacCrann} N},
  \bibinfo{author}{{Troxel} MA} and  \bibinfo{author}{{Fang} X}
  (\bibinfo{year}{2019}), \bibinfo{month}{Nov.}
\bibinfo{title}{{Beyond linear galaxy alignments}}.
\bibinfo{journal}{{\em \prd}} \bibinfo{volume}{100} (\bibinfo{number}{10}),
  \bibinfo{eid}{103506}. \bibinfo{doi}{\doi{10.1103/PhysRevD.100.103506}}.
\eprint{1708.09247}.

\bibtype{Article}%
\bibitem[Bridle and King(2007)]{Bridle2007}
\bibinfo{author}{Bridle S} and  \bibinfo{author}{King L}
  (\bibinfo{year}{2007}), \bibinfo{month}{dec}.
\bibinfo{title}{{Dark energy constraints from cosmic shear power spectra:
  impact of intrinsic alignments on photometric redshift requirements}}.
\bibinfo{journal}{{\em New J. Phys.}} \bibinfo{volume}{9}
  (\bibinfo{number}{12}): \bibinfo{pages}{444--444}.
ISSN \bibinfo{issn}{1367-2630}.
  \bibinfo{doi}{\doi{10.1088/1367-2630/9/12/444}}.
\bibinfo{url}{\url{http://stacks.iop.org/1367-2630/9/i=12/a=444?key=crossref.963e3bacba42a32c810320e17f938ce6}}.

\bibtype{Article}%
\bibitem[Brown et al.(2002)]{Brown2002}
\bibinfo{author}{Brown ML}, \bibinfo{author}{Taylor AN},
  \bibinfo{author}{Hambly NC} and  \bibinfo{author}{Dye S}
  (\bibinfo{year}{2002}), \bibinfo{month}{jul}.
\bibinfo{title}{{Measurement of intrinsic alignments in galaxy ellipticities}}.
\bibinfo{journal}{{\em \mnras}} \bibinfo{volume}{333} (\bibinfo{number}{3}):
  \bibinfo{pages}{501--509}.
ISSN \bibinfo{issn}{0035-8711}.
  \bibinfo{doi}{\doi{10.1046/j.1365-8711.2002.05354.x}}.
\bibinfo{url}{\url{http://mnras.oxfordjournals.org/cgi/doi/10.1046/j.1365-8711.2002.05354.x}}.

\bibtype{Article}%
\bibitem[{Dalal} et al.(2023)]{Dalal2023}
\bibinfo{author}{{Dalal} R}, \bibinfo{author}{{Li} X},
  \bibinfo{author}{{Nicola} A}, \bibinfo{author}{{Zuntz} J},
  \bibinfo{author}{{Strauss} MA}, \bibinfo{author}{{Sugiyama} S},
  \bibinfo{author}{{Zhang} T}, \bibinfo{author}{{Rau} MM},
  \bibinfo{author}{{Mandelbaum} R}, \bibinfo{author}{{Takada} M},
  \bibinfo{author}{{More} S}, \bibinfo{author}{{Miyatake} H},
  \bibinfo{author}{{Kannawadi} A}, \bibinfo{author}{{Shirasaki} M},
  \bibinfo{author}{{Taniguchi} T}, \bibinfo{author}{{Takahashi} R},
  \bibinfo{author}{{Osato} K}, \bibinfo{author}{{Hamana} T},
  \bibinfo{author}{{Oguri} M}, \bibinfo{author}{{Nishizawa} AJ},
  \bibinfo{author}{{Malag{\'o}n} AAP}, \bibinfo{author}{{Sunayama} T},
  \bibinfo{author}{{Alonso} D}, \bibinfo{author}{{Slosar} A},
  \bibinfo{author}{{Luo} W}, \bibinfo{author}{{Armstrong} R},
  \bibinfo{author}{{Bosch} J}, \bibinfo{author}{{Hsieh} BC},
  \bibinfo{author}{{Komiyama} Y}, \bibinfo{author}{{Lupton} RH},
  \bibinfo{author}{{Lust} NB}, \bibinfo{author}{{MacArthur} LA},
  \bibinfo{author}{{Miyazaki} S}, \bibinfo{author}{{Murayama} H},
  \bibinfo{author}{{Nishimichi} T}, \bibinfo{author}{{Okura} Y},
  \bibinfo{author}{{Price} PA}, \bibinfo{author}{{Tait} PJ},
  \bibinfo{author}{{Tanaka} M} and  \bibinfo{author}{{Wang} SY}
  (\bibinfo{year}{2023}), \bibinfo{month}{Dec.}
\bibinfo{title}{{Hyper Suprime-Cam Year 3 results: Cosmology from cosmic shear
  power spectra}}.
\bibinfo{journal}{{\em \prd}} \bibinfo{volume}{108} (\bibinfo{number}{12}),
  \bibinfo{eid}{123519}. \bibinfo{doi}{\doi{10.1103/PhysRevD.108.123519}}.
\eprint{2304.00701}.

\bibtype{Article}%
\bibitem[de~Jong et al.(2013)]{deJong2013}
\bibinfo{author}{de~Jong JTA}, \bibinfo{author}{{Verdoes Kleijn} GA},
  \bibinfo{author}{Kuijken KH} and  \bibinfo{author}{Valentijn EA}
  (\bibinfo{year}{2013}), \bibinfo{month}{jan}.
\bibinfo{title}{{The Kilo-Degree Survey}}.
\bibinfo{journal}{{\em Exp. Astron.}} \bibinfo{volume}{35}
  (\bibinfo{number}{1-2}): \bibinfo{pages}{25--44}.
ISSN \bibinfo{issn}{0922-6435}. \bibinfo{doi}{\doi{10.1007/s10686-012-9306-1}}.
\bibinfo{url}{\url{http://link.springer.com/10.1007/s10686-012-9306-1}}.

\bibtype{Book}%
\bibitem[{Dodelson}(2017)]{Dodelson2017}
\bibinfo{author}{{Dodelson} S} (\bibinfo{year}{2017}).
\bibinfo{title}{Gravitational lensing}.

\bibtype{Article}%
\bibitem[{Eifler} et al.(2021)]{Eifler2021}
\bibinfo{author}{{Eifler} T}, \bibinfo{author}{{Miyatake} H},
  \bibinfo{author}{{Krause} E}, \bibinfo{author}{{Heinrich} C},
  \bibinfo{author}{{Miranda} V}, \bibinfo{author}{{Hirata} C},
  \bibinfo{author}{{Xu} J}, \bibinfo{author}{{Hemmati} S},
  \bibinfo{author}{{Simet} M}, \bibinfo{author}{{Capak} P},
  \bibinfo{author}{{Choi} A}, \bibinfo{author}{{Dor{\'e}} O},
  \bibinfo{author}{{Doux} C}, \bibinfo{author}{{Fang} X},
  \bibinfo{author}{{Hounsell} R}, \bibinfo{author}{{Huff} E},
  \bibinfo{author}{{Huang} HJ}, \bibinfo{author}{{Jarvis} M},
  \bibinfo{author}{{Kruk} J}, \bibinfo{author}{{Masters} D},
  \bibinfo{author}{{Rozo} E}, \bibinfo{author}{{Scolnic} D},
  \bibinfo{author}{{Spergel} DN}, \bibinfo{author}{{Troxel} M},
  \bibinfo{author}{{von der Linden} A}, \bibinfo{author}{{Wang} Y},
  \bibinfo{author}{{Weinberg} DH}, \bibinfo{author}{{Wenzl} L} and
  \bibinfo{author}{{Wu} HY} (\bibinfo{year}{2021}), \bibinfo{month}{Oct.}
\bibinfo{title}{{Cosmology with the Roman Space Telescope - multiprobe
  strategies}}.
\bibinfo{journal}{{\em \mnras}} \bibinfo{volume}{507} (\bibinfo{number}{2}):
  \bibinfo{pages}{1746--1761}. \bibinfo{doi}{\doi{10.1093/mnras/stab1762}}.
\eprint{2004.05271}.

\bibtype{Article}%
\bibitem[{Elvin-Poole} et al.(2023)]{Elvin-Poole2023}
\bibinfo{author}{{Elvin-Poole} J}, \bibinfo{author}{{MacCrann} N},
  \bibinfo{author}{{Everett} S}, \bibinfo{author}{{Prat} J},
  \bibinfo{author}{{Rykoff} ES}, \bibinfo{author}{{De Vicente} J},
  \bibinfo{author}{{Yanny} B}, \bibinfo{author}{{Herner} K},
  \bibinfo{author}{{Fert{\'e}} A}, \bibinfo{author}{{Di Valentino} E},
  \bibinfo{author}{{Choi} A}, \bibinfo{author}{{Burke} DL},
  \bibinfo{author}{{Sevilla-Noarbe} I}, \bibinfo{author}{{Alarcon} A},
  \bibinfo{author}{{Alves} O}, \bibinfo{author}{{Amon} A},
  \bibinfo{author}{{Andrade-Oliveira} F}, \bibinfo{author}{{Baxter} E},
  \bibinfo{author}{{Bechtol} K}, \bibinfo{author}{{Becker} MR},
  \bibinfo{author}{{Bernstein} GM}, \bibinfo{author}{{Blazek} J},
  \bibinfo{author}{{Camacho} H}, \bibinfo{author}{{Campos} A},
  \bibinfo{author}{{Carnero Rosell} A}, \bibinfo{author}{{Carrasco Kind} M},
  \bibinfo{author}{{Cawthon} R}, \bibinfo{author}{{Chang} C},
  \bibinfo{author}{{Chen} R}, \bibinfo{author}{{Cordero} J},
  \bibinfo{author}{{Crocce} M}, \bibinfo{author}{{Davis} C},
  \bibinfo{author}{{DeRose} J}, \bibinfo{author}{{Diehl} HT},
  \bibinfo{author}{{Dodelson} S}, \bibinfo{author}{{Doux} C},
  \bibinfo{author}{{Drlica-Wagner} A}, \bibinfo{author}{{Eckert} K},
  \bibinfo{author}{{Eifler} TF}, \bibinfo{author}{{Elsner} F},
  \bibinfo{author}{{Fang} X}, \bibinfo{author}{{Fosalba} P},
  \bibinfo{author}{{Friedrich} O}, \bibinfo{author}{{Gatti} M},
  \bibinfo{author}{{Giannini} G}, \bibinfo{author}{{Gruen} D},
  \bibinfo{author}{{Gruendl} RA}, \bibinfo{author}{{Harrison} I},
  \bibinfo{author}{{Hartley} WG}, \bibinfo{author}{{Huang} H},
  \bibinfo{author}{{Huff} EM}, \bibinfo{author}{{Huterer} D},
  \bibinfo{author}{{Krause} E}, \bibinfo{author}{{Kuropatkin} N},
  \bibinfo{author}{{Leget} PF}, \bibinfo{author}{{Lemos} P},
  \bibinfo{author}{{Liddle} AR}, \bibinfo{author}{{McCullough} J},
  \bibinfo{author}{{Muir} J}, \bibinfo{author}{{Myles} J},
  \bibinfo{author}{{Navarro-Alsina} A}, \bibinfo{author}{{Pandey} S},
  \bibinfo{author}{{Park} Y}, \bibinfo{author}{{Porredon} A},
  \bibinfo{author}{{Raveri} M}, \bibinfo{author}{{Rodriguez-Monroy} M},
  \bibinfo{author}{{Rollins} RP}, \bibinfo{author}{{Roodman} A},
  \bibinfo{author}{{Rosenfeld} R}, \bibinfo{author}{{Ross} AJ},
  \bibinfo{author}{{S{\'a}nchez} C}, \bibinfo{author}{{Sanchez} J},
  \bibinfo{author}{{Secco} LF}, \bibinfo{author}{{Sheldon} E},
  \bibinfo{author}{{Shin} T}, \bibinfo{author}{{Troxel} MA},
  \bibinfo{author}{{Tutusaus} I}, \bibinfo{author}{{Varga} TN},
  \bibinfo{author}{{Weaverdyck} N}, \bibinfo{author}{{Wechsler} RH},
  \bibinfo{author}{{Yin} B}, \bibinfo{author}{{Zhang} Y},
  \bibinfo{author}{{Zuntz} J}, \bibinfo{author}{{Aguena} M},
  \bibinfo{author}{{Avila} S}, \bibinfo{author}{{Bacon} D},
  \bibinfo{author}{{Bertin} E}, \bibinfo{author}{{Bocquet} S},
  \bibinfo{author}{{Brooks} D}, \bibinfo{author}{{Garc{\'\i}a-Bellido} J},
  \bibinfo{author}{{Honscheid} K}, \bibinfo{author}{{Jarvis} M},
  \bibinfo{author}{{Li} TS}, \bibinfo{author}{{Mena-Fern{\'a}ndez} J},
  \bibinfo{author}{{To} C}, \bibinfo{author}{{Wilkinson} RD} and
  \bibinfo{author}{{DES Collaboration}} (\bibinfo{year}{2023}),
  \bibinfo{month}{Aug.}
\bibinfo{title}{{Dark Energy Survey Year 3 results: magnification modelling and
  impact on cosmological constraints from galaxy clustering and galaxy-galaxy
  lensing}}.
\bibinfo{journal}{{\em \mnras}} \bibinfo{volume}{523} (\bibinfo{number}{3}):
  \bibinfo{pages}{3649--3670}. \bibinfo{doi}{\doi{10.1093/mnras/stad1594}}.
\eprint{2209.09782}.

\bibtype{Article}%
\bibitem[{Euclid Collaboration} et al.(2024)]{Euclid2024}
\bibinfo{author}{{Euclid Collaboration}}, \bibinfo{author}{{Mellier} Y},
  \bibinfo{author}{{Abdurro'uf}}, \bibinfo{author}{{Acevedo Barroso} JA},
  \bibinfo{author}{{Ach{\'u}carro} A}, \bibinfo{author}{{Adamek} J},
  \bibinfo{author}{{Adam} R}, \bibinfo{author}{{Addison} GE},
  \bibinfo{author}{{Aghanim} N}, \bibinfo{author}{{Aguena} M},
  \bibinfo{author}{{Ajani} V}, \bibinfo{author}{{Akrami} Y},
  \bibinfo{author}{{Al-Bahlawan} A}, \bibinfo{author}{{Alavi} A},
  \bibinfo{author}{{Albuquerque} IS}, \bibinfo{author}{{Alestas} G},
  \bibinfo{author}{{Alguero} G}, \bibinfo{author}{{Allaoui} A},
  \bibinfo{author}{{Allen} SW}, \bibinfo{author}{{Allevato} V},
  \bibinfo{author}{{Alonso-Tetilla} AV}, \bibinfo{author}{{Altieri} B},
  \bibinfo{author}{{Alvarez-Candal} A}, \bibinfo{author}{{Amara} A},
  \bibinfo{author}{{Amendola} L}, \bibinfo{author}{{Amiaux} J},
  \bibinfo{author}{{Andika} IT}, \bibinfo{author}{{Andreon} S},
  \bibinfo{author}{{Andrews} A}, \bibinfo{author}{{Angora} G},
  \bibinfo{author}{{Angulo} RE}, \bibinfo{author}{{Annibali} F},
  \bibinfo{author}{{Anselmi} A}, \bibinfo{author}{{Anselmi} S},
  \bibinfo{author}{{Arcari} S}, \bibinfo{author}{{Archidiacono} M},
  \bibinfo{author}{{Aric{\`o}} G}, \bibinfo{author}{{Arnaud} M},
  \bibinfo{author}{{Arnouts} S}, \bibinfo{author}{{Asgari} M},
  \bibinfo{author}{{Asorey} J}, \bibinfo{author}{{Atayde} L},
  \bibinfo{author}{{Atek} H}, \bibinfo{author}{{Atrio-Barandela} F},
  \bibinfo{author}{{Aubert} M}, \bibinfo{author}{{Aubourg} E},
  \bibinfo{author}{{Auphan} T}, \bibinfo{author}{{Auricchio} N},
  \bibinfo{author}{{Aussel} B}, \bibinfo{author}{{Aussel} H},
  \bibinfo{author}{{Avelino} PP}, \bibinfo{author}{{Avgoustidis} A},
  \bibinfo{author}{{Avila} S}, \bibinfo{author}{{Awan} S},
  \bibinfo{author}{{Azzollini} R}, \bibinfo{author}{{Baccigalupi} C},
  \bibinfo{author}{{Bachelet} E}, \bibinfo{author}{{Bacon} D},
  \bibinfo{author}{{Baes} M}, \bibinfo{author}{{Bagley} MB},
  \bibinfo{author}{{Bahr-Kalus} B}, \bibinfo{author}{{Balaguera-Antolinez} A},
  \bibinfo{author}{{Balbinot} E}, \bibinfo{author}{{Balcells} M},
  \bibinfo{author}{{Baldi} M}, \bibinfo{author}{{Baldry} I},
  \bibinfo{author}{{Balestra} A}, \bibinfo{author}{{Ballardini} M},
  \bibinfo{author}{{Ballester} O}, \bibinfo{author}{{Balogh} M},
  \bibinfo{author}{{Ba{\~n}ados} E}, \bibinfo{author}{{Barbier} R},
  \bibinfo{author}{{Bardelli} S}, \bibinfo{author}{{Barreiro} T},
  \bibinfo{author}{{Barriere} JC}, \bibinfo{author}{{Barros} BJ},
  \bibinfo{author}{{Barthelemy} A}, \bibinfo{author}{{Bartolo} N},
  \bibinfo{author}{{Basset} A}, \bibinfo{author}{{Battaglia} P},
  \bibinfo{author}{{Battisti} AJ}, \bibinfo{author}{{Baugh} CM},
  \bibinfo{author}{{Baumont} L}, \bibinfo{author}{{Bazzanini} L},
  \bibinfo{author}{{Beaulieu} JP}, \bibinfo{author}{{Beckmann} V},
  \bibinfo{author}{{Belikov} AN}, \bibinfo{author}{{Bel} J},
  \bibinfo{author}{{Bellagamba} F}, \bibinfo{author}{{Bella} M},
  \bibinfo{author}{{Bellini} E}, \bibinfo{author}{{Benabed} K},
  \bibinfo{author}{{Bender} R}, \bibinfo{author}{{Benevento} G},
  \bibinfo{author}{{Bennett} CL}, \bibinfo{author}{{Benson} K},
  \bibinfo{author}{{Bergamini} P}, \bibinfo{author}{{Bermejo-Climent} JR},
  \bibinfo{author}{{Bernardeau} F}, \bibinfo{author}{{Bertacca} D},
  \bibinfo{author}{{Berthe} M}, \bibinfo{author}{{Berthier} J},
  \bibinfo{author}{{Bethermin} M}, \bibinfo{author}{{Beutler} F},
  \bibinfo{author}{{Bevillon} C}, \bibinfo{author}{{Bhargava} S},
  \bibinfo{author}{{Bhatawdekar} R}, \bibinfo{author}{{Bisigello} L},
  \bibinfo{author}{{Biviano} A}, \bibinfo{author}{{Blake} RP},
  \bibinfo{author}{{Blanchard} A}, \bibinfo{author}{{Blazek} J},
  \bibinfo{author}{{Blot} L}, \bibinfo{author}{{Bosco} A},
  \bibinfo{author}{{Bodendorf} C}, \bibinfo{author}{{Boenke} T},
  \bibinfo{author}{{B{\"o}hringer} H}, \bibinfo{author}{{Bolzonella} M},
  \bibinfo{author}{{Bonchi} A}, \bibinfo{author}{{Bonici} M},
  \bibinfo{author}{{Bonino} D}, \bibinfo{author}{{Bonino} L},
  \bibinfo{author}{{Bonvin} C}, \bibinfo{author}{{Bon} W},
  \bibinfo{author}{{Booth} JT}, \bibinfo{author}{{Borgani} S},
  \bibinfo{author}{{Borlaff} AS}, \bibinfo{author}{{Borsato} E},
  \bibinfo{author}{{Bosco} A}, \bibinfo{author}{{Bose} B},
  \bibinfo{author}{{Botticella} MT}, \bibinfo{author}{{Boucaud} A},
  \bibinfo{author}{{Bouche} F}, \bibinfo{author}{{Boucher} JS},
  \bibinfo{author}{{Boutigny} D}, \bibinfo{author}{{Bouvard} T},
  \bibinfo{author}{{Bouy} H}, \bibinfo{author}{{Bowler} RAA},
  \bibinfo{author}{{Bozza} V}, \bibinfo{author}{{Bozzo} E},
  \bibinfo{author}{{Branchini} E}, \bibinfo{author}{{Brau-Nogue} S},
  \bibinfo{author}{{Brekke} P}, \bibinfo{author}{{Bremer} MN},
  \bibinfo{author}{{Brescia} M}, \bibinfo{author}{{Breton} MA},
  \bibinfo{author}{{Brinchmann} J}, \bibinfo{author}{{Brinckmann} T},
  \bibinfo{author}{{Brockley-Blatt} C}, \bibinfo{author}{{Brodwin} M},
  \bibinfo{author}{{Brouard} L}, \bibinfo{author}{{Brown} ML},
  \bibinfo{author}{{Bruton} S}, \bibinfo{author}{{Bucko} J},
  \bibinfo{author}{{Buddelmeijer} H}, \bibinfo{author}{{Buenadicha} G},
  \bibinfo{author}{{Buitrago} F}, \bibinfo{author}{{Burger} P},
  \bibinfo{author}{{Burigana} C}, \bibinfo{author}{{Busillo} V},
  \bibinfo{author}{{Busonero} D}, \bibinfo{author}{{Cabanac} R},
  \bibinfo{author}{{Cabayol-Garcia} L}, \bibinfo{author}{{Cagliari} MS},
  \bibinfo{author}{{Caillat} A}, \bibinfo{author}{{Caillat} L},
  \bibinfo{author}{{Calabrese} M}, \bibinfo{author}{{Calabro} A},
  \bibinfo{author}{{Calderone} G}, \bibinfo{author}{{Calura} F},
  \bibinfo{author}{{Camacho Quevedo} B}, \bibinfo{author}{{Camera} S},
  \bibinfo{author}{{Campos} L}, \bibinfo{author}{{Canas-Herrera} G},
  \bibinfo{author}{{Candini} GP}, \bibinfo{author}{{Cantiello} M},
  \bibinfo{author}{{Capobianco} V}, \bibinfo{author}{{Cappellaro} E},
  \bibinfo{author}{{Cappelluti} N}, \bibinfo{author}{{Cappi} A},
  \bibinfo{author}{{Caputi} KI}, \bibinfo{author}{{Cara} C},
  \bibinfo{author}{{Carbone} C}, \bibinfo{author}{{Cardone} VF},
  \bibinfo{author}{{Carella} E}, \bibinfo{author}{{Carlberg} RG},
  \bibinfo{author}{{Carle} M}, \bibinfo{author}{{Carminati} L},
  \bibinfo{author}{{Caro} F}, \bibinfo{author}{{Carrasco} JM},
  \bibinfo{author}{{Carretero} J}, \bibinfo{author}{{Carrilho} P},
  \bibinfo{author}{{Carron Duque} J}, \bibinfo{author}{{Carry} B},
  \bibinfo{author}{{Carvalho} A}, \bibinfo{author}{{Carvalho} CS},
  \bibinfo{author}{{Casas} R}, \bibinfo{author}{{Casas} S},
  \bibinfo{author}{{Casenove} P}, \bibinfo{author}{{Casey} CM},
  \bibinfo{author}{{Cassata} P}, \bibinfo{author}{{Castander} FJ},
  \bibinfo{author}{{Castelao} D}, \bibinfo{author}{{Castellano} M},
  \bibinfo{author}{{Castiblanco} L}, \bibinfo{author}{{Castignani} G},
  \bibinfo{author}{{Castro} T}, \bibinfo{author}{{Cavet} C},
  \bibinfo{author}{{Cavuoti} S}, \bibinfo{author}{{Chabaud} PY},
  \bibinfo{author}{{Chambers} KC}, \bibinfo{author}{{Charles} Y},
  \bibinfo{author}{{Charlot} S}, \bibinfo{author}{{Chartab} N},
  \bibinfo{author}{{Chary} R}, \bibinfo{author}{{Chaumeil} F},
  \bibinfo{author}{{Cho} H}, \bibinfo{author}{{Chon} G},
  \bibinfo{author}{{Ciancetta} E}, \bibinfo{author}{{Ciliegi} P},
  \bibinfo{author}{{Cimatti} A}, \bibinfo{author}{{Cimino} M},
  \bibinfo{author}{{Cioni} MRL}, \bibinfo{author}{{Claydon} R},
  \bibinfo{author}{{Cleland} C}, \bibinfo{author}{{Cl{\'e}ment} B},
  \bibinfo{author}{{Clements} DL}, \bibinfo{author}{{Clerc} N},
  \bibinfo{author}{{Clesse} S}, \bibinfo{author}{{Codis} S},
  \bibinfo{author}{{Cogato} F}, \bibinfo{author}{{Colbert} J},
  \bibinfo{author}{{Cole} RE}, \bibinfo{author}{{Coles} P},
  \bibinfo{author}{{Collett} TE}, \bibinfo{author}{{Collins} RS},
  \bibinfo{author}{{Colodro-Conde} C}, \bibinfo{author}{{Colombo} C},
  \bibinfo{author}{{Combes} F}, \bibinfo{author}{{Conforti} V},
  \bibinfo{author}{{Congedo} G}, \bibinfo{author}{{Conseil} S},
  \bibinfo{author}{{Conselice} CJ}, \bibinfo{author}{{Contarini} S},
  \bibinfo{author}{{Contini} T}, \bibinfo{author}{{Conversi} L},
  \bibinfo{author}{{Cooray} AR}, \bibinfo{author}{{Copin} Y},
  \bibinfo{author}{{Corasaniti} PS}, \bibinfo{author}{{Corcho-Caballero} P},
  \bibinfo{author}{{Corcione} L}, \bibinfo{author}{{Cordes} O},
  \bibinfo{author}{{Corpace} O}, \bibinfo{author}{{Correnti} M},
  \bibinfo{author}{{Costanzi} M}, \bibinfo{author}{{Costille} A},
  \bibinfo{author}{{Courbin} F}, \bibinfo{author}{{Courcoult Mifsud} L},
  \bibinfo{author}{{Courtois} HM}, \bibinfo{author}{{Cousinou} MC},
  \bibinfo{author}{{Covone} G}, \bibinfo{author}{{Cowell} T},
  \bibinfo{author}{{Cragg} C}, \bibinfo{author}{{Cresci} G},
  \bibinfo{author}{{Cristiani} S}, \bibinfo{author}{{Crocce} M},
  \bibinfo{author}{{Cropper} M}, \bibinfo{author}{{E Crouzet} P},
  \bibinfo{author}{{Csizi} B}, \bibinfo{author}{{Cuby} JG},
  \bibinfo{author}{{Cucchetti} E}, \bibinfo{author}{{Cucciati} O},
  \bibinfo{author}{{Cuillandre} JC}, \bibinfo{author}{{Cunha} PAC},
  \bibinfo{author}{{Cuozzo} V}, \bibinfo{author}{{Daddi} E},
  \bibinfo{author}{{D'Addona} M}, \bibinfo{author}{{Dafonte} C},
  \bibinfo{author}{{Dagoneau} N}, \bibinfo{author}{{Dalessandro} E},
  \bibinfo{author}{{Dalton} GB}, \bibinfo{author}{{D'Amico} G},
  \bibinfo{author}{{Dannerbauer} H}, \bibinfo{author}{{Danto} P},
  \bibinfo{author}{{Das} I}, \bibinfo{author}{{Da Silva} A},
  \bibinfo{author}{{da Silva} R}, \bibinfo{author}{{Daste} G},
  \bibinfo{author}{{Davies} JE}, \bibinfo{author}{{Davini} S},
  \bibinfo{author}{{de Boer} T}, \bibinfo{author}{{Decarli} R},
  \bibinfo{author}{{De Caro} B}, \bibinfo{author}{{Degaudenzi} H},
  \bibinfo{author}{{Degni} G}, \bibinfo{author}{{de Jong} JTA},
  \bibinfo{author}{{de la Bella} LF}, \bibinfo{author}{{de la Torre} S},
  \bibinfo{author}{{Delhaise} F}, \bibinfo{author}{{Delley} D},
  \bibinfo{author}{{Delucchi} G}, \bibinfo{author}{{De Lucia} G},
  \bibinfo{author}{{Denniston} J}, \bibinfo{author}{{De Paolis} F},
  \bibinfo{author}{{De Petris} M}, \bibinfo{author}{{Derosa} A},
  \bibinfo{author}{{Desai} S}, \bibinfo{author}{{Desjacques} V},
  \bibinfo{author}{{Despali} G}, \bibinfo{author}{{Desprez} G},
  \bibinfo{author}{{De Vicente-Albendea} J}, \bibinfo{author}{{Deville} Y},
  \bibinfo{author}{{Dias} JDF}, \bibinfo{author}{{D{\'\i}az-S{\'a}nchez} A},
  \bibinfo{author}{{Diaz} JJ}, \bibinfo{author}{{Di Domizio} S},
  \bibinfo{author}{{Diego} JM}, \bibinfo{author}{{Di Ferdinando} D},
  \bibinfo{author}{{Di Giorgio} AM}, \bibinfo{author}{{Dimauro} P},
  \bibinfo{author}{{Dinis} J}, \bibinfo{author}{{Dolag} K},
  \bibinfo{author}{{Dolding} C}, \bibinfo{author}{{Dole} H},
  \bibinfo{author}{{Dom{\'\i}nguez S{\'a}nchez} H}, \bibinfo{author}{{Dor{\'e}}
  O}, \bibinfo{author}{{Dournac} F}, \bibinfo{author}{{Douspis} M},
  \bibinfo{author}{{Dreihahn} H}, \bibinfo{author}{{Droge} B},
  \bibinfo{author}{{Dryer} B}, \bibinfo{author}{{Dubath} F},
  \bibinfo{author}{{Duc} PA}, \bibinfo{author}{{Ducret} F},
  \bibinfo{author}{{Duffy} C}, \bibinfo{author}{{Dufresne} F},
  \bibinfo{author}{{Duncan} CAJ}, \bibinfo{author}{{Dupac} X},
  \bibinfo{author}{{Duret} V}, \bibinfo{author}{{Durrer} R},
  \bibinfo{author}{{Durret} F}, \bibinfo{author}{{Dusini} S},
  \bibinfo{author}{{Ealet} A}, \bibinfo{author}{{Eggemeier} A},
  \bibinfo{author}{{Eisenhardt} PRM}, \bibinfo{author}{{Elbaz} D},
  \bibinfo{author}{{Elkhashab} MY}, \bibinfo{author}{{Ellien} A},
  \bibinfo{author}{{Endicott} J}, \bibinfo{author}{{Enia} A},
  \bibinfo{author}{{Erben} T}, \bibinfo{author}{{Escartin Vigo} JA},
  \bibinfo{author}{{Escoffier} S}, \bibinfo{author}{{Escudero Sanz} I},
  \bibinfo{author}{{Essert} J}, \bibinfo{author}{{Ettori} S},
  \bibinfo{author}{{Ezziati} M}, \bibinfo{author}{{Fabbian} G},
  \bibinfo{author}{{Fabricius} M}, \bibinfo{author}{{Fang} Y},
  \bibinfo{author}{{Farina} A}, \bibinfo{author}{{Farina} M},
  \bibinfo{author}{{Farinelli} R}, \bibinfo{author}{{Farrens} S},
  \bibinfo{author}{{Faustini} F}, \bibinfo{author}{{Feltre} A},
  \bibinfo{author}{{Ferguson} AMN}, \bibinfo{author}{{Ferrando} P},
  \bibinfo{author}{{Ferrari} AG}, \bibinfo{author}{{Ferr{\'e}-Mateu} A},
  \bibinfo{author}{{Ferreira} PG}, \bibinfo{author}{{Ferreras} I},
  \bibinfo{author}{{Ferrero} I}, \bibinfo{author}{{Ferriol} S},
  \bibinfo{author}{{Ferruit} P}, \bibinfo{author}{{Filleul} D},
  \bibinfo{author}{{Finelli} F}, \bibinfo{author}{{Finkelstein} SL},
  \bibinfo{author}{{Finoguenov} A}, \bibinfo{author}{{Fiorini} B},
  \bibinfo{author}{{Flentge} F}, \bibinfo{author}{{Focardi} P},
  \bibinfo{author}{{Fonseca} J}, \bibinfo{author}{{Fontana} A},
  \bibinfo{author}{{Fontanot} F}, \bibinfo{author}{{Fornari} F},
  \bibinfo{author}{{Fosalba} P}, \bibinfo{author}{{Fossati} M},
  \bibinfo{author}{{Fotopoulou} S}, \bibinfo{author}{{Fouchez} D},
  \bibinfo{author}{{Fourmanoit} N}, \bibinfo{author}{{Frailis} M},
  \bibinfo{author}{{Fraix-Burnet} D}, \bibinfo{author}{{Franceschi} E},
  \bibinfo{author}{{Franco} A}, \bibinfo{author}{{Franzetti} P},
  \bibinfo{author}{{Freihoefer} J}, \bibinfo{author}{{Frittoli} G},
  \bibinfo{author}{{Frugier} PA}, \bibinfo{author}{{Frusciante} N},
  \bibinfo{author}{{Fumagalli} A}, \bibinfo{author}{{Fumagalli} M},
  \bibinfo{author}{{Fumana} M}, \bibinfo{author}{{Fu} Y},
  \bibinfo{author}{{Gabarra} L}, \bibinfo{author}{{Galeotta} S},
  \bibinfo{author}{{Galluccio} L}, \bibinfo{author}{{Ganga} K},
  \bibinfo{author}{{Gao} H}, \bibinfo{author}{{Garc{\'\i}a-Bellido} J},
  \bibinfo{author}{{Garcia} K}, \bibinfo{author}{{Gardner} JP},
  \bibinfo{author}{{Garilli} B}, \bibinfo{author}{{Gaspar-Venancio} LM},
  \bibinfo{author}{{Gasparetto} T}, \bibinfo{author}{{Gautard} V},
  \bibinfo{author}{{Gavazzi} R}, \bibinfo{author}{{Gaztanaga} E},
  \bibinfo{author}{{Genolet} L}, \bibinfo{author}{{Genova Santos} R},
  \bibinfo{author}{{Gentile} F}, \bibinfo{author}{{George} K},
  \bibinfo{author}{{Ghaffari} Z}, \bibinfo{author}{{Giacomini} F},
  \bibinfo{author}{{Gianotti} F}, \bibinfo{author}{{Gibb} GPS},
  \bibinfo{author}{{Gillard} W}, \bibinfo{author}{{Gillis} B},
  \bibinfo{author}{{Ginolfi} M}, \bibinfo{author}{{Giocoli} C},
  \bibinfo{author}{{Girardi} M}, \bibinfo{author}{{Giri} SK},
  \bibinfo{author}{{Goh} LWK}, \bibinfo{author}{{G{\'o}mez-Alvarez} P},
  \bibinfo{author}{{Gonzalez} AH}, \bibinfo{author}{{Gonzalez} EJ},
  \bibinfo{author}{{Gonzalez} JC}, \bibinfo{author}{{Gouyou Beauchamps} S},
  \bibinfo{author}{{Gozaliasl} G}, \bibinfo{author}{{Gracia-Carpio} J},
  \bibinfo{author}{{Grandis} S}, \bibinfo{author}{{Granett} BR},
  \bibinfo{author}{{Granvik} M}, \bibinfo{author}{{Grazian} A},
  \bibinfo{author}{{Gregorio} A}, \bibinfo{author}{{Grenet} C},
  \bibinfo{author}{{Grillo} C}, \bibinfo{author}{{Grupp} F},
  \bibinfo{author}{{Gruppioni} C}, \bibinfo{author}{{Gruppuso} A},
  \bibinfo{author}{{Guerbuez} C}, \bibinfo{author}{{Guerrini} S},
  \bibinfo{author}{{Guidi} M}, \bibinfo{author}{{Guillard} P},
  \bibinfo{author}{{Gutierrez} CM}, \bibinfo{author}{{Guttridge} P},
  \bibinfo{author}{{Guzzo} L}, \bibinfo{author}{{Gwyn} S},
  \bibinfo{author}{{Haapala} J}, \bibinfo{author}{{Haase} J},
  \bibinfo{author}{{Haddow} CR}, \bibinfo{author}{{Hailey} M},
  \bibinfo{author}{{Hall} A}, \bibinfo{author}{{Hall} D},
  \bibinfo{author}{{Hamaus} N}, \bibinfo{author}{{Haridasu} BS},
  \bibinfo{author}{{Harnois-D{\'e}raps} J}, \bibinfo{author}{{Harper} C},
  \bibinfo{author}{{Hartley} WG}, \bibinfo{author}{{Hasinger} G},
  \bibinfo{author}{{Hassani} F}, \bibinfo{author}{{Hatch} NA},
  \bibinfo{author}{{Haugan} SVH}, \bibinfo{author}{{H{\"a}u{\ss}ler} B},
  \bibinfo{author}{{Heavens} A}, \bibinfo{author}{{Heisenberg} L},
  \bibinfo{author}{{Helmi} A}, \bibinfo{author}{{Helou} G},
  \bibinfo{author}{{Hemmati} S}, \bibinfo{author}{{Henares} K},
  \bibinfo{author}{{Herent} O}, \bibinfo{author}{{Hern{\'a}ndez-Monteagudo} C},
  \bibinfo{author}{{Heuberger} T}, \bibinfo{author}{{Hewett} PC},
  \bibinfo{author}{{Heydenreich} S}, \bibinfo{author}{{Hildebrandt} H},
  \bibinfo{author}{{Hirschmann} M}, \bibinfo{author}{{Hjorth} J},
  \bibinfo{author}{{Hoar} J}, \bibinfo{author}{{Hoekstra} H},
  \bibinfo{author}{{Holland} AD}, \bibinfo{author}{{Holliman} MS},
  \bibinfo{author}{{Holmes} W}, \bibinfo{author}{{Hook} I},
  \bibinfo{author}{{Horeau} B}, \bibinfo{author}{{Hormuth} F},
  \bibinfo{author}{{Hornstrup} A}, \bibinfo{author}{{Hosseini} S},
  \bibinfo{author}{{Hu} D}, \bibinfo{author}{{Hudelot} P},
  \bibinfo{author}{{Hudson} MJ}, \bibinfo{author}{{Huertas-Company} M},
  \bibinfo{author}{{Huff} EM}, \bibinfo{author}{{Hughes} ACN},
  \bibinfo{author}{{Humphrey} A}, \bibinfo{author}{{Hunt} LK},
  \bibinfo{author}{{Huynh} DD}, \bibinfo{author}{{Ibata} R},
  \bibinfo{author}{{Ichikawa} K}, \bibinfo{author}{{Iglesias-Groth} S},
  \bibinfo{author}{{Ilbert} O}, \bibinfo{author}{{Ili{\'c}} S},
  \bibinfo{author}{{Ingoglia} L}, \bibinfo{author}{{Iodice} E},
  \bibinfo{author}{{Israel} H}, \bibinfo{author}{{Israelsson} UE},
  \bibinfo{author}{{Izzo} L}, \bibinfo{author}{{Jablonka} P},
  \bibinfo{author}{{Jackson} N}, \bibinfo{author}{{Jacobson} J},
  \bibinfo{author}{{Jafariyazani} M}, \bibinfo{author}{{Jahnke} K},
  \bibinfo{author}{{Jansen} H}, \bibinfo{author}{{Jarvis} MJ},
  \bibinfo{author}{{Jasche} J}, \bibinfo{author}{{Jauzac} M},
  \bibinfo{author}{{Jeffrey} N}, \bibinfo{author}{{Jhabvala} M},
  \bibinfo{author}{{Jimenez-Teja} Y}, \bibinfo{author}{{Jimenez Mu{\~n}oz} A},
  \bibinfo{author}{{Joachimi} B}, \bibinfo{author}{{Johansson} PH},
  \bibinfo{author}{{Joudaki} S}, \bibinfo{author}{{Jullo} E},
  \bibinfo{author}{{Kajava} JJE}, \bibinfo{author}{{Kang} Y},
  \bibinfo{author}{{Kannawadi} A}, \bibinfo{author}{{Kansal} V},
  \bibinfo{author}{{Karagiannis} D}, \bibinfo{author}{{K{\"a}rcher} M},
  \bibinfo{author}{{Kashlinsky} A}, \bibinfo{author}{{Kazandjian} MV},
  \bibinfo{author}{{Keck} F}, \bibinfo{author}{{Keih{\"a}nen} E},
  \bibinfo{author}{{Kerins} E}, \bibinfo{author}{{Kermiche} S},
  \bibinfo{author}{{Khalil} A}, \bibinfo{author}{{Kiessling} A},
  \bibinfo{author}{{Kiiveri} K}, \bibinfo{author}{{Kilbinger} M},
  \bibinfo{author}{{Kim} J}, \bibinfo{author}{{King} R},
  \bibinfo{author}{{Kirkpatrick} CC}, \bibinfo{author}{{Kitching} T},
  \bibinfo{author}{{Kluge} M}, \bibinfo{author}{{Knabenhans} M},
  \bibinfo{author}{{Knapen} JH}, \bibinfo{author}{{Knebe} A},
  \bibinfo{author}{{Kneib} JP}, \bibinfo{author}{{Kohley} R},
  \bibinfo{author}{{Koopmans} LVE}, \bibinfo{author}{{Koskinen} H},
  \bibinfo{author}{{Koulouridis} E}, \bibinfo{author}{{Kou} R},
  \bibinfo{author}{{Kov{\'a}cs} A}, \bibinfo{author}{{Kova\{{\v{c}}\}i{\'c}}
  I}, \bibinfo{author}{{Kowalczyk} A}, \bibinfo{author}{{Koyama} K},
  \bibinfo{author}{{Kraljic} K}, \bibinfo{author}{{Krause} O},
  \bibinfo{author}{{Kruk} S}, \bibinfo{author}{{Kubik} B},
  \bibinfo{author}{{Kuchner} U}, \bibinfo{author}{{Kuijken} K},
  \bibinfo{author}{{K{\"u}mmel} M}, \bibinfo{author}{{Kunz} M},
  \bibinfo{author}{{Kurki-Suonio} H}, \bibinfo{author}{{Lacasa} F},
  \bibinfo{author}{{Lacey} CG}, \bibinfo{author}{{La Franca} F},
  \bibinfo{author}{{Lagarde} N}, \bibinfo{author}{{Lahav} O},
  \bibinfo{author}{{Laigle} C}, \bibinfo{author}{{La Marca} A},
  \bibinfo{author}{{La Marle} O}, \bibinfo{author}{{Lamine} B},
  \bibinfo{author}{{Lam} MC}, \bibinfo{author}{{Lan{\c{c}}on} A},
  \bibinfo{author}{{Landt} H}, \bibinfo{author}{{Langer} M},
  \bibinfo{author}{{Lapi} A}, \bibinfo{author}{{Larcheveque} C},
  \bibinfo{author}{{Larsen} SS}, \bibinfo{author}{{Lattanzi} M},
  \bibinfo{author}{{Laudisio} F}, \bibinfo{author}{{Laugier} D},
  \bibinfo{author}{{Laureijs} R}, \bibinfo{author}{{Lavaux} G},
  \bibinfo{author}{{Lawrenson} A}, \bibinfo{author}{{Lazanu} A},
  \bibinfo{author}{{Lazeyras} T}, \bibinfo{author}{{Le Boulc'h} Q},
  \bibinfo{author}{{Le Brun} AMC}, \bibinfo{author}{{Le Brun} V},
  \bibinfo{author}{{Leclercq} F}, \bibinfo{author}{{Lee} S},
  \bibinfo{author}{{Le Graet} J}, \bibinfo{author}{{Legrand} L},
  \bibinfo{author}{{Leirvik} KN}, \bibinfo{author}{{Le Jeune} M},
  \bibinfo{author}{{Lembo} M}, \bibinfo{author}{{Le Mignant} D},
  \bibinfo{author}{{Lepinzan} MD}, \bibinfo{author}{{Lepori} F},
  \bibinfo{author}{{Lesci} GF}, \bibinfo{author}{{Lesgourgues} J},
  \bibinfo{author}{{Leuzzi} L}, \bibinfo{author}{{Levi} ME},
  \bibinfo{author}{{Liaudat} TI}, \bibinfo{author}{{Libet} G},
  \bibinfo{author}{{Liebing} P}, \bibinfo{author}{{Ligori} S},
  \bibinfo{author}{{Lilje} PB}, \bibinfo{author}{{Lin} CC},
  \bibinfo{author}{{Linde} D}, \bibinfo{author}{{Linder} E},
  \bibinfo{author}{{Lindholm} V}, \bibinfo{author}{{Linke} L},
  \bibinfo{author}{{Li} SS}, \bibinfo{author}{{Liu} SJ},
  \bibinfo{author}{{Lloro} I}, \bibinfo{author}{{Lobo} FSN},
  \bibinfo{author}{{Lodieu} N}, \bibinfo{author}{{Lombardi} M},
  \bibinfo{author}{{Lombriser} L}, \bibinfo{author}{{Lonare} P},
  \bibinfo{author}{{Longo} G}, \bibinfo{author}{{L{\'o}pez-Caniego} M},
  \bibinfo{author}{{Lopez Lopez} X}, \bibinfo{author}{{Alvarez} JL},
  \bibinfo{author}{{Loureiro} A}, \bibinfo{author}{{Loveday} J},
  \bibinfo{author}{{Lusso} E}, \bibinfo{author}{{Macias-Perez} J},
  \bibinfo{author}{{Maciaszek} T}, \bibinfo{author}{{Magliocchetti} M},
  \bibinfo{author}{{Magnard} F}, \bibinfo{author}{{Magnier} EA},
  \bibinfo{author}{{Magro} A}, \bibinfo{author}{{Mahler} G},
  \bibinfo{author}{{Mainetti} G}, \bibinfo{author}{{Maino} D},
  \bibinfo{author}{{Maiorano} E}, \bibinfo{author}{{Maiorano} E},
  \bibinfo{author}{{Malavasi} N}, \bibinfo{author}{{Mamon} GA},
  \bibinfo{author}{{Mancini} C}, \bibinfo{author}{{Mandelbaum} R},
  \bibinfo{author}{{Manera} M}, \bibinfo{author}{{Manj{\'o}n-Garc{\'\i}a} A},
  \bibinfo{author}{{Mannucci} F}, \bibinfo{author}{{Mansutti} O},
  \bibinfo{author}{{Manteiga Outeiro} M}, \bibinfo{author}{{Maoli} R},
  \bibinfo{author}{{Maraston} C}, \bibinfo{author}{{Marcin} S},
  \bibinfo{author}{{Marcos-Arenal} P}, \bibinfo{author}{{Margalef-Bentabol} B},
  \bibinfo{author}{{Marggraf} O}, \bibinfo{author}{{Marinucci} D},
  \bibinfo{author}{{Marinucci} M}, \bibinfo{author}{{Markovic} K},
  \bibinfo{author}{{Marleau} FR}, \bibinfo{author}{{Marpaud} J},
  \bibinfo{author}{{Martignac} J}, \bibinfo{author}{{Mart{\'\i}n-Fleitas} J},
  \bibinfo{author}{{Martin-Moruno} P}, \bibinfo{author}{{Martin} EL},
  \bibinfo{author}{{Martinelli} M}, \bibinfo{author}{{Martinet} N},
  \bibinfo{author}{{Martin} H}, \bibinfo{author}{{Martins} CJAP},
  \bibinfo{author}{{Marulli} F}, \bibinfo{author}{{Massari} D},
  \bibinfo{author}{{Massey} R}, \bibinfo{author}{{Masters} DC},
  \bibinfo{author}{{Matarrese} S}, \bibinfo{author}{{Matsuoka} Y},
  \bibinfo{author}{{Matthew} S}, \bibinfo{author}{{Maughan} BJ},
  \bibinfo{author}{{Mauri} N}, \bibinfo{author}{{Maurin} L},
  \bibinfo{author}{{Maurogordato} S}, \bibinfo{author}{{McCarthy} K},
  \bibinfo{author}{{McConnachie} AW}, \bibinfo{author}{{McCracken} HJ},
  \bibinfo{author}{{McDonald} I}, \bibinfo{author}{{McEwen} JD},
  \bibinfo{author}{{McPartland} CJR}, \bibinfo{author}{{Medinaceli} E},
  \bibinfo{author}{{Mehta} V}, \bibinfo{author}{{Mei} S},
  \bibinfo{author}{{Melchior} M}, \bibinfo{author}{{Melin} JB},
  \bibinfo{author}{{M{\'e}nard} B}, \bibinfo{author}{{Mendes} J},
  \bibinfo{author}{{Mendez-Abreu} J}, \bibinfo{author}{{Meneghetti} M},
  \bibinfo{author}{{Mercurio} A}, \bibinfo{author}{{Merlin} E},
  \bibinfo{author}{{Metcalf} RB}, \bibinfo{author}{{Meylan} G},
  \bibinfo{author}{{Migliaccio} M}, \bibinfo{author}{{Mignoli} M},
  \bibinfo{author}{{Miller} L}, \bibinfo{author}{{Miluzio} M},
  \bibinfo{author}{{Milvang-Jensen} B}, \bibinfo{author}{{Mimoso} JP},
  \bibinfo{author}{{Miquel} R}, \bibinfo{author}{{Miyatake} H},
  \bibinfo{author}{{Mobasher} B}, \bibinfo{author}{{Mohr} JJ},
  \bibinfo{author}{{Monaco} P}, \bibinfo{author}{{Mongui{\'o}} M},
  \bibinfo{author}{{Montoro} A}, \bibinfo{author}{{Mora} A},
  \bibinfo{author}{{Moradinezhad Dizgah} A}, \bibinfo{author}{{Moresco} M},
  \bibinfo{author}{{Moretti} C}, \bibinfo{author}{{Morgante} G},
  \bibinfo{author}{{Morisset} N}, \bibinfo{author}{{Moriya} TJ},
  \bibinfo{author}{{Morris} PW}, \bibinfo{author}{{Mortlock} DJ},
  \bibinfo{author}{{Moscardini} L}, \bibinfo{author}{{Mota} DF},
  \bibinfo{author}{{Moustakas} LA}, \bibinfo{author}{{Moutard} T},
  \bibinfo{author}{{M{\"u}ller} T}, \bibinfo{author}{{Munari} E},
  \bibinfo{author}{{Murphree} G}, \bibinfo{author}{{Murray} C},
  \bibinfo{author}{{Murray} N}, \bibinfo{author}{{Musi} P},
  \bibinfo{author}{{Nadathur} S}, \bibinfo{author}{{Nagam} BC},
  \bibinfo{author}{{Nagao} T}, \bibinfo{author}{{Naidoo} K},
  \bibinfo{author}{{Nakajima} R}, \bibinfo{author}{{Nally} C},
  \bibinfo{author}{{Natoli} P}, \bibinfo{author}{{Navarro-Alsina} A},
  \bibinfo{author}{{Navarro Girones} D}, \bibinfo{author}{{Neissner} C},
  \bibinfo{author}{{Nersesian} A}, \bibinfo{author}{{Nesseris} S},
  \bibinfo{author}{{Nguyen-Kim} HN}, \bibinfo{author}{{Nicastro} L},
  \bibinfo{author}{{Nichol} RC}, \bibinfo{author}{{Nielbock} M},
  \bibinfo{author}{{Niemi} SM}, \bibinfo{author}{{Nieto} S},
  \bibinfo{author}{{Nilsson} K}, \bibinfo{author}{{Noller} J},
  \bibinfo{author}{{Norberg} P}, \bibinfo{author}{{Nourizonoz} A},
  \bibinfo{author}{{Ntelis} P}, \bibinfo{author}{{Nucita} AA},
  \bibinfo{author}{{Nugent} P}, \bibinfo{author}{{Nunes} NJ},
  \bibinfo{author}{{Nutma} T}, \bibinfo{author}{{Ocampo} I},
  \bibinfo{author}{{Odier} J}, \bibinfo{author}{{Oesch} PA},
  \bibinfo{author}{{Oguri} M}, \bibinfo{author}{{Magalhaes Oliveira} D},
  \bibinfo{author}{{Onoue} M}, \bibinfo{author}{{Oosterbroek} T},
  \bibinfo{author}{{Oppizzi} F}, \bibinfo{author}{{Ordenovic} C},
  \bibinfo{author}{{Osato} K}, \bibinfo{author}{{Pacaud} F},
  \bibinfo{author}{{Pace} F}, \bibinfo{author}{{Padilla} C},
  \bibinfo{author}{{Paech} K}, \bibinfo{author}{{Pagano} L},
  \bibinfo{author}{{Page} MJ}, \bibinfo{author}{{Palazzi} E},
  \bibinfo{author}{{Paltani} S}, \bibinfo{author}{{Pamuk} S},
  \bibinfo{author}{{Pandolfi} S}, \bibinfo{author}{{Paoletti} D},
  \bibinfo{author}{{Paolillo} M}, \bibinfo{author}{{Papaderos} P},
  \bibinfo{author}{{Pardede} K}, \bibinfo{author}{{Parimbelli} G},
  \bibinfo{author}{{Parmar} A}, \bibinfo{author}{{Partmann} C},
  \bibinfo{author}{{Pasian} F}, \bibinfo{author}{{Passalacqua} F},
  \bibinfo{author}{{Paterson} K}, \bibinfo{author}{{Patrizii} L},
  \bibinfo{author}{{Pattison} C}, \bibinfo{author}{{Paulino-Afonso} A},
  \bibinfo{author}{{Paviot} R}, \bibinfo{author}{{Peacock} JA},
  \bibinfo{author}{{Pearce} FR}, \bibinfo{author}{{Pedersen} K},
  \bibinfo{author}{{Peel} A}, \bibinfo{author}{{Peletier} RF},
  \bibinfo{author}{{Pellejero Ibanez} M}, \bibinfo{author}{{Pello} R},
  \bibinfo{author}{{Penny} MT}, \bibinfo{author}{{Percival} WJ},
  \bibinfo{author}{{Perez-Garrido} A}, \bibinfo{author}{{Perotto} L},
  \bibinfo{author}{{Pettorino} V}, \bibinfo{author}{{Pezzotta} A},
  \bibinfo{author}{{Pezzuto} S}, \bibinfo{author}{{Philippon} A},
  \bibinfo{author}{{Piersanti} O}, \bibinfo{author}{{Pietroni} M},
  \bibinfo{author}{{Piga} L}, \bibinfo{author}{{Pilo} L},
  \bibinfo{author}{{Pires} S}, \bibinfo{author}{{Pisani} A},
  \bibinfo{author}{{Pizzella} A}, \bibinfo{author}{{Pizzuti} L},
  \bibinfo{author}{{Plana} C}, \bibinfo{author}{{Polenta} G},
  \bibinfo{author}{{Pollack} JE}, \bibinfo{author}{{Poncet} M},
  \bibinfo{author}{{P{\"o}ntinen} M}, \bibinfo{author}{{Pool} P},
  \bibinfo{author}{{Popa} LA}, \bibinfo{author}{{Popa} V},
  \bibinfo{author}{{Popp} J}, \bibinfo{author}{{Porciani} C},
  \bibinfo{author}{{Porth} L}, \bibinfo{author}{{Potter} D},
  \bibinfo{author}{{Poulain} M}, \bibinfo{author}{{Pourtsidou} A},
  \bibinfo{author}{{Pozzetti} L}, \bibinfo{author}{{Prandoni} I},
  \bibinfo{author}{{Pratt} GW}, \bibinfo{author}{{Prezelus} S},
  \bibinfo{author}{{Prieto} E}, \bibinfo{author}{{Pugno} A},
  \bibinfo{author}{{Quai} S}, \bibinfo{author}{{Quilley} L},
  \bibinfo{author}{{Racca} GD}, \bibinfo{author}{{Raccanelli} A},
  \bibinfo{author}{{R{\'a}cz} G}, \bibinfo{author}{{Radinovi{\'c}} S},
  \bibinfo{author}{{Radovich} M}, \bibinfo{author}{{Ragagnin} A},
  \bibinfo{author}{{Ragnit} U}, \bibinfo{author}{{Raison} F},
  \bibinfo{author}{{Ramos-Chernenko} N}, \bibinfo{author}{{Ranc} C},
  \bibinfo{author}{{Raylet} N}, \bibinfo{author}{{Rebolo} R},
  \bibinfo{author}{{Refregier} A}, \bibinfo{author}{{Reimberg} P},
  \bibinfo{author}{{Reiprich} TH}, \bibinfo{author}{{Renk} F},
  \bibinfo{author}{{Renzi} A}, \bibinfo{author}{{Retre} J},
  \bibinfo{author}{{Revaz} Y}, \bibinfo{author}{{Reyl{\'e}} C},
  \bibinfo{author}{{Reynolds} L}, \bibinfo{author}{{Rhodes} J},
  \bibinfo{author}{{Ricci} F}, \bibinfo{author}{{Ricci} M},
  \bibinfo{author}{{Riccio} G}, \bibinfo{author}{{Ricken} SO},
  \bibinfo{author}{{Rissanen} S}, \bibinfo{author}{{Risso} I},
  \bibinfo{author}{{Rix} HW}, \bibinfo{author}{{Robin} AC},
  \bibinfo{author}{{Rocca-Volmerange} B}, \bibinfo{author}{{Rocci} PF},
  \bibinfo{author}{{Rodenhuis} M}, \bibinfo{author}{{Rodighiero} G},
  \bibinfo{author}{{Rodriguez Monroy} M}, \bibinfo{author}{{Rollins} RP},
  \bibinfo{author}{{Romanello} M}, \bibinfo{author}{{Roman} J},
  \bibinfo{author}{{Romelli} E}, \bibinfo{author}{{Romero-Gomez} M},
  \bibinfo{author}{{Roncarelli} M}, \bibinfo{author}{{Rosati} P},
  \bibinfo{author}{{Rosset} C}, \bibinfo{author}{{Rossetti} E},
  \bibinfo{author}{{Roster} W}, \bibinfo{author}{{Rottgering} HJA},
  \bibinfo{author}{{Rozas-Fern{\'a}ndez} A}, \bibinfo{author}{{Ruane} K},
  \bibinfo{author}{{Rubino-Martin} JA}, \bibinfo{author}{{Rudolph} A},
  \bibinfo{author}{{Ruppin} F}, \bibinfo{author}{{Rusholme} B},
  \bibinfo{author}{{Sacquegna} S}, \bibinfo{author}{{S{\'a}ez-Casares} I},
  \bibinfo{author}{{Saga} S}, \bibinfo{author}{{Saglia} R},
  \bibinfo{author}{{Sahl{\'e}n} M}, \bibinfo{author}{{Saifollahi} T},
  \bibinfo{author}{{Sakr} Z}, \bibinfo{author}{{Salvalaggio} J},
  \bibinfo{author}{{Salvaterra} R}, \bibinfo{author}{{Salvati} L},
  \bibinfo{author}{{Salvato} M}, \bibinfo{author}{{Salvignol} JC},
  \bibinfo{author}{{S{\'a}nchez} AG}, \bibinfo{author}{{Sanchez} E},
  \bibinfo{author}{{Sanders} DB}, \bibinfo{author}{{Sapone} D},
  \bibinfo{author}{{Saponara} M}, \bibinfo{author}{{Sarpa} E},
  \bibinfo{author}{{Sarron} F}, \bibinfo{author}{{Sartori} S},
  \bibinfo{author}{{Sassolas} B}, \bibinfo{author}{{Sauniere} L},
  \bibinfo{author}{{Sauvage} M}, \bibinfo{author}{{Sawicki} M},
  \bibinfo{author}{{Scaramella} R}, \bibinfo{author}{{Scarlata} C},
  \bibinfo{author}{{Scharr{\'e}} L}, \bibinfo{author}{{Schaye} J},
  \bibinfo{author}{{Schewtschenko} JA}, \bibinfo{author}{{Schindler} JT},
  \bibinfo{author}{{Schinnerer} E}, \bibinfo{author}{{Schirmer} M},
  \bibinfo{author}{{Schmidt} F}, \bibinfo{author}{{Schmidt} F},
  \bibinfo{author}{{Schmidt} M}, \bibinfo{author}{{Schneider} A},
  \bibinfo{author}{{Schneider} M}, \bibinfo{author}{{Schneider} P},
  \bibinfo{author}{{Sch{\"o}neberg} N}, \bibinfo{author}{{Schrabback} T},
  \bibinfo{author}{{Schultheis} M}, \bibinfo{author}{{Schulz} S},
  \bibinfo{author}{{Schwartz} J}, \bibinfo{author}{{Sciotti} D},
  \bibinfo{author}{{Scodeggio} M}, \bibinfo{author}{{Scognamiglio} D},
  \bibinfo{author}{{Scott} D}, \bibinfo{author}{{Scottez} V},
  \bibinfo{author}{{Secroun} A}, \bibinfo{author}{{Sefusatti} E},
  \bibinfo{author}{{Seidel} G}, \bibinfo{author}{{Seiffert} M},
  \bibinfo{author}{{Sellentin} E}, \bibinfo{author}{{Selwood} M},
  \bibinfo{author}{{Semboloni} E}, \bibinfo{author}{{Sereno} M},
  \bibinfo{author}{{Serjeant} S}, \bibinfo{author}{{Serrano} S},
  \bibinfo{author}{{Shankar} F}, \bibinfo{author}{{Sharples} RM},
  \bibinfo{author}{{Short} A}, \bibinfo{author}{{Shulevski} A},
  \bibinfo{author}{{Shuntov} M}, \bibinfo{author}{{Sias} M},
  \bibinfo{author}{{Sikkema} G}, \bibinfo{author}{{Silvestri} A},
  \bibinfo{author}{{Simon} P}, \bibinfo{author}{{Sirignano} C},
  \bibinfo{author}{{Sirri} G}, \bibinfo{author}{{Skottfelt} J},
  \bibinfo{author}{{Slezak} E}, \bibinfo{author}{{Sluse} D},
  \bibinfo{author}{{Smith} GP}, \bibinfo{author}{{Smith} LC},
  \bibinfo{author}{{Smith} RE}, \bibinfo{author}{{Smit} SJA},
  \bibinfo{author}{{Soldano} F}, \bibinfo{author}{{Solheim} BGB},
  \bibinfo{author}{{Sorce} JG}, \bibinfo{author}{{Sorrenti} F},
  \bibinfo{author}{{Soubrie} E}, \bibinfo{author}{{Spinoglio} L},
  \bibinfo{author}{{Spurio Mancini} A}, \bibinfo{author}{{Stadel} J},
  \bibinfo{author}{{Stagnaro} L}, \bibinfo{author}{{Stanco} L},
  \bibinfo{author}{{Stanford} SA}, \bibinfo{author}{{Starck} JL},
  \bibinfo{author}{{Stassi} P}, \bibinfo{author}{{Steinwagner} J},
  \bibinfo{author}{{Stern} D}, \bibinfo{author}{{Stone} C},
  \bibinfo{author}{{Strada} P}, \bibinfo{author}{{Strafella} F},
  \bibinfo{author}{{Stramaccioni} D}, \bibinfo{author}{{Surace} C},
  \bibinfo{author}{{Sureau} F}, \bibinfo{author}{{Suyu} SH},
  \bibinfo{author}{{Swindells} I}, \bibinfo{author}{{Szafraniec} M},
  \bibinfo{author}{{Szapudi} I}, \bibinfo{author}{{Taamoli} S},
  \bibinfo{author}{{Talia} M}, \bibinfo{author}{{Tallada-Cresp{\'\i}} P},
  \bibinfo{author}{{Tanidis} K}, \bibinfo{author}{{Tao} C},
  \bibinfo{author}{{Tarr{\'\i}o} P}, \bibinfo{author}{{Tavagnacco} D},
  \bibinfo{author}{{Taylor} AN}, \bibinfo{author}{{Taylor} JE},
  \bibinfo{author}{{Taylor} PL}, \bibinfo{author}{{Teixeira} EM},
  \bibinfo{author}{{Tenti} M}, \bibinfo{author}{{Teodoro Idiago} P},
  \bibinfo{author}{{Teplitz} HI}, \bibinfo{author}{{Tereno} I},
  \bibinfo{author}{{Tessore} N}, \bibinfo{author}{{Testa} V},
  \bibinfo{author}{{Testera} G}, \bibinfo{author}{{Tewes} M},
  \bibinfo{author}{{Teyssier} R}, \bibinfo{author}{{Theret} N},
  \bibinfo{author}{{Thizy} C}, \bibinfo{author}{{Thomas} PD},
  \bibinfo{author}{{Toba} Y}, \bibinfo{author}{{Toft} S},
  \bibinfo{author}{{Toledo-Moreo} R}, \bibinfo{author}{{Tolstoy} E},
  \bibinfo{author}{{Tommasi} E}, \bibinfo{author}{{Torbaniuk} O},
  \bibinfo{author}{{Torradeflot} F}, \bibinfo{author}{{Tortora} C},
  \bibinfo{author}{{Tosi} S}, \bibinfo{author}{{Tosti} S},
  \bibinfo{author}{{Trifoglio} M}, \bibinfo{author}{{Troja} A},
  \bibinfo{author}{{Trombetti} T}, \bibinfo{author}{{Tronconi} A},
  \bibinfo{author}{{Tsedrik} M}, \bibinfo{author}{{Tsyganov} A},
  \bibinfo{author}{{Tucci} M}, \bibinfo{author}{{Tutusaus} I},
  \bibinfo{author}{{Uhlemann} C}, \bibinfo{author}{{Ulivi} L},
  \bibinfo{author}{{Urbano} M}, \bibinfo{author}{{Vacher} L},
  \bibinfo{author}{{Vaillon} L}, \bibinfo{author}{{Valdes} I},
  \bibinfo{author}{{Valentijn} EA}, \bibinfo{author}{{Valenziano} L},
  \bibinfo{author}{{Valieri} C}, \bibinfo{author}{{Valiviita} J},
  \bibinfo{author}{{Van den Broeck} M}, \bibinfo{author}{{Vassallo} T},
  \bibinfo{author}{{Vavrek} R}, \bibinfo{author}{{Venemans} B},
  \bibinfo{author}{{Venhola} A}, \bibinfo{author}{{Ventura} S},
  \bibinfo{author}{{Verdoes Kleijn} G}, \bibinfo{author}{{Vergani} D},
  \bibinfo{author}{{Verma} A}, \bibinfo{author}{{Vernizzi} F},
  \bibinfo{author}{{Veropalumbo} A}, \bibinfo{author}{{Verza} G},
  \bibinfo{author}{{Vescovi} C}, \bibinfo{author}{{Vibert} D},
  \bibinfo{author}{{Viel} M}, \bibinfo{author}{{Vielzeuf} P},
  \bibinfo{author}{{Viglione} C}, \bibinfo{author}{{Viitanen} A},
  \bibinfo{author}{{Villaescusa-Navarro} F}, \bibinfo{author}{{Vinciguerra} S},
  \bibinfo{author}{{Visticot} F}, \bibinfo{author}{{Voggel} K},
  \bibinfo{author}{{von Wietersheim-Kramsta} M}, \bibinfo{author}{{Vriend} WJ},
  \bibinfo{author}{{Wachter} S}, \bibinfo{author}{{Walmsley} M},
  \bibinfo{author}{{Walth} G}, \bibinfo{author}{{Walton} DM},
  \bibinfo{author}{{Walton} NA}, \bibinfo{author}{{Wander} M},
  \bibinfo{author}{{Wang} L}, \bibinfo{author}{{Wang} Y},
  \bibinfo{author}{{Weaver} JR}, \bibinfo{author}{{Weller} J},
  \bibinfo{author}{{Whalen} DJ}, \bibinfo{author}{{Wiesmann} M},
  \bibinfo{author}{{Wilde} J}, \bibinfo{author}{{Williams} OR},
  \bibinfo{author}{{Winther} HA}, \bibinfo{author}{{Wittje} A},
  \bibinfo{author}{{Wong} JHW}, \bibinfo{author}{{Wright} AH},
  \bibinfo{author}{{Yankelevich} V}, \bibinfo{author}{{Yeung} HW},
  \bibinfo{author}{{Youles} S}, \bibinfo{author}{{Yung} LYA},
  \bibinfo{author}{{Zacchei} A}, \bibinfo{author}{{Zalesky} L},
  \bibinfo{author}{{Zamorani} G}, \bibinfo{author}{{Zamorano Vitorelli} A},
  \bibinfo{author}{{Zanoni Marc} M}, \bibinfo{author}{{Zennaro} M},
  \bibinfo{author}{{Zerbi} FM}, \bibinfo{author}{{Zinchenko} IA},
  \bibinfo{author}{{Zoubian} J}, \bibinfo{author}{{Zucca} E} and
  \bibinfo{author}{{Zumalacarregui} M} (\bibinfo{year}{2024}),
  \bibinfo{month}{May}.
\bibinfo{title}{{Euclid. I. Overview of the Euclid mission}}.
\bibinfo{journal}{{\em arXiv e-prints}} ,
  \bibinfo{eid}{arXiv:2405.13491}\bibinfo{doi}{\doi{10.48550/arXiv.2405.13491}}.
\eprint{2405.13491}.

\bibtype{Article}%
\bibitem[{Flaugher}(2005)]{Flaugher2005}
\bibinfo{author}{{Flaugher} B} (\bibinfo{year}{2005}), \bibinfo{month}{Jan.}
\bibinfo{title}{{The Dark Energy Survey}}.
\bibinfo{journal}{{\em International Journal of Modern Physics A}}
  \bibinfo{volume}{20} (\bibinfo{number}{14}): \bibinfo{pages}{3121--3123}.
  \bibinfo{doi}{\doi{10.1142/S0217751X05025917}}.

\bibtype{Article}%
\bibitem[{Gatti} et al.(2021)]{Gatti2021}
\bibinfo{author}{{Gatti} M}, \bibinfo{author}{{Sheldon} E},
  \bibinfo{author}{{Amon} A}, \bibinfo{author}{{Becker} M},
  \bibinfo{author}{{Troxel} M}, \bibinfo{author}{{Choi} A},
  \bibinfo{author}{{Doux} C}, \bibinfo{author}{{MacCrann} N},
  \bibinfo{author}{{Navarro-Alsina} A}, \bibinfo{author}{{Harrison} I},
  \bibinfo{author}{{Gruen} D}, \bibinfo{author}{{Bernstein} G},
  \bibinfo{author}{{Jarvis} M}, \bibinfo{author}{{Secco} LF},
  \bibinfo{author}{{Fert{\'e}} A}, \bibinfo{author}{{Shin} T},
  \bibinfo{author}{{McCullough} J}, \bibinfo{author}{{Rollins} RP},
  \bibinfo{author}{{Chen} R}, \bibinfo{author}{{Chang} C},
  \bibinfo{author}{{Pandey} S}, \bibinfo{author}{{Tutusaus} I},
  \bibinfo{author}{{Prat} J}, \bibinfo{author}{{Elvin-Poole} J},
  \bibinfo{author}{{Sanchez} C}, \bibinfo{author}{{Plazas} AA},
  \bibinfo{author}{{Roodman} A}, \bibinfo{author}{{Zuntz} J},
  \bibinfo{author}{{Abbott} TMC}, \bibinfo{author}{{Aguena} M},
  \bibinfo{author}{{Allam} S}, \bibinfo{author}{{Annis} J},
  \bibinfo{author}{{Avila} S}, \bibinfo{author}{{Bacon} D},
  \bibinfo{author}{{Bertin} E}, \bibinfo{author}{{Bhargava} S},
  \bibinfo{author}{{Brooks} D}, \bibinfo{author}{{Burke} DL},
  \bibinfo{author}{{Carnero Rosell} A}, \bibinfo{author}{{Carrasco Kind} M},
  \bibinfo{author}{{Carretero} J}, \bibinfo{author}{{Castander} FJ},
  \bibinfo{author}{{Conselice} C}, \bibinfo{author}{{Costanzi} M},
  \bibinfo{author}{{Crocce} M}, \bibinfo{author}{{da Costa} LN},
  \bibinfo{author}{{Davis} TM}, \bibinfo{author}{{De Vicente} J},
  \bibinfo{author}{{Desai} S}, \bibinfo{author}{{Diehl} HT},
  \bibinfo{author}{{Dietrich} JP}, \bibinfo{author}{{Doel} P},
  \bibinfo{author}{{Drlica-Wagner} A}, \bibinfo{author}{{Eckert} K},
  \bibinfo{author}{{Everett} S}, \bibinfo{author}{{Ferrero} I},
  \bibinfo{author}{{Frieman} J}, \bibinfo{author}{{Garc{\'\i}a-Bellido} J},
  \bibinfo{author}{{Gerdes} DW}, \bibinfo{author}{{Giannantonio} T},
  \bibinfo{author}{{Gruendl} RA}, \bibinfo{author}{{Gschwend} J},
  \bibinfo{author}{{Gutierrez} G}, \bibinfo{author}{{Hartley} WG},
  \bibinfo{author}{{Hinton} SR}, \bibinfo{author}{{Hollowood} DL},
  \bibinfo{author}{{Honscheid} K}, \bibinfo{author}{{Hoyle} B},
  \bibinfo{author}{{Huff} EM}, \bibinfo{author}{{Huterer} D},
  \bibinfo{author}{{Jain} B}, \bibinfo{author}{{James} DJ},
  \bibinfo{author}{{Jeltema} T}, \bibinfo{author}{{Krause} E},
  \bibinfo{author}{{Kron} R}, \bibinfo{author}{{Kuropatkin} N},
  \bibinfo{author}{{Lima} M}, \bibinfo{author}{{Maia} MAG},
  \bibinfo{author}{{Marshall} JL}, \bibinfo{author}{{Miquel} R},
  \bibinfo{author}{{Morgan} R}, \bibinfo{author}{{Myles} J},
  \bibinfo{author}{{Palmese} A}, \bibinfo{author}{{Paz-Chinch{\'o}n} F},
  \bibinfo{author}{{Rykoff} ES}, \bibinfo{author}{{Samuroff} S},
  \bibinfo{author}{{Sanchez} E}, \bibinfo{author}{{Scarpine} V},
  \bibinfo{author}{{Schubnell} M}, \bibinfo{author}{{Serrano} S},
  \bibinfo{author}{{Sevilla-Noarbe} I}, \bibinfo{author}{{Smith} M},
  \bibinfo{author}{{Soares-Santos} M}, \bibinfo{author}{{Suchyta} E},
  \bibinfo{author}{{Swanson} MEC}, \bibinfo{author}{{Tarle} G},
  \bibinfo{author}{{Thomas} D}, \bibinfo{author}{{To} C},
  \bibinfo{author}{{Tucker} DL}, \bibinfo{author}{{Varga} TN},
  \bibinfo{author}{{Wechsler} RH}, \bibinfo{author}{{Weller} J},
  \bibinfo{author}{{Wester} W} and  \bibinfo{author}{{Wilkinson} RD}
  (\bibinfo{year}{2021}), \bibinfo{month}{Jul.}
\bibinfo{title}{{Dark energy survey year 3 results: weak lensing shape
  catalogue}}.
\bibinfo{journal}{{\em \mnras}} \bibinfo{volume}{504} (\bibinfo{number}{3}):
  \bibinfo{pages}{4312--4336}. \bibinfo{doi}{\doi{10.1093/mnras/stab918}}.
\eprint{2011.03408}.

\bibtype{Article}%
\bibitem[{Gatti} et al.(2024{\natexlab{a}})]{Gatti2024}
\bibinfo{author}{{Gatti} M}, \bibinfo{author}{{Campailla} G},
  \bibinfo{author}{{Jeffrey} N}, \bibinfo{author}{{Whiteway} L},
  \bibinfo{author}{{Porredon} A}, \bibinfo{author}{{Prat} J},
  \bibinfo{author}{{Williamson} J}, \bibinfo{author}{{Raveri} M},
  \bibinfo{author}{{Jain} B}, \bibinfo{author}{{Ajani} V},
  \bibinfo{author}{{Giannini} G}, \bibinfo{author}{{Yamamoto} M},
  \bibinfo{author}{{Zhou} C}, \bibinfo{author}{{Blazek} J},
  \bibinfo{author}{{Anbajagane} D}, \bibinfo{author}{{Samuroff} S},
  \bibinfo{author}{{Kacprzak} T}, \bibinfo{author}{{Alarcon} A},
  \bibinfo{author}{{Amon} A}, \bibinfo{author}{{Bechtol} K},
  \bibinfo{author}{{Becker} M}, \bibinfo{author}{{Bernstein} G},
  \bibinfo{author}{{Campos} A}, \bibinfo{author}{{Chang} C},
  \bibinfo{author}{{Chen} R}, \bibinfo{author}{{Choi} A},
  \bibinfo{author}{{Davis} C}, \bibinfo{author}{{Derose} J},
  \bibinfo{author}{{Diehl} HT}, \bibinfo{author}{{Dodelson} S},
  \bibinfo{author}{{Doux} C}, \bibinfo{author}{{Eckert} K},
  \bibinfo{author}{{Elvin-Poole} J}, \bibinfo{author}{{Everett} S},
  \bibinfo{author}{{Ferte} A}, \bibinfo{author}{{Gruen} D},
  \bibinfo{author}{{Gruendl} R}, \bibinfo{author}{{Harrison} I},
  \bibinfo{author}{{Hartley} WG}, \bibinfo{author}{{Herner} K},
  \bibinfo{author}{{Huff} EM}, \bibinfo{author}{{Jarvis} M},
  \bibinfo{author}{{Kuropatkin} N}, \bibinfo{author}{{Leget} PF},
  \bibinfo{author}{{MacCrann} N}, \bibinfo{author}{{McCullough} J},
  \bibinfo{author}{{Myles} J}, \bibinfo{author}{{Navarro-Alsina} A},
  \bibinfo{author}{{Pandey} S}, \bibinfo{author}{{Rollins} RP},
  \bibinfo{author}{{Roodman} A}, \bibinfo{author}{{Sanchez} C},
  \bibinfo{author}{{Secco} LF}, \bibinfo{author}{{Sevilla-Noarbe} I},
  \bibinfo{author}{{Sheldon} E}, \bibinfo{author}{{Shin} T},
  \bibinfo{author}{{Troxel} M}, \bibinfo{author}{{Tutusaus} I},
  \bibinfo{author}{{Varga} TN}, \bibinfo{author}{{Yanny} B},
  \bibinfo{author}{{Yin} B}, \bibinfo{author}{{Zhang} Y},
  \bibinfo{author}{{Zuntz} J}, \bibinfo{author}{{Abbott} TMC},
  \bibinfo{author}{{Aguena} M}, \bibinfo{author}{{Allam} SS},
  \bibinfo{author}{{Alves} O}, \bibinfo{author}{{Andrade-Oliveira} F},
  \bibinfo{author}{{Bacon} D}, \bibinfo{author}{{Bocquet} S},
  \bibinfo{author}{{Brooks} D}, \bibinfo{author}{{Carnero Rosell} A},
  \bibinfo{author}{{Carretero} J}, \bibinfo{author}{{da Costa} LN},
  \bibinfo{author}{{Pereira} MES}, \bibinfo{author}{{De Vicente} J},
  \bibinfo{author}{{Ferrero} I}, \bibinfo{author}{{Frieman} J},
  \bibinfo{author}{{Garc{\'\i}a-Bellido} J}, \bibinfo{author}{{Gaztanaga} E},
  \bibinfo{author}{{Gutierrez} G}, \bibinfo{author}{{Hinton} SR},
  \bibinfo{author}{{Hollowood} DL}, \bibinfo{author}{{Honscheid} K},
  \bibinfo{author}{{James} DJ}, \bibinfo{author}{{Kuehn} K},
  \bibinfo{author}{{Lahav} O}, \bibinfo{author}{{Lee} S},
  \bibinfo{author}{{Marshall} JL}, \bibinfo{author}{{Mena-Fern{\'a}ndez} J},
  \bibinfo{author}{{Miquel} R}, \bibinfo{author}{{Pieres} A},
  \bibinfo{author}{{Plazas Malag{\'o}n} AA}, \bibinfo{author}{{Sanchez} E},
  \bibinfo{author}{{Sanchez Cid} D}, \bibinfo{author}{{Schubnell} M},
  \bibinfo{author}{{Smith} M}, \bibinfo{author}{{Suchyta} E},
  \bibinfo{author}{{Tarle} G}, \bibinfo{author}{{Weaverdyck} N},
  \bibinfo{author}{{Weller} J} and  \bibinfo{author}{{Wiseman} P}
  (\bibinfo{year}{2024}{\natexlab{a}}), \bibinfo{month}{May}.
\bibinfo{title}{{Dark Energy Survey Year 3 results: simulation-based
  cosmological inference with wavelet harmonics, scattering transforms, and
  moments of weak lensing mass maps II. Cosmological results}}.
\bibinfo{journal}{{\em arXiv e-prints}} ,
  \bibinfo{eid}{arXiv:2405.10881}\bibinfo{doi}{\doi{10.48550/arXiv.2405.10881}}.
\eprint{2405.10881}.

\bibtype{Article}%
\bibitem[{Gatti} et al.(2024{\natexlab{b}})]{Gatti2024sims}
\bibinfo{author}{{Gatti} M}, \bibinfo{author}{{Jeffrey} N},
  \bibinfo{author}{{Whiteway} L}, \bibinfo{author}{{Williamson} J},
  \bibinfo{author}{{Jain} B}, \bibinfo{author}{{Ajani} V},
  \bibinfo{author}{{Anbajagane} D}, \bibinfo{author}{{Giannini} G},
  \bibinfo{author}{{Zhou} C}, \bibinfo{author}{{Porredon} A},
  \bibinfo{author}{{Prat} J}, \bibinfo{author}{{Yamamoto} M},
  \bibinfo{author}{{Blazek} J}, \bibinfo{author}{{Kacprzak} T},
  \bibinfo{author}{{Samuroff} S}, \bibinfo{author}{{Alarcon} A},
  \bibinfo{author}{{Amon} A}, \bibinfo{author}{{Bechtol} K},
  \bibinfo{author}{{Becker} M}, \bibinfo{author}{{Bernstein} G},
  \bibinfo{author}{{Campos} A}, \bibinfo{author}{{Chang} C},
  \bibinfo{author}{{Chen} R}, \bibinfo{author}{{Choi} A},
  \bibinfo{author}{{Davis} C}, \bibinfo{author}{{Derose} J},
  \bibinfo{author}{{Diehl} HT}, \bibinfo{author}{{Dodelson} S},
  \bibinfo{author}{{Doux} C}, \bibinfo{author}{{Eckert} K},
  \bibinfo{author}{{Elvin-Poole} J}, \bibinfo{author}{{Everett} S},
  \bibinfo{author}{{Ferte} A}, \bibinfo{author}{{Gruen} D},
  \bibinfo{author}{{Gruendl} R}, \bibinfo{author}{{Harrison} I},
  \bibinfo{author}{{Hartley} WG}, \bibinfo{author}{{Herner} K},
  \bibinfo{author}{{Huff} EM}, \bibinfo{author}{{Jarvis} M},
  \bibinfo{author}{{Kuropatkin} N}, \bibinfo{author}{{Leget} PF},
  \bibinfo{author}{{MacCrann} N}, \bibinfo{author}{{McCullough} J},
  \bibinfo{author}{{Myles} J}, \bibinfo{author}{{Navarro-Alsina} A},
  \bibinfo{author}{{Pandey} S}, \bibinfo{author}{{Raveri} M},
  \bibinfo{author}{{Rollins} RP}, \bibinfo{author}{{Roodman} A},
  \bibinfo{author}{{Sanchez} C}, \bibinfo{author}{{Secco} LF},
  \bibinfo{author}{{Sevilla-Noarbe} I}, \bibinfo{author}{{Sheldon} E},
  \bibinfo{author}{{Shin} T}, \bibinfo{author}{{Troxel} M},
  \bibinfo{author}{{Tutusaus} I}, \bibinfo{author}{{Varga} TN},
  \bibinfo{author}{{Yanny} B}, \bibinfo{author}{{Yin} B},
  \bibinfo{author}{{Zhang} Y}, \bibinfo{author}{{Zuntz} J},
  \bibinfo{author}{{Aguena} M}, \bibinfo{author}{{Alves} O},
  \bibinfo{author}{{Annis} J}, \bibinfo{author}{{Brooks} D},
  \bibinfo{author}{{Carretero} J}, \bibinfo{author}{{Castander} FJ},
  \bibinfo{author}{{Cawthon} R}, \bibinfo{author}{{Costanzi} M},
  \bibinfo{author}{{da Costa} LN}, \bibinfo{author}{{Pereira} MES},
  \bibinfo{author}{{Evrard} AE}, \bibinfo{author}{{Flaugher} B},
  \bibinfo{author}{{Fosalba} P}, \bibinfo{author}{{Frieman} J},
  \bibinfo{author}{{Garc{\'\i}a-Bellido} J}, \bibinfo{author}{{Gerdes} DW},
  \bibinfo{author}{{Gruen} D}, \bibinfo{author}{{Gruendl} RA},
  \bibinfo{author}{{Gschwend} J}, \bibinfo{author}{{Gutierrez} G},
  \bibinfo{author}{{Hollowood} DL}, \bibinfo{author}{{Honscheid} K},
  \bibinfo{author}{{James} DJ}, \bibinfo{author}{{Kuehn} K},
  \bibinfo{author}{{Lahav} O}, \bibinfo{author}{{Lee} S},
  \bibinfo{author}{{Marshall} JL}, \bibinfo{author}{{Mena-Fern{\'a}ndez} J},
  \bibinfo{author}{{Menanteau} F}, \bibinfo{author}{{Miquel} R},
  \bibinfo{author}{{Ogando} RLC}, \bibinfo{author}{{Pereira} MES},
  \bibinfo{author}{{Pieres} A}, \bibinfo{author}{{Plazas Malag{\'o}n} AA},
  \bibinfo{author}{{Sanchez} E}, \bibinfo{author}{{Smith} M},
  \bibinfo{author}{{Suchyta} E}, \bibinfo{author}{{Swanson} MEC},
  \bibinfo{author}{{Tarle} G}, \bibinfo{author}{{Weaverdyck} N},
  \bibinfo{author}{{Weller} J}, \bibinfo{author}{{Wiseman} P} and
  \bibinfo{author}{{DES Collaboration}} (\bibinfo{year}{2024}{\natexlab{b}}),
  \bibinfo{month}{Mar.}
\bibinfo{title}{{Dark Energy Survey Year 3 results: Simulation-based
  cosmological inference with wavelet harmonics, scattering transforms, and
  moments of weak lensing mass maps. Validation on simulations}}.
\bibinfo{journal}{{\em \prd}} \bibinfo{volume}{109} (\bibinfo{number}{6}),
  \bibinfo{eid}{063534}. \bibinfo{doi}{\doi{10.1103/PhysRevD.109.063534}}.
\eprint{2310.17557}.

\bibtype{Article}%
\bibitem[{Giblin} et al.(2021)]{Giblin2021}
\bibinfo{author}{{Giblin} B}, \bibinfo{author}{{Heymans} C},
  \bibinfo{author}{{Asgari} M}, \bibinfo{author}{{Hildebrandt} H},
  \bibinfo{author}{{Hoekstra} H}, \bibinfo{author}{{Joachimi} B},
  \bibinfo{author}{{Kannawadi} A}, \bibinfo{author}{{Kuijken} K},
  \bibinfo{author}{{Lin} CA}, \bibinfo{author}{{Miller} L},
  \bibinfo{author}{{Tr{\"o}ster} T}, \bibinfo{author}{{van den Busch} JL},
  \bibinfo{author}{{Wright} AH}, \bibinfo{author}{{Bilicki} M},
  \bibinfo{author}{{Blake} C}, \bibinfo{author}{{de Jong} J},
  \bibinfo{author}{{Dvornik} A}, \bibinfo{author}{{Erben} T},
  \bibinfo{author}{{Getman} F}, \bibinfo{author}{{Napolitano} NR},
  \bibinfo{author}{{Schneider} P}, \bibinfo{author}{{Shan} H} and
  \bibinfo{author}{{Valentijn} E} (\bibinfo{year}{2021}), \bibinfo{month}{Jan.}
\bibinfo{title}{{KiDS-1000 catalogue: Weak gravitational lensing shear
  measurements}}.
\bibinfo{journal}{{\em \aap}} \bibinfo{volume}{645}, \bibinfo{eid}{A105}.
  \bibinfo{doi}{\doi{10.1051/0004-6361/202038850}}.
\eprint{2007.01845}.

\bibtype{Article}%
\bibitem[{Heymans} et al.(2012)]{Heymans2012}
\bibinfo{author}{{Heymans} C}, \bibinfo{author}{{Van Waerbeke} L},
  \bibinfo{author}{{Miller} L}, \bibinfo{author}{{Erben} T},
  \bibinfo{author}{{Hildebrandt} H}, \bibinfo{author}{{Hoekstra} H},
  \bibinfo{author}{{Kitching} TD}, \bibinfo{author}{{Mellier} Y},
  \bibinfo{author}{{Simon} P}, \bibinfo{author}{{Bonnett} C},
  \bibinfo{author}{{Coupon} J}, \bibinfo{author}{{Fu} L},
  \bibinfo{author}{{Harnois D{\'e}raps} J}, \bibinfo{author}{{Hudson} MJ},
  \bibinfo{author}{{Kilbinger} M}, \bibinfo{author}{{Kuijken} K},
  \bibinfo{author}{{Rowe} B}, \bibinfo{author}{{Schrabback} T},
  \bibinfo{author}{{Semboloni} E}, \bibinfo{author}{{van Uitert} E},
  \bibinfo{author}{{Vafaei} S} and  \bibinfo{author}{{Velander} M}
  (\bibinfo{year}{2012}), \bibinfo{month}{Nov.}
\bibinfo{title}{{CFHTLenS: the Canada-France-Hawaii Telescope Lensing Survey}}.
\bibinfo{journal}{{\em \mnras}} \bibinfo{volume}{427} (\bibinfo{number}{1}):
  \bibinfo{pages}{146--166}.
  \bibinfo{doi}{\doi{10.1111/j.1365-2966.2012.21952.x}}.
\eprint{1210.0032}.

\bibtype{Article}%
\bibitem[{Heymans} et al.(2021)]{kids1000}
\bibinfo{author}{{Heymans} C}, \bibinfo{author}{{Tr{\"o}ster} T},
  \bibinfo{author}{{Asgari} M}, \bibinfo{author}{{Blake} C},
  \bibinfo{author}{{Hildebrandt} H}, \bibinfo{author}{{Joachimi} B},
  \bibinfo{author}{{Kuijken} K}, \bibinfo{author}{{Lin} CA},
  \bibinfo{author}{{S{\'a}nchez} AG}, \bibinfo{author}{{van den Busch} JL},
  \bibinfo{author}{{Wright} AH}, \bibinfo{author}{{Amon} A},
  \bibinfo{author}{{Bilicki} M}, \bibinfo{author}{{de Jong} J},
  \bibinfo{author}{{Crocce} M}, \bibinfo{author}{{Dvornik} A},
  \bibinfo{author}{{Erben} T}, \bibinfo{author}{{Fortuna} MC},
  \bibinfo{author}{{Getman} F}, \bibinfo{author}{{Giblin} B},
  \bibinfo{author}{{Glazebrook} K}, \bibinfo{author}{{Hoekstra} H},
  \bibinfo{author}{{Joudaki} S}, \bibinfo{author}{{Kannawadi} A},
  \bibinfo{author}{{K{\"o}hlinger} F}, \bibinfo{author}{{Lidman} C},
  \bibinfo{author}{{Miller} L}, \bibinfo{author}{{Napolitano} NR},
  \bibinfo{author}{{Parkinson} D}, \bibinfo{author}{{Schneider} P},
  \bibinfo{author}{{Shan} H}, \bibinfo{author}{{Valentijn} EA},
  \bibinfo{author}{{Verdoes Kleijn} G} and  \bibinfo{author}{{Wolf} C}
  (\bibinfo{year}{2021}), \bibinfo{month}{Feb.}
\bibinfo{title}{{KiDS-1000 Cosmology: Multi-probe weak gravitational lensing
  and spectroscopic galaxy clustering constraints}}.
\bibinfo{journal}{{\em \aap}} \bibinfo{volume}{646}, \bibinfo{eid}{A140}.
  \bibinfo{doi}{\doi{10.1051/0004-6361/202039063}}.
\eprint{2007.15632}.

\bibtype{Book}%
\bibitem[Hobson et al.(2006)]{Hobson:2006se}
\bibinfo{author}{Hobson MP}, \bibinfo{author}{Efstathiou GP} and
  \bibinfo{author}{Lasenby AN} (\bibinfo{year}{2006}).
\bibinfo{title}{{General relativity: An introduction for physicists}}.

\bibtype{Article}%
\bibitem[Hu and Jain(2004)]{Hu:2003pt}
\bibinfo{author}{Hu W} and  \bibinfo{author}{Jain B} (\bibinfo{year}{2004}).
\bibinfo{title}{{Joint galaxy - lensing observables and the dark energy}}.
\bibinfo{journal}{{\em Phys. Rev.}} \bibinfo{volume}{D70}:
  \bibinfo{pages}{043009}. \bibinfo{doi}{\doi{10.1103/PhysRevD.70.043009}}.
\eprint{astro-ph/0312395}.

\bibtype{Article}%
\bibitem[Huff and Mandelbaum(2017)]{Huff2017}
\bibinfo{author}{Huff E} and  \bibinfo{author}{Mandelbaum R}
  (\bibinfo{year}{2017}), \bibinfo{month}{feb}.
\bibinfo{title}{{Metacalibration: Direct Self-Calibration of Biases in Shear
  Measurement}}.
\bibinfo{journal}{{\em arXiv:1702.02600}} \eprint{1702.02600},
  \bibinfo{url}{\url{http://arxiv.org/abs/1702.02600}}.

\bibtype{Article}%
\bibitem[{Ivezi{\'c}} et al.(2019)]{Ivezic2019}
\bibinfo{author}{{Ivezi{\'c}} {\v{Z}}}, \bibinfo{author}{{Kahn} SM},
  \bibinfo{author}{{Tyson} JA}, \bibinfo{author}{{Abel} B},
  \bibinfo{author}{{Acosta} E}, \bibinfo{author}{{Allsman} R},
  \bibinfo{author}{{Alonso} D}, \bibinfo{author}{{AlSayyad} Y},
  \bibinfo{author}{{Anderson} SF}, \bibinfo{author}{{Andrew} J},
  \bibinfo{author}{{Angel} JRP}, \bibinfo{author}{{Angeli} GZ},
  \bibinfo{author}{{Ansari} R}, \bibinfo{author}{{Antilogus} P},
  \bibinfo{author}{{Araujo} C}, \bibinfo{author}{{Armstrong} R},
  \bibinfo{author}{{Arndt} KT}, \bibinfo{author}{{Astier} P},
  \bibinfo{author}{{Aubourg} {\'E}}, \bibinfo{author}{{Auza} N},
  \bibinfo{author}{{Axelrod} TS}, \bibinfo{author}{{Bard} DJ},
  \bibinfo{author}{{Barr} JD}, \bibinfo{author}{{Barrau} A},
  \bibinfo{author}{{Bartlett} JG}, \bibinfo{author}{{Bauer} AE},
  \bibinfo{author}{{Bauman} BJ}, \bibinfo{author}{{Baumont} S},
  \bibinfo{author}{{Bechtol} E}, \bibinfo{author}{{Bechtol} K},
  \bibinfo{author}{{Becker} AC}, \bibinfo{author}{{Becla} J},
  \bibinfo{author}{{Beldica} C}, \bibinfo{author}{{Bellavia} S},
  \bibinfo{author}{{Bianco} FB}, \bibinfo{author}{{Biswas} R},
  \bibinfo{author}{{Blanc} G}, \bibinfo{author}{{Blazek} J},
  \bibinfo{author}{{Bland ford} RD}, \bibinfo{author}{{Bloom} JS},
  \bibinfo{author}{{Bogart} J}, \bibinfo{author}{{Bond} TW},
  \bibinfo{author}{{Booth} MT}, \bibinfo{author}{{Borgland} AW},
  \bibinfo{author}{{Borne} K}, \bibinfo{author}{{Bosch} JF},
  \bibinfo{author}{{Boutigny} D}, \bibinfo{author}{{Brackett} CA},
  \bibinfo{author}{{Bradshaw} A}, \bibinfo{author}{{Brand t} WN},
  \bibinfo{author}{{Brown} ME}, \bibinfo{author}{{Bullock} JS},
  \bibinfo{author}{{Burchat} P}, \bibinfo{author}{{Burke} DL},
  \bibinfo{author}{{Cagnoli} G}, \bibinfo{author}{{Calabrese} D},
  \bibinfo{author}{{Callahan} S}, \bibinfo{author}{{Callen} AL},
  \bibinfo{author}{{Carlin} JL}, \bibinfo{author}{{Carlson} EL},
  \bibinfo{author}{{Chand rasekharan} S}, \bibinfo{author}{{Charles-Emerson}
  G}, \bibinfo{author}{{Chesley} S}, \bibinfo{author}{{Cheu} EC},
  \bibinfo{author}{{Chiang} HF}, \bibinfo{author}{{Chiang} J},
  \bibinfo{author}{{Chirino} C}, \bibinfo{author}{{Chow} D},
  \bibinfo{author}{{Ciardi} DR}, \bibinfo{author}{{Claver} CF},
  \bibinfo{author}{{Cohen-Tanugi} J}, \bibinfo{author}{{Cockrum} JJ},
  \bibinfo{author}{{Coles} R}, \bibinfo{author}{{Connolly} AJ},
  \bibinfo{author}{{Cook} KH}, \bibinfo{author}{{Cooray} A},
  \bibinfo{author}{{Covey} KR}, \bibinfo{author}{{Cribbs} C},
  \bibinfo{author}{{Cui} W}, \bibinfo{author}{{Cutri} R},
  \bibinfo{author}{{Daly} PN}, \bibinfo{author}{{Daniel} SF},
  \bibinfo{author}{{Daruich} F}, \bibinfo{author}{{Daubard} G},
  \bibinfo{author}{{Daues} G}, \bibinfo{author}{{Dawson} W},
  \bibinfo{author}{{Delgado} F}, \bibinfo{author}{{Dellapenna} A},
  \bibinfo{author}{{de Peyster} R}, \bibinfo{author}{{de Val-Borro} M},
  \bibinfo{author}{{Digel} SW}, \bibinfo{author}{{Doherty} P},
  \bibinfo{author}{{Dubois} R}, \bibinfo{author}{{Dubois-Felsmann} GP},
  \bibinfo{author}{{Durech} J}, \bibinfo{author}{{Economou} F},
  \bibinfo{author}{{Eifler} T}, \bibinfo{author}{{Eracleous} M},
  \bibinfo{author}{{Emmons} BL}, \bibinfo{author}{{Fausti Neto} A},
  \bibinfo{author}{{Ferguson} H}, \bibinfo{author}{{Figueroa} E},
  \bibinfo{author}{{Fisher-Levine} M}, \bibinfo{author}{{Focke} W},
  \bibinfo{author}{{Foss} MD}, \bibinfo{author}{{Frank} J},
  \bibinfo{author}{{Freemon} MD}, \bibinfo{author}{{Gangler} E},
  \bibinfo{author}{{Gawiser} E}, \bibinfo{author}{{Geary} JC},
  \bibinfo{author}{{Gee} P}, \bibinfo{author}{{Geha} M},
  \bibinfo{author}{{Gessner} CJB}, \bibinfo{author}{{Gibson} RR},
  \bibinfo{author}{{Gilmore} DK}, \bibinfo{author}{{Glanzman} T},
  \bibinfo{author}{{Glick} W}, \bibinfo{author}{{Goldina} T},
  \bibinfo{author}{{Goldstein} DA}, \bibinfo{author}{{Goodenow} I},
  \bibinfo{author}{{Graham} ML}, \bibinfo{author}{{Gressler} WJ},
  \bibinfo{author}{{Gris} P}, \bibinfo{author}{{Guy} LP},
  \bibinfo{author}{{Guyonnet} A}, \bibinfo{author}{{Haller} G},
  \bibinfo{author}{{Harris} R}, \bibinfo{author}{{Hascall} PA},
  \bibinfo{author}{{Haupt} J}, \bibinfo{author}{{Hernand ez} F},
  \bibinfo{author}{{Herrmann} S}, \bibinfo{author}{{Hileman} E},
  \bibinfo{author}{{Hoblitt} J}, \bibinfo{author}{{Hodgson} JA},
  \bibinfo{author}{{Hogan} C}, \bibinfo{author}{{Howard} JD},
  \bibinfo{author}{{Huang} D}, \bibinfo{author}{{Huffer} ME},
  \bibinfo{author}{{Ingraham} P}, \bibinfo{author}{{Innes} WR},
  \bibinfo{author}{{Jacoby} SH}, \bibinfo{author}{{Jain} B},
  \bibinfo{author}{{Jammes} F}, \bibinfo{author}{{Jee} MJ},
  \bibinfo{author}{{Jenness} T}, \bibinfo{author}{{Jernigan} G},
  \bibinfo{author}{{Jevremovi{\'c}} D}, \bibinfo{author}{{Johns} K},
  \bibinfo{author}{{Johnson} AS}, \bibinfo{author}{{Johnson} MWG},
  \bibinfo{author}{{Jones} RL}, \bibinfo{author}{{Juramy-Gilles} C},
  \bibinfo{author}{{Juri{\'c}} M}, \bibinfo{author}{{Kalirai} JS},
  \bibinfo{author}{{Kallivayalil} NJ}, \bibinfo{author}{{Kalmbach} B},
  \bibinfo{author}{{Kantor} JP}, \bibinfo{author}{{Karst} P},
  \bibinfo{author}{{Kasliwal} MM}, \bibinfo{author}{{Kelly} H},
  \bibinfo{author}{{Kessler} R}, \bibinfo{author}{{Kinnison} V},
  \bibinfo{author}{{Kirkby} D}, \bibinfo{author}{{Knox} L},
  \bibinfo{author}{{Kotov} IV}, \bibinfo{author}{{Krabbendam} VL},
  \bibinfo{author}{{Krughoff} KS}, \bibinfo{author}{{Kub{\'a}nek} P},
  \bibinfo{author}{{Kuczewski} J}, \bibinfo{author}{{Kulkarni} S},
  \bibinfo{author}{{Ku} J}, \bibinfo{author}{{Kurita} NR},
  \bibinfo{author}{{Lage} CS}, \bibinfo{author}{{Lambert} R},
  \bibinfo{author}{{Lange} T}, \bibinfo{author}{{Langton} JB},
  \bibinfo{author}{{Le Guillou} L}, \bibinfo{author}{{Levine} D},
  \bibinfo{author}{{Liang} M}, \bibinfo{author}{{Lim} KT},
  \bibinfo{author}{{Lintott} CJ}, \bibinfo{author}{{Long} KE},
  \bibinfo{author}{{Lopez} M}, \bibinfo{author}{{Lotz} PJ},
  \bibinfo{author}{{Lupton} RH}, \bibinfo{author}{{Lust} NB},
  \bibinfo{author}{{MacArthur} LA}, \bibinfo{author}{{Mahabal} A},
  \bibinfo{author}{{Mand elbaum} R}, \bibinfo{author}{{Markiewicz} TW},
  \bibinfo{author}{{Marsh} DS}, \bibinfo{author}{{Marshall} PJ},
  \bibinfo{author}{{Marshall} S}, \bibinfo{author}{{May} M},
  \bibinfo{author}{{McKercher} R}, \bibinfo{author}{{McQueen} M},
  \bibinfo{author}{{Meyers} J}, \bibinfo{author}{{Migliore} M},
  \bibinfo{author}{{Miller} M}, \bibinfo{author}{{Mills} DJ},
  \bibinfo{author}{{Miraval} C}, \bibinfo{author}{{Moeyens} J},
  \bibinfo{author}{{Moolekamp} FE}, \bibinfo{author}{{Monet} DG},
  \bibinfo{author}{{Moniez} M}, \bibinfo{author}{{Monkewitz} S},
  \bibinfo{author}{{Montgomery} C}, \bibinfo{author}{{Morrison} CB},
  \bibinfo{author}{{Mueller} F}, \bibinfo{author}{{Muller} GP},
  \bibinfo{author}{{Mu{\~n}oz Arancibia} F}, \bibinfo{author}{{Neill} DR},
  \bibinfo{author}{{Newbry} SP}, \bibinfo{author}{{Nief} JY},
  \bibinfo{author}{{Nomerotski} A}, \bibinfo{author}{{Nordby} M},
  \bibinfo{author}{{O'Connor} P}, \bibinfo{author}{{Oliver} J},
  \bibinfo{author}{{Olivier} SS}, \bibinfo{author}{{Olsen} K},
  \bibinfo{author}{{O'Mullane} W}, \bibinfo{author}{{Ortiz} S},
  \bibinfo{author}{{Osier} S}, \bibinfo{author}{{Owen} RE},
  \bibinfo{author}{{Pain} R}, \bibinfo{author}{{Palecek} PE},
  \bibinfo{author}{{Parejko} JK}, \bibinfo{author}{{Parsons} JB},
  \bibinfo{author}{{Pease} NM}, \bibinfo{author}{{Peterson} JM},
  \bibinfo{author}{{Peterson} JR}, \bibinfo{author}{{Petravick} DL},
  \bibinfo{author}{{Libby Petrick} ME}, \bibinfo{author}{{Petry} CE},
  \bibinfo{author}{{Pierfederici} F}, \bibinfo{author}{{Pietrowicz} S},
  \bibinfo{author}{{Pike} R}, \bibinfo{author}{{Pinto} PA},
  \bibinfo{author}{{Plante} R}, \bibinfo{author}{{Plate} S},
  \bibinfo{author}{{Plutchak} JP}, \bibinfo{author}{{Price} PA},
  \bibinfo{author}{{Prouza} M}, \bibinfo{author}{{Radeka} V},
  \bibinfo{author}{{Rajagopal} J}, \bibinfo{author}{{Rasmussen} AP},
  \bibinfo{author}{{Regnault} N}, \bibinfo{author}{{Reil} KA},
  \bibinfo{author}{{Reiss} DJ}, \bibinfo{author}{{Reuter} MA},
  \bibinfo{author}{{Ridgway} ST}, \bibinfo{author}{{Riot} VJ},
  \bibinfo{author}{{Ritz} S}, \bibinfo{author}{{Robinson} S},
  \bibinfo{author}{{Roby} W}, \bibinfo{author}{{Roodman} A},
  \bibinfo{author}{{Rosing} W}, \bibinfo{author}{{Roucelle} C},
  \bibinfo{author}{{Rumore} MR}, \bibinfo{author}{{Russo} S},
  \bibinfo{author}{{Saha} A}, \bibinfo{author}{{Sassolas} B},
  \bibinfo{author}{{Schalk} TL}, \bibinfo{author}{{Schellart} P},
  \bibinfo{author}{{Schindler} RH}, \bibinfo{author}{{Schmidt} S},
  \bibinfo{author}{{Schneider} DP}, \bibinfo{author}{{Schneider} MD},
  \bibinfo{author}{{Schoening} W}, \bibinfo{author}{{Schumacher} G},
  \bibinfo{author}{{Schwamb} ME}, \bibinfo{author}{{Sebag} J},
  \bibinfo{author}{{Selvy} B}, \bibinfo{author}{{Sembroski} GH},
  \bibinfo{author}{{Seppala} LG}, \bibinfo{author}{{Serio} A},
  \bibinfo{author}{{Serrano} E}, \bibinfo{author}{{Shaw} RA},
  \bibinfo{author}{{Shipsey} I}, \bibinfo{author}{{Sick} J},
  \bibinfo{author}{{Silvestri} N}, \bibinfo{author}{{Slater} CT},
  \bibinfo{author}{{Smith} JA}, \bibinfo{author}{{Smith} RC},
  \bibinfo{author}{{Sobhani} S}, \bibinfo{author}{{Soldahl} C},
  \bibinfo{author}{{Storrie-Lombardi} L}, \bibinfo{author}{{Stover} E},
  \bibinfo{author}{{Strauss} MA}, \bibinfo{author}{{Street} RA},
  \bibinfo{author}{{Stubbs} CW}, \bibinfo{author}{{Sullivan} IS},
  \bibinfo{author}{{Sweeney} D}, \bibinfo{author}{{Swinbank} JD},
  \bibinfo{author}{{Szalay} A}, \bibinfo{author}{{Takacs} P},
  \bibinfo{author}{{Tether} SA}, \bibinfo{author}{{Thaler} JJ},
  \bibinfo{author}{{Thayer} JG}, \bibinfo{author}{{Thomas} S},
  \bibinfo{author}{{Thornton} AJ}, \bibinfo{author}{{Thukral} V},
  \bibinfo{author}{{Tice} J}, \bibinfo{author}{{Trilling} DE},
  \bibinfo{author}{{Turri} M}, \bibinfo{author}{{Van Berg} R},
  \bibinfo{author}{{Vanden Berk} D}, \bibinfo{author}{{Vetter} K},
  \bibinfo{author}{{Virieux} F}, \bibinfo{author}{{Vucina} T},
  \bibinfo{author}{{Wahl} W}, \bibinfo{author}{{Walkowicz} L},
  \bibinfo{author}{{Walsh} B}, \bibinfo{author}{{Walter} CW},
  \bibinfo{author}{{Wang} DL}, \bibinfo{author}{{Wang} SY},
  \bibinfo{author}{{Warner} M}, \bibinfo{author}{{Wiecha} O},
  \bibinfo{author}{{Willman} B}, \bibinfo{author}{{Winters} SE},
  \bibinfo{author}{{Wittman} D}, \bibinfo{author}{{Wolff} SC},
  \bibinfo{author}{{Wood-Vasey} WM}, \bibinfo{author}{{Wu} X},
  \bibinfo{author}{{Xin} B}, \bibinfo{author}{{Yoachim} P} and
  \bibinfo{author}{{Zhan} H} (\bibinfo{year}{2019}), \bibinfo{month}{Mar.}
\bibinfo{title}{{LSST: From Science Drivers to Reference Design and Anticipated
  Data Products}}.
\bibinfo{journal}{{\em \apj}} \bibinfo{volume}{873} (\bibinfo{number}{2}),
  \bibinfo{eid}{111}. \bibinfo{doi}{\doi{10.3847/1538-4357/ab042c}}.
\eprint{0805.2366}.

\bibtype{Article}%
\bibitem[{Jeffrey} et al.(2021)]{Jeffrey2021}
\bibinfo{author}{{Jeffrey} N}, \bibinfo{author}{{Alsing} J} and
  \bibinfo{author}{{Lanusse} F} (\bibinfo{year}{2021}), \bibinfo{month}{Feb.}
\bibinfo{title}{{Likelihood-free inference with neural compression of DES SV
  weak lensing map statistics}}.
\bibinfo{journal}{{\em \mnras}} \bibinfo{volume}{501} (\bibinfo{number}{1}):
  \bibinfo{pages}{954--969}. \bibinfo{doi}{\doi{10.1093/mnras/staa3594}}.
\eprint{2009.08459}.

\bibtype{Article}%
\bibitem[{Jeffrey} et al.(2024)]{Jeffrey2024}
\bibinfo{author}{{Jeffrey} N}, \bibinfo{author}{{Whiteway} L},
  \bibinfo{author}{{Gatti} M}, \bibinfo{author}{{Williamson} J},
  \bibinfo{author}{{Alsing} J}, \bibinfo{author}{{Porredon} A},
  \bibinfo{author}{{Prat} J}, \bibinfo{author}{{Doux} C},
  \bibinfo{author}{{Jain} B}, \bibinfo{author}{{Chang} C},
  \bibinfo{author}{{Cheng} TY}, \bibinfo{author}{{Kacprzak} T},
  \bibinfo{author}{{Lemos} P}, \bibinfo{author}{{Alarcon} A},
  \bibinfo{author}{{Amon} A}, \bibinfo{author}{{Bechtol} K},
  \bibinfo{author}{{Becker} MR}, \bibinfo{author}{{Bernstein} GM},
  \bibinfo{author}{{Campos} A}, \bibinfo{author}{{Carnero Rosell} A},
  \bibinfo{author}{{Chen} R}, \bibinfo{author}{{Choi} A},
  \bibinfo{author}{{DeRose} J}, \bibinfo{author}{{Drlica-Wagner} A},
  \bibinfo{author}{{Eckert} K}, \bibinfo{author}{{Everett} S},
  \bibinfo{author}{{Fert{\'e}} A}, \bibinfo{author}{{Gruen} D},
  \bibinfo{author}{{Gruendl} RA}, \bibinfo{author}{{Herner} K},
  \bibinfo{author}{{Jarvis} M}, \bibinfo{author}{{McCullough} J},
  \bibinfo{author}{{Myles} J}, \bibinfo{author}{{Navarro-Alsina} A},
  \bibinfo{author}{{Pandey} S}, \bibinfo{author}{{Raveri} M},
  \bibinfo{author}{{Rollins} RP}, \bibinfo{author}{{Rykoff} ES},
  \bibinfo{author}{{S{\'a}nchez} C}, \bibinfo{author}{{Secco} LF},
  \bibinfo{author}{{Sevilla-Noarbe} I}, \bibinfo{author}{{Sheldon} E},
  \bibinfo{author}{{Shin} T}, \bibinfo{author}{{Troxel} MA},
  \bibinfo{author}{{Tutusaus} I}, \bibinfo{author}{{Varga} TN},
  \bibinfo{author}{{Yanny} B}, \bibinfo{author}{{Yin} B},
  \bibinfo{author}{{Zuntz} J}, \bibinfo{author}{{Aguena} M},
  \bibinfo{author}{{Allam} SS}, \bibinfo{author}{{Alves} O},
  \bibinfo{author}{{Bacon} D}, \bibinfo{author}{{Bocquet} S},
  \bibinfo{author}{{Brooks} D}, \bibinfo{author}{{da Costa} LN},
  \bibinfo{author}{{Davis} TM}, \bibinfo{author}{{De Vicente} J},
  \bibinfo{author}{{Desai} S}, \bibinfo{author}{{Diehl} HT},
  \bibinfo{author}{{Ferrero} I}, \bibinfo{author}{{Frieman} J},
  \bibinfo{author}{{Garc{\'\i}a-Bellido} J}, \bibinfo{author}{{Gaztanaga} E},
  \bibinfo{author}{{Giannini} G}, \bibinfo{author}{{Gutierrez} G},
  \bibinfo{author}{{Hinton} SR}, \bibinfo{author}{{Hollowood} DL},
  \bibinfo{author}{{Honscheid} K}, \bibinfo{author}{{Huterer} D},
  \bibinfo{author}{{James} DJ}, \bibinfo{author}{{Lahav} O},
  \bibinfo{author}{{Lee} S}, \bibinfo{author}{{Marshall} JL},
  \bibinfo{author}{{Mena-Fern{\'a}ndez} J}, \bibinfo{author}{{Miquel} R},
  \bibinfo{author}{{Pieres} A}, \bibinfo{author}{{Plazas Malag{\'o}n} AA},
  \bibinfo{author}{{Roodman} A}, \bibinfo{author}{{Sako} M},
  \bibinfo{author}{{Sanchez} E}, \bibinfo{author}{{Sanchez Cid} D},
  \bibinfo{author}{{Smith} M}, \bibinfo{author}{{Suchyta} E},
  \bibinfo{author}{{Swanson} MEC}, \bibinfo{author}{{Tarle} G},
  \bibinfo{author}{{Tucker} DL}, \bibinfo{author}{{Weaverdyck} N},
  \bibinfo{author}{{Weller} J}, \bibinfo{author}{{Wiseman} P} and
  \bibinfo{author}{{Yamamoto} M} (\bibinfo{year}{2024}), \bibinfo{month}{Mar.}
\bibinfo{title}{{Dark Energy Survey Year 3 results: likelihood-free,
  simulation-based $w$CDM inference with neural compression of weak-lensing map
  statistics}}.
\bibinfo{journal}{{\em arXiv e-prints}} ,
  \bibinfo{eid}{arXiv:2403.02314}\bibinfo{doi}{\doi{10.48550/arXiv.2403.02314}}.
\eprint{2403.02314}.

\bibtype{Article}%
\bibitem[Joachimi and Bridle(2010)]{Joachimi:2009ez}
\bibinfo{author}{Joachimi B} and  \bibinfo{author}{Bridle SL}
  (\bibinfo{year}{2010}).
\bibinfo{title}{{Simultaneous measurement of cosmology and intrinsic alignments
  using joint cosmic shear and galaxy number density correlations}}.
\bibinfo{journal}{{\em Astron. Astrophys.}} \bibinfo{volume}{523}:
  \bibinfo{pages}{A1}. \bibinfo{doi}{\doi{10.1051/0004-6361/200913657}}.
\eprint{0911.2454}.

\bibtype{Article}%
\bibitem[{Kaiser} and {Squires}(1993)]{KaiserSquires}
\bibinfo{author}{{Kaiser} N} and  \bibinfo{author}{{Squires} G}
  (\bibinfo{year}{1993}), \bibinfo{month}{Feb}.
\bibinfo{title}{{Mapping the Dark Matter with Weak Gravitational Lensing}}.
\bibinfo{journal}{{\em \apj}} \bibinfo{volume}{404}: \bibinfo{pages}{441}.
  \bibinfo{doi}{\doi{10.1086/172297}}.

\bibtype{Article}%
\bibitem[{Kilbinger} et al.(2017)]{Kilbinger2017}
\bibinfo{author}{{Kilbinger} M}, \bibinfo{author}{{Heymans} C},
  \bibinfo{author}{{Asgari} M}, \bibinfo{author}{{Joudaki} S},
  \bibinfo{author}{{Schneider} P}, \bibinfo{author}{{Simon} P},
  \bibinfo{author}{{Van Waerbeke} L}, \bibinfo{author}{{Harnois-D{\'e}raps} J},
  \bibinfo{author}{{Hildebrandt} H}, \bibinfo{author}{{K{\"o}hlinger} F},
  \bibinfo{author}{{Kuijken} K} and  \bibinfo{author}{{Viola} M}
  (\bibinfo{year}{2017}), \bibinfo{month}{Dec.}
\bibinfo{title}{{Precision calculations of the cosmic shear power spectrum
  projection}}.
\bibinfo{journal}{{\em \mnras}} \bibinfo{volume}{472} (\bibinfo{number}{2}):
  \bibinfo{pages}{2126--2141}. \bibinfo{doi}{\doi{10.1093/mnras/stx2082}}.
\eprint{1702.05301}.

\bibtype{Article}%
\bibitem[{Krause} et al.(2021)]{Krause2021}
\bibinfo{author}{{Krause} E}, \bibinfo{author}{{Fang} X},
  \bibinfo{author}{{Pandey} S}, \bibinfo{author}{{Secco} LF},
  \bibinfo{author}{{Alves} O}, \bibinfo{author}{{Huang} H},
  \bibinfo{author}{{Blazek} J}, \bibinfo{author}{{Prat} J},
  \bibinfo{author}{{Zuntz} J}, \bibinfo{author}{{Eifler} TF},
  \bibinfo{author}{{MacCrann} N}, \bibinfo{author}{{DeRose} J},
  \bibinfo{author}{{Crocce} M}, \bibinfo{author}{{Porredon} A},
  \bibinfo{author}{{Jain} B}, \bibinfo{author}{{Troxel} MA},
  \bibinfo{author}{{Dodelson} S}, \bibinfo{author}{{Huterer} D},
  \bibinfo{author}{{Liddle} AR}, \bibinfo{author}{{Leonard} CD},
  \bibinfo{author}{{Amon} A}, \bibinfo{author}{{Chen} A},
  \bibinfo{author}{{Elvin-Poole} J}, \bibinfo{author}{{Fert{\'e}} A},
  \bibinfo{author}{{Muir} J}, \bibinfo{author}{{Park} Y},
  \bibinfo{author}{{Samuroff} S}, \bibinfo{author}{{Brandao-Souza} A},
  \bibinfo{author}{{Weaverdyck} N}, \bibinfo{author}{{Zacharegkas} G},
  \bibinfo{author}{{Rosenfeld} R}, \bibinfo{author}{{Campos} A},
  \bibinfo{author}{{Chintalapati} P}, \bibinfo{author}{{Choi} A},
  \bibinfo{author}{{Di Valentino} E}, \bibinfo{author}{{Doux} C},
  \bibinfo{author}{{Herner} K}, \bibinfo{author}{{Lemos} P},
  \bibinfo{author}{{Mena-Fern{\'a}ndez} J}, \bibinfo{author}{{Omori} Y},
  \bibinfo{author}{{Paterno} M}, \bibinfo{author}{{Rodriguez-Monroy} M},
  \bibinfo{author}{{Rogozenski} P}, \bibinfo{author}{{Rollins} RP},
  \bibinfo{author}{{Troja} A}, \bibinfo{author}{{Tutusaus} I},
  \bibinfo{author}{{Wechsler} RH}, \bibinfo{author}{{Abbott} TMC},
  \bibinfo{author}{{Aguena} M}, \bibinfo{author}{{Allam} S},
  \bibinfo{author}{{Andrade-Oliveira} F}, \bibinfo{author}{{Annis} J},
  \bibinfo{author}{{Bacon} D}, \bibinfo{author}{{Baxter} E},
  \bibinfo{author}{{Bechtol} K}, \bibinfo{author}{{Bernstein} GM},
  \bibinfo{author}{{Brooks} D}, \bibinfo{author}{{Buckley-Geer} E},
  \bibinfo{author}{{Burke} DL}, \bibinfo{author}{{Carnero Rosell} A},
  \bibinfo{author}{{Carrasco Kind} M}, \bibinfo{author}{{Carretero} J},
  \bibinfo{author}{{Castander} FJ}, \bibinfo{author}{{Cawthon} R},
  \bibinfo{author}{{Chang} C}, \bibinfo{author}{{Costanzi} M},
  \bibinfo{author}{{da Costa} LN}, \bibinfo{author}{{Pereira} MES},
  \bibinfo{author}{{De Vicente} J}, \bibinfo{author}{{Desai} S},
  \bibinfo{author}{{Diehl} HT}, \bibinfo{author}{{Doel} P},
  \bibinfo{author}{{Everett} S}, \bibinfo{author}{{Evrard} AE},
  \bibinfo{author}{{Ferrero} I}, \bibinfo{author}{{Flaugher} B},
  \bibinfo{author}{{Fosalba} P}, \bibinfo{author}{{Frieman} J},
  \bibinfo{author}{{Garc{\'\i}a-Bellido} J}, \bibinfo{author}{{Gaztanaga} E},
  \bibinfo{author}{{Gerdes} DW}, \bibinfo{author}{{Giannantonio} T},
  \bibinfo{author}{{Gruen} D}, \bibinfo{author}{{Gruendl} RA},
  \bibinfo{author}{{Gschwend} J}, \bibinfo{author}{{Gutierrez} G},
  \bibinfo{author}{{Hartley} WG}, \bibinfo{author}{{Hinton} SR},
  \bibinfo{author}{{Hollowood} DL}, \bibinfo{author}{{Honscheid} K},
  \bibinfo{author}{{Hoyle} B}, \bibinfo{author}{{Huff} EM},
  \bibinfo{author}{{James} DJ}, \bibinfo{author}{{Kuehn} K},
  \bibinfo{author}{{Kuropatkin} N}, \bibinfo{author}{{Lahav} O},
  \bibinfo{author}{{Lima} M}, \bibinfo{author}{{Maia} MAG},
  \bibinfo{author}{{Marshall} JL}, \bibinfo{author}{{Martini} P},
  \bibinfo{author}{{Melchior} P}, \bibinfo{author}{{Menanteau} F},
  \bibinfo{author}{{Miquel} R}, \bibinfo{author}{{Mohr} JJ},
  \bibinfo{author}{{Morgan} R}, \bibinfo{author}{{Myles} J},
  \bibinfo{author}{{Palmese} A}, \bibinfo{author}{{Paz-Chinch{\'o}n} F},
  \bibinfo{author}{{Petravick} D}, \bibinfo{author}{{Pieres} A},
  \bibinfo{author}{{Plazas Malag{\'o}n} AA}, \bibinfo{author}{{Sanchez} E},
  \bibinfo{author}{{Scarpine} V}, \bibinfo{author}{{Schubnell} M},
  \bibinfo{author}{{Serrano} S}, \bibinfo{author}{{Sevilla-Noarbe} I},
  \bibinfo{author}{{Smith} M}, \bibinfo{author}{{Soares-Santos} M},
  \bibinfo{author}{{Suchyta} E}, \bibinfo{author}{{Tarle} G},
  \bibinfo{author}{{Thomas} D}, \bibinfo{author}{{To} C},
  \bibinfo{author}{{Varga} TN} and  \bibinfo{author}{{Weller} J}
  (\bibinfo{year}{2021}), \bibinfo{month}{May}.
\bibinfo{title}{{Dark Energy Survey Year 3 Results: Multi-Probe Modeling
  Strategy and Validation}}.
\bibinfo{journal}{{\em arXiv e-prints}} ,
  \bibinfo{eid}{arXiv:2105.13548}\bibinfo{doi}{\doi{10.48550/arXiv.2105.13548}}.
\eprint{2105.13548}.

\bibtype{Article}%
\bibitem[{Lamman} et al.(2024)]{Lamman2024}
\bibinfo{author}{{Lamman} C}, \bibinfo{author}{{Tsaprazi} E},
  \bibinfo{author}{{Shi} J}, \bibinfo{author}{{{\v{S}}ar{\v{c}}evi{\'c}} NN},
  \bibinfo{author}{{Pyne} S}, \bibinfo{author}{{Legnani} E} and
  \bibinfo{author}{{Ferreira} T} (\bibinfo{year}{2024}), \bibinfo{month}{Feb.}
\bibinfo{title}{{The IA Guide: A Breakdown of Intrinsic Alignment Formalisms}}.
\bibinfo{journal}{{\em The Open Journal of Astrophysics}} \bibinfo{volume}{7},
  \bibinfo{eid}{14}. \bibinfo{doi}{\doi{10.21105/astro.2309.08605}}.
\eprint{2309.08605}.

\bibtype{Article}%
\bibitem[{MacCrann} et al.(2022)]{MacCrann2022}
\bibinfo{author}{{MacCrann} N}, \bibinfo{author}{{Becker} MR},
  \bibinfo{author}{{McCullough} J}, \bibinfo{author}{{Amon} A},
  \bibinfo{author}{{Gruen} D}, \bibinfo{author}{{Jarvis} M},
  \bibinfo{author}{{Choi} A}, \bibinfo{author}{{Troxel} MA},
  \bibinfo{author}{{Sheldon} E}, \bibinfo{author}{{Yanny} B},
  \bibinfo{author}{{Herner} K}, \bibinfo{author}{{Dodelson} S},
  \bibinfo{author}{{Zuntz} J}, \bibinfo{author}{{Eckert} K},
  \bibinfo{author}{{Rollins} RP}, \bibinfo{author}{{Varga} TN},
  \bibinfo{author}{{Bernstein} GM}, \bibinfo{author}{{Gruendl} RA},
  \bibinfo{author}{{Harrison} I}, \bibinfo{author}{{Hartley} WG},
  \bibinfo{author}{{Sevilla-Noarbe} I}, \bibinfo{author}{{Pieres} A},
  \bibinfo{author}{{Bridle} SL}, \bibinfo{author}{{Myles} J},
  \bibinfo{author}{{Alarcon} A}, \bibinfo{author}{{Everett} S},
  \bibinfo{author}{{S{\'a}nchez} C}, \bibinfo{author}{{Huff} EM},
  \bibinfo{author}{{Tarsitano} F}, \bibinfo{author}{{Gatti} M},
  \bibinfo{author}{{Secco} LF}, \bibinfo{author}{{Abbott} TMC},
  \bibinfo{author}{{Aguena} M}, \bibinfo{author}{{Allam} S},
  \bibinfo{author}{{Annis} J}, \bibinfo{author}{{Bacon} D},
  \bibinfo{author}{{Bertin} E}, \bibinfo{author}{{Brooks} D},
  \bibinfo{author}{{Burke} DL}, \bibinfo{author}{{Carnero Rosell} A},
  \bibinfo{author}{{Carrasco Kind} M}, \bibinfo{author}{{Carretero} J},
  \bibinfo{author}{{Costanzi} M}, \bibinfo{author}{{Crocce} M},
  \bibinfo{author}{{Pereira} MES}, \bibinfo{author}{{De Vicente} J},
  \bibinfo{author}{{Desai} S}, \bibinfo{author}{{Diehl} HT},
  \bibinfo{author}{{Dietrich} JP}, \bibinfo{author}{{Doel} P},
  \bibinfo{author}{{Eifler} TF}, \bibinfo{author}{{Ferrero} I},
  \bibinfo{author}{{Fert{\'e}} A}, \bibinfo{author}{{Flaugher} B},
  \bibinfo{author}{{Fosalba} P}, \bibinfo{author}{{Frieman} J},
  \bibinfo{author}{{Garc{\'\i}a-Bellido} J}, \bibinfo{author}{{Gaztanaga} E},
  \bibinfo{author}{{Gerdes} DW}, \bibinfo{author}{{Giannantonio} T},
  \bibinfo{author}{{Gschwend} J}, \bibinfo{author}{{Gutierrez} G},
  \bibinfo{author}{{Hinton} SR}, \bibinfo{author}{{Hollowood} DL},
  \bibinfo{author}{{Honscheid} K}, \bibinfo{author}{{James} DJ},
  \bibinfo{author}{{Lahav} O}, \bibinfo{author}{{Lima} M},
  \bibinfo{author}{{Maia} MAG}, \bibinfo{author}{{March} M},
  \bibinfo{author}{{Marshall} JL}, \bibinfo{author}{{Martini} P},
  \bibinfo{author}{{Melchior} P}, \bibinfo{author}{{Menanteau} F},
  \bibinfo{author}{{Miquel} R}, \bibinfo{author}{{Mohr} JJ},
  \bibinfo{author}{{Morgan} R}, \bibinfo{author}{{Muir} J},
  \bibinfo{author}{{Ogando} RLC}, \bibinfo{author}{{Palmese} A},
  \bibinfo{author}{{Paz-Chinch{\'o}n} F}, \bibinfo{author}{{Plazas} AA},
  \bibinfo{author}{{Rodriguez-Monroy} M}, \bibinfo{author}{{Roodman} A},
  \bibinfo{author}{{Samuroff} S}, \bibinfo{author}{{Sanchez} E},
  \bibinfo{author}{{Scarpine} V}, \bibinfo{author}{{Serrano} S},
  \bibinfo{author}{{Smith} M}, \bibinfo{author}{{Soares-Santos} M},
  \bibinfo{author}{{Suchyta} E}, \bibinfo{author}{{Swanson} MEC},
  \bibinfo{author}{{Tarle} G}, \bibinfo{author}{{Thomas} D},
  \bibinfo{author}{{To} C}, \bibinfo{author}{{Wilkinson} RD},
  \bibinfo{author}{{Wilkinson} RD} and  \bibinfo{author}{{DES Collaboration}}
  (\bibinfo{year}{2022}), \bibinfo{month}{Jan.}
\bibinfo{title}{{Dark Energy Survey Y3 results: blending shear and redshift
  biases in image simulations}}.
\bibinfo{journal}{{\em \mnras}} \bibinfo{volume}{509} (\bibinfo{number}{3}):
  \bibinfo{pages}{3371--3394}. \bibinfo{doi}{\doi{10.1093/mnras/stab2870}}.
\eprint{2012.08567}.

\bibtype{Article}%
\bibitem[{Mead} et al.(2021)]{Mead2021}
\bibinfo{author}{{Mead} AJ}, \bibinfo{author}{{Brieden} S},
  \bibinfo{author}{{Tr{\"o}ster} T} and  \bibinfo{author}{{Heymans} C}
  (\bibinfo{year}{2021}), \bibinfo{month}{Mar.}
\bibinfo{title}{{HMCODE-2020: improved modelling of non-linear cosmological
  power spectra with baryonic feedback}}.
\bibinfo{journal}{{\em \mnras}} \bibinfo{volume}{502} (\bibinfo{number}{1}):
  \bibinfo{pages}{1401--1422}. \bibinfo{doi}{\doi{10.1093/mnras/stab082}}.
\eprint{2009.01858}.

\bibtype{Article}%
\bibitem[Miller et al.(2007)]{Miller2007}
\bibinfo{author}{Miller L}, \bibinfo{author}{Kitching TD},
  \bibinfo{author}{Heymans C}, \bibinfo{author}{Heavens AF} and
  \bibinfo{author}{{Van Waerbeke} L} (\bibinfo{year}{2007}),
  \bibinfo{month}{nov}.
\bibinfo{title}{{Bayesian galaxy shape measurement for weak lensing surveys -
  I. Methodology and a fast-fitting algorithm}}.
\bibinfo{journal}{{\em \mnras}} \bibinfo{volume}{382} (\bibinfo{number}{1}):
  \bibinfo{pages}{315--324}.
ISSN \bibinfo{issn}{0035-8711}.
  \bibinfo{doi}{\doi{10.1111/j.1365-2966.2007.12363.x}}.
\eprint{0708.2340},
  \bibinfo{url}{\url{http://mnras.oxfordjournals.org/cgi/doi/10.1111/j.1365-2966.2007.12363.x}}.

\bibtype{Article}%
\bibitem[{Myles} et al.(2021)]{Myles2021}
\bibinfo{author}{{Myles} J}, \bibinfo{author}{{Alarcon} A},
  \bibinfo{author}{{Amon} A}, \bibinfo{author}{{S{\'a}nchez} C},
  \bibinfo{author}{{Everett} S}, \bibinfo{author}{{DeRose} J},
  \bibinfo{author}{{McCullough} J}, \bibinfo{author}{{Gruen} D},
  \bibinfo{author}{{Bernstein} GM}, \bibinfo{author}{{Troxel} MA},
  \bibinfo{author}{{Dodelson} S}, \bibinfo{author}{{Campos} A},
  \bibinfo{author}{{MacCrann} N}, \bibinfo{author}{{Yin} B},
  \bibinfo{author}{{Raveri} M}, \bibinfo{author}{{Amara} A},
  \bibinfo{author}{{Becker} MR}, \bibinfo{author}{{Choi} A},
  \bibinfo{author}{{Cordero} J}, \bibinfo{author}{{Eckert} K},
  \bibinfo{author}{{Gatti} M}, \bibinfo{author}{{Giannini} G},
  \bibinfo{author}{{Gschwend} J}, \bibinfo{author}{{Gruendl} RA},
  \bibinfo{author}{{Harrison} I}, \bibinfo{author}{{Hartley} WG},
  \bibinfo{author}{{Huff} EM}, \bibinfo{author}{{Kuropatkin} N},
  \bibinfo{author}{{Lin} H}, \bibinfo{author}{{Masters} D},
  \bibinfo{author}{{Miquel} R}, \bibinfo{author}{{Prat} J},
  \bibinfo{author}{{Roodman} A}, \bibinfo{author}{{Rykoff} ES},
  \bibinfo{author}{{Sevilla-Noarbe} I}, \bibinfo{author}{{Sheldon} E},
  \bibinfo{author}{{Wechsler} RH}, \bibinfo{author}{{Yanny} B},
  \bibinfo{author}{{Abbott} TMC}, \bibinfo{author}{{Aguena} M},
  \bibinfo{author}{{Allam} S}, \bibinfo{author}{{Annis} J},
  \bibinfo{author}{{Bacon} D}, \bibinfo{author}{{Bertin} E},
  \bibinfo{author}{{Bhargava} S}, \bibinfo{author}{{Bridle} SL},
  \bibinfo{author}{{Brooks} D}, \bibinfo{author}{{Burke} DL},
  \bibinfo{author}{{Carnero Rosell} A}, \bibinfo{author}{{Carrasco Kind} M},
  \bibinfo{author}{{Carretero} J}, \bibinfo{author}{{Castander} FJ},
  \bibinfo{author}{{Conselice} C}, \bibinfo{author}{{Costanzi} M},
  \bibinfo{author}{{Crocce} M}, \bibinfo{author}{{da Costa} LN},
  \bibinfo{author}{{Pereira} MES}, \bibinfo{author}{{Desai} S},
  \bibinfo{author}{{Diehl} HT}, \bibinfo{author}{{Eifler} TF},
  \bibinfo{author}{{Elvin-Poole} J}, \bibinfo{author}{{Evrard} AE},
  \bibinfo{author}{{Ferrero} I}, \bibinfo{author}{{Fert{\'e}} A},
  \bibinfo{author}{{Flaugher} B}, \bibinfo{author}{{Fosalba} P},
  \bibinfo{author}{{Frieman} J}, \bibinfo{author}{{Garc{\'\i}a-Bellido} J},
  \bibinfo{author}{{Gaztanaga} E}, \bibinfo{author}{{Giannantonio} T},
  \bibinfo{author}{{Hinton} SR}, \bibinfo{author}{{Hollowood} DL},
  \bibinfo{author}{{Honscheid} K}, \bibinfo{author}{{Hoyle} B},
  \bibinfo{author}{{Huterer} D}, \bibinfo{author}{{James} DJ},
  \bibinfo{author}{{Krause} E}, \bibinfo{author}{{Kuehn} K},
  \bibinfo{author}{{Lahav} O}, \bibinfo{author}{{Lima} M},
  \bibinfo{author}{{Maia} MAG}, \bibinfo{author}{{Marshall} JL},
  \bibinfo{author}{{Martini} P}, \bibinfo{author}{{Melchior} P},
  \bibinfo{author}{{Menanteau} F}, \bibinfo{author}{{Mohr} JJ},
  \bibinfo{author}{{Morgan} R}, \bibinfo{author}{{Muir} J},
  \bibinfo{author}{{Ogando} RLC}, \bibinfo{author}{{Palmese} A},
  \bibinfo{author}{{Paz-Chinch{\'o}n} F}, \bibinfo{author}{{Plazas} AA},
  \bibinfo{author}{{Rodriguez-Monroy} M}, \bibinfo{author}{{Samuroff} S},
  \bibinfo{author}{{Sanchez} E}, \bibinfo{author}{{Scarpine} V},
  \bibinfo{author}{{Secco} LF}, \bibinfo{author}{{Serrano} S},
  \bibinfo{author}{{Smith} M}, \bibinfo{author}{{Soares-Santos} M},
  \bibinfo{author}{{Suchyta} E}, \bibinfo{author}{{Swanson} MEC},
  \bibinfo{author}{{Tarle} G}, \bibinfo{author}{{Thomas} D},
  \bibinfo{author}{{To} C}, \bibinfo{author}{{Varga} TN},
  \bibinfo{author}{{Weller} J} and  \bibinfo{author}{{Wester} W}
  (\bibinfo{year}{2021}), \bibinfo{month}{Aug.}
\bibinfo{title}{{Dark Energy Survey Year 3 results: redshift calibration of the
  weak lensing source galaxies}}.
\bibinfo{journal}{{\em \mnras}} \bibinfo{volume}{505} (\bibinfo{number}{3}):
  \bibinfo{pages}{4249--4277}. \bibinfo{doi}{\doi{10.1093/mnras/stab1515}}.
\eprint{2012.08566}.

\bibtype{Article}%
\bibitem[Nicola et al.(2016)]{Nicola:2016eua}
\bibinfo{author}{Nicola A}, \bibinfo{author}{Refregier A} and
  \bibinfo{author}{Amara A} (\bibinfo{year}{2016}).
\bibinfo{title}{{Integrated approach to cosmology: Combining CMB, large-scale
  structure and weak lensing}}.
\bibinfo{journal}{{\em Phys. Rev.}} \bibinfo{volume}{D94}
  (\bibinfo{number}{8}): \bibinfo{pages}{083517}.
  \bibinfo{doi}{\doi{10.1103/PhysRevD.94.083517}}.
\eprint{1607.01014}.

\bibtype{Article}%
\bibitem[{Planck Collaboration}(2018)]{Planck2018}
\bibinfo{author}{{Planck Collaboration}} (\bibinfo{year}{2018}),
  \bibinfo{month}{Jul.}
\bibinfo{title}{{Planck 2018 results. VI. Cosmological parameters}}.
\bibinfo{journal}{{\em arXiv:1807.06209}} \eprint{1807.06209}.

\bibtype{Article}%
\bibitem[{Prat} et al.(2022)]{Prat2022}
\bibinfo{author}{{Prat} J}, \bibinfo{author}{{Blazek} J},
  \bibinfo{author}{{S{\'a}nchez} C}, \bibinfo{author}{{Tutusaus} I},
  \bibinfo{author}{{Pandey} S}, \bibinfo{author}{{Elvin-Poole} J},
  \bibinfo{author}{{Krause} E}, \bibinfo{author}{{Troxel} MA},
  \bibinfo{author}{{Secco} LF}, \bibinfo{author}{{Amon} A},
  \bibinfo{author}{{DeRose} J}, \bibinfo{author}{{Zacharegkas} G},
  \bibinfo{author}{{Chang} C}, \bibinfo{author}{{Jain} B},
  \bibinfo{author}{{MacCrann} N}, \bibinfo{author}{{Park} Y},
  \bibinfo{author}{{Sheldon} E}, \bibinfo{author}{{Giannini} G},
  \bibinfo{author}{{Bocquet} S}, \bibinfo{author}{{To} C},
  \bibinfo{author}{{Alarcon} A}, \bibinfo{author}{{Alves} O},
  \bibinfo{author}{{Andrade-Oliveira} F}, \bibinfo{author}{{Baxter} E},
  \bibinfo{author}{{Bechtol} K}, \bibinfo{author}{{Becker} MR},
  \bibinfo{author}{{Bernstein} GM}, \bibinfo{author}{{Camacho} H},
  \bibinfo{author}{{Campos} A}, \bibinfo{author}{{Carnero Rosell} A},
  \bibinfo{author}{{Carrasco Kind} M}, \bibinfo{author}{{Cawthon} R},
  \bibinfo{author}{{Chen} R}, \bibinfo{author}{{Choi} A},
  \bibinfo{author}{{Cordero} J}, \bibinfo{author}{{Crocce} M},
  \bibinfo{author}{{Davis} C}, \bibinfo{author}{{De Vicente} J},
  \bibinfo{author}{{Diehl} HT}, \bibinfo{author}{{Dodelson} S},
  \bibinfo{author}{{Doux} C}, \bibinfo{author}{{Drlica-Wagner} A},
  \bibinfo{author}{{Eckert} K}, \bibinfo{author}{{Eifler} TF},
  \bibinfo{author}{{Elsner} F}, \bibinfo{author}{{Everett} S},
  \bibinfo{author}{{Fang} X}, \bibinfo{author}{{Farahi} A},
  \bibinfo{author}{{Fert{\'e}} A}, \bibinfo{author}{{Fosalba} P},
  \bibinfo{author}{{Friedrich} O}, \bibinfo{author}{{Gatti} M},
  \bibinfo{author}{{Gruen} D}, \bibinfo{author}{{Gruendl} RA},
  \bibinfo{author}{{Harrison} I}, \bibinfo{author}{{Hartley} WG},
  \bibinfo{author}{{Herner} K}, \bibinfo{author}{{Huang} H},
  \bibinfo{author}{{Huff} EM}, \bibinfo{author}{{Huterer} D},
  \bibinfo{author}{{Jarvis} M}, \bibinfo{author}{{Kuropatkin} N},
  \bibinfo{author}{{Leget} PF}, \bibinfo{author}{{Lemos} P},
  \bibinfo{author}{{Liddle} AR}, \bibinfo{author}{{McCullough} J},
  \bibinfo{author}{{Muir} J}, \bibinfo{author}{{Myles} J},
  \bibinfo{author}{{Navarro-Alsina} A}, \bibinfo{author}{{Porredon} A},
  \bibinfo{author}{{Raveri} M}, \bibinfo{author}{{Rodriguez-Monroy} M},
  \bibinfo{author}{{Rollins} RP}, \bibinfo{author}{{Roodman} A},
  \bibinfo{author}{{Rosenfeld} R}, \bibinfo{author}{{Ross} AJ},
  \bibinfo{author}{{Rykoff} ES}, \bibinfo{author}{{Sanchez} J},
  \bibinfo{author}{{Sevilla-Noarbe} I}, \bibinfo{author}{{Shin} T},
  \bibinfo{author}{{Troja} A}, \bibinfo{author}{{Varga} TN},
  \bibinfo{author}{{Weaverdyck} N}, \bibinfo{author}{{Wechsler} RH},
  \bibinfo{author}{{Yanny} B}, \bibinfo{author}{{Yin} B},
  \bibinfo{author}{{Zuntz} J}, \bibinfo{author}{{Abbott} TMC},
  \bibinfo{author}{{Aguena} M}, \bibinfo{author}{{Allam} S},
  \bibinfo{author}{{Annis} J}, \bibinfo{author}{{Bacon} D},
  \bibinfo{author}{{Brooks} D}, \bibinfo{author}{{Burke} DL},
  \bibinfo{author}{{Carretero} J}, \bibinfo{author}{{Conselice} C},
  \bibinfo{author}{{Costanzi} M}, \bibinfo{author}{{da Costa} LN},
  \bibinfo{author}{{Pereira} MES}, \bibinfo{author}{{Desai} S},
  \bibinfo{author}{{Dietrich} JP}, \bibinfo{author}{{Doel} P},
  \bibinfo{author}{{Evrard} AE}, \bibinfo{author}{{Ferrero} I},
  \bibinfo{author}{{Flaugher} B}, \bibinfo{author}{{Frieman} J},
  \bibinfo{author}{{Garc{\'\i}a-Bellido} J}, \bibinfo{author}{{Gaztanaga} E},
  \bibinfo{author}{{Gerdes} DW}, \bibinfo{author}{{Giannantonio} T},
  \bibinfo{author}{{Gschwend} J}, \bibinfo{author}{{Gutierrez} G},
  \bibinfo{author}{{Hinton} SR}, \bibinfo{author}{{Hollowood} DL},
  \bibinfo{author}{{Honscheid} K}, \bibinfo{author}{{James} DJ},
  \bibinfo{author}{{Kuehn} K}, \bibinfo{author}{{Lahav} O},
  \bibinfo{author}{{Lin} H}, \bibinfo{author}{{Maia} MAG},
  \bibinfo{author}{{Marshall} JL}, \bibinfo{author}{{Martini} P},
  \bibinfo{author}{{Melchior} P}, \bibinfo{author}{{Menanteau} F},
  \bibinfo{author}{{Miller} CJ}, \bibinfo{author}{{Miquel} R},
  \bibinfo{author}{{Mohr} JJ}, \bibinfo{author}{{Morgan} R},
  \bibinfo{author}{{Ogando} RLC}, \bibinfo{author}{{Palmese} A},
  \bibinfo{author}{{Paz-Chinch{\'o}n} F}, \bibinfo{author}{{Petravick} D},
  \bibinfo{author}{{Plazas Malag{\'o}n} AA}, \bibinfo{author}{{Sanchez} E},
  \bibinfo{author}{{Serrano} S}, \bibinfo{author}{{Smith} M},
  \bibinfo{author}{{Soares-Santos} M}, \bibinfo{author}{{Suchyta} E},
  \bibinfo{author}{{Tarle} G}, \bibinfo{author}{{Thomas} D},
  \bibinfo{author}{{Weller} J} and  \bibinfo{author}{{DES Collaboration}}
  (\bibinfo{year}{2022}), \bibinfo{month}{Apr.}
\bibinfo{title}{{Dark energy survey year 3 results: High-precision measurement
  and modeling of galaxy-galaxy lensing}}.
\bibinfo{journal}{{\em \prd}} \bibinfo{volume}{105} (\bibinfo{number}{8}),
  \bibinfo{eid}{083528}. \bibinfo{doi}{\doi{10.1103/PhysRevD.105.083528}}.
\eprint{2105.13541}.

\bibtype{Article}%
\bibitem[{Rodr{\'\i}guez-Monroy} et al.(2022)]{Monroy2022}
\bibinfo{author}{{Rodr{\'\i}guez-Monroy} M}, \bibinfo{author}{{Weaverdyck} N},
  \bibinfo{author}{{Elvin-Poole} J}, \bibinfo{author}{{Crocce} M},
  \bibinfo{author}{{Carnero Rosell} A}, \bibinfo{author}{{Andrade-Oliveira} F},
  \bibinfo{author}{{Avila} S}, \bibinfo{author}{{Bechtol} K},
  \bibinfo{author}{{Bernstein} GM}, \bibinfo{author}{{Blazek} J},
  \bibinfo{author}{{Camacho} H}, \bibinfo{author}{{Cawthon} R},
  \bibinfo{author}{{De Vicente} J}, \bibinfo{author}{{DeRose} J},
  \bibinfo{author}{{Dodelson} S}, \bibinfo{author}{{Everett} S},
  \bibinfo{author}{{Fang} X}, \bibinfo{author}{{Ferrero} I},
  \bibinfo{author}{{Fert{\'e}} A}, \bibinfo{author}{{Friedrich} O},
  \bibinfo{author}{{Gaztanaga} E}, \bibinfo{author}{{Giannini} G},
  \bibinfo{author}{{Gruendl} RA}, \bibinfo{author}{{Hartley} WG},
  \bibinfo{author}{{Herner} K}, \bibinfo{author}{{Huff} EM},
  \bibinfo{author}{{Jarvis} M}, \bibinfo{author}{{Krause} E},
  \bibinfo{author}{{MacCrann} N}, \bibinfo{author}{{Mena-Fern{\'a}ndez} J},
  \bibinfo{author}{{Muir} J}, \bibinfo{author}{{Pandey} S},
  \bibinfo{author}{{Park} Y}, \bibinfo{author}{{Porredon} A},
  \bibinfo{author}{{Prat} J}, \bibinfo{author}{{Rosenfeld} R},
  \bibinfo{author}{{Ross} AJ}, \bibinfo{author}{{Rozo} E},
  \bibinfo{author}{{Rykoff} ES}, \bibinfo{author}{{Sanchez} E},
  \bibinfo{author}{{Sanchez Cid} D}, \bibinfo{author}{{Sevilla-Noarbe} I},
  \bibinfo{author}{{Tabbutt} M}, \bibinfo{author}{{To} C},
  \bibinfo{author}{{Wagoner} EL}, \bibinfo{author}{{Wechsler} RH},
  \bibinfo{author}{{Aguena} M}, \bibinfo{author}{{Allam} S},
  \bibinfo{author}{{Amon} A}, \bibinfo{author}{{Annis} J},
  \bibinfo{author}{{Bacon} D}, \bibinfo{author}{{Baxter} E},
  \bibinfo{author}{{Bertin} E}, \bibinfo{author}{{Bhargava} S},
  \bibinfo{author}{{Brooks} D}, \bibinfo{author}{{Burke} DL},
  \bibinfo{author}{{Carrasco Kind} M}, \bibinfo{author}{{Carretero} J},
  \bibinfo{author}{{Castander} FJ}, \bibinfo{author}{{Choi} A},
  \bibinfo{author}{{Conselice} C}, \bibinfo{author}{{Costanzi} M},
  \bibinfo{author}{{da Costa} LN}, \bibinfo{author}{{Pereira} MES},
  \bibinfo{author}{{Desai} S}, \bibinfo{author}{{Diehl} HT},
  \bibinfo{author}{{Flaugher} B}, \bibinfo{author}{{Fosalba} P},
  \bibinfo{author}{{Frieman} J}, \bibinfo{author}{{Garc{\'\i}a-Bellido} J},
  \bibinfo{author}{{Giannantonio} T}, \bibinfo{author}{{Gruen} D},
  \bibinfo{author}{{Gschwend} J}, \bibinfo{author}{{Gutierrez} G},
  \bibinfo{author}{{Hinton} SR}, \bibinfo{author}{{Hollowood} DL},
  \bibinfo{author}{{Honscheid} K}, \bibinfo{author}{{Huterer} D},
  \bibinfo{author}{{Jain} B}, \bibinfo{author}{{James} DJ},
  \bibinfo{author}{{Kuehn} K}, \bibinfo{author}{{Kuropatkin} N},
  \bibinfo{author}{{Lima} M}, \bibinfo{author}{{Maia} MAG},
  \bibinfo{author}{{March} M}, \bibinfo{author}{{Marshall} JL},
  \bibinfo{author}{{Melchior} P}, \bibinfo{author}{{Menanteau} F},
  \bibinfo{author}{{Miller} CJ}, \bibinfo{author}{{Miquel} R},
  \bibinfo{author}{{Mohr} JJ}, \bibinfo{author}{{Morgan} R},
  \bibinfo{author}{{Palmese} A}, \bibinfo{author}{{Paz-Chinch{\'o}n} F},
  \bibinfo{author}{{Pieres} A}, \bibinfo{author}{{Plazas Malag{\'o}n} AA},
  \bibinfo{author}{{Roodman} A}, \bibinfo{author}{{Scarpine} V},
  \bibinfo{author}{{Serrano} S}, \bibinfo{author}{{Smith} M},
  \bibinfo{author}{{Soares-Santos} M}, \bibinfo{author}{{Suchyta} E},
  \bibinfo{author}{{Tarle} G}, \bibinfo{author}{{Thomas} D},
  \bibinfo{author}{{Varga} TN} and  \bibinfo{author}{{DES Collaboration}}
  (\bibinfo{year}{2022}), \bibinfo{month}{Apr.}
\bibinfo{title}{{Dark Energy Survey Year 3 results: galaxy clustering and
  systematics treatment for lens galaxy samples}}.
\bibinfo{journal}{{\em \mnras}} \bibinfo{volume}{511} (\bibinfo{number}{2}):
  \bibinfo{pages}{2665--2687}. \bibinfo{doi}{\doi{10.1093/mnras/stac104}}.
\eprint{2105.13540}.

\bibtype{Article}%
\bibitem[{S{\'a}nchez} et al.(2022)]{Sanchez2022}
\bibinfo{author}{{S{\'a}nchez} C}, \bibinfo{author}{{Prat} J},
  \bibinfo{author}{{Zacharegkas} G}, \bibinfo{author}{{Pandey} S},
  \bibinfo{author}{{Baxter} E}, \bibinfo{author}{{Bernstein} GM},
  \bibinfo{author}{{Blazek} J}, \bibinfo{author}{{Cawthon} R},
  \bibinfo{author}{{Chang} C}, \bibinfo{author}{{Krause} E},
  \bibinfo{author}{{Lemos} P}, \bibinfo{author}{{Park} Y},
  \bibinfo{author}{{Raveri} M}, \bibinfo{author}{{Sanchez} J},
  \bibinfo{author}{{Troxel} MA}, \bibinfo{author}{{Amon} A},
  \bibinfo{author}{{Fang} X}, \bibinfo{author}{{Friedrich} O},
  \bibinfo{author}{{Gruen} D}, \bibinfo{author}{{Porredon} A},
  \bibinfo{author}{{Secco} LF}, \bibinfo{author}{{Samuroff} S},
  \bibinfo{author}{{Alarcon} A}, \bibinfo{author}{{Alves} O},
  \bibinfo{author}{{Andrade-Oliveira} F}, \bibinfo{author}{{Bechtol} K},
  \bibinfo{author}{{Becker} MR}, \bibinfo{author}{{Camacho} H},
  \bibinfo{author}{{Campos} A}, \bibinfo{author}{{Carnero Rosell} A},
  \bibinfo{author}{{Carrasco Kind} M}, \bibinfo{author}{{Chen} R},
  \bibinfo{author}{{Choi} A}, \bibinfo{author}{{Crocce} M},
  \bibinfo{author}{{Davis} C}, \bibinfo{author}{{De Vicente} J},
  \bibinfo{author}{{DeRose} J}, \bibinfo{author}{{Di Valentino} E},
  \bibinfo{author}{{Diehl} HT}, \bibinfo{author}{{Dodelson} S},
  \bibinfo{author}{{Doux} C}, \bibinfo{author}{{Drlica-Wagner} A},
  \bibinfo{author}{{Eckert} K}, \bibinfo{author}{{Eifler} TF},
  \bibinfo{author}{{Elsner} F}, \bibinfo{author}{{Elvin-Poole} J},
  \bibinfo{author}{{Everett} S}, \bibinfo{author}{{Fert{\'e}} A},
  \bibinfo{author}{{Fosalba} P}, \bibinfo{author}{{Gatti} M},
  \bibinfo{author}{{Giannini} G}, \bibinfo{author}{{Gruendl} RA},
  \bibinfo{author}{{Harrison} I}, \bibinfo{author}{{Hartley} WG},
  \bibinfo{author}{{Herner} K}, \bibinfo{author}{{Huff} EM},
  \bibinfo{author}{{Huterer} D}, \bibinfo{author}{{Jarvis} M},
  \bibinfo{author}{{Jain} B}, \bibinfo{author}{{Kuropatkin} N},
  \bibinfo{author}{{Leget} PF}, \bibinfo{author}{{MacCrann} N},
  \bibinfo{author}{{McCullough} J}, \bibinfo{author}{{Muir} J},
  \bibinfo{author}{{Myles} J}, \bibinfo{author}{{Navarro-Alsina} A},
  \bibinfo{author}{{Rollins} RP}, \bibinfo{author}{{Roodman} A},
  \bibinfo{author}{{Rosenfeld} R}, \bibinfo{author}{{Rykoff} ES},
  \bibinfo{author}{{Sevilla-Noarbe} I}, \bibinfo{author}{{Sheldon} E},
  \bibinfo{author}{{Shin} T}, \bibinfo{author}{{Troja} A},
  \bibinfo{author}{{Tutusaus} I}, \bibinfo{author}{{Varga} TN},
  \bibinfo{author}{{Wechsler} RH}, \bibinfo{author}{{Yanny} B},
  \bibinfo{author}{{Yin} B}, \bibinfo{author}{{Zhang} Y},
  \bibinfo{author}{{Zuntz} J}, \bibinfo{author}{{Abbott} TMC},
  \bibinfo{author}{{Aguena} M}, \bibinfo{author}{{Allam} S},
  \bibinfo{author}{{Bacon} D}, \bibinfo{author}{{Bertin} E},
  \bibinfo{author}{{Bhargava} S}, \bibinfo{author}{{Brooks} D},
  \bibinfo{author}{{Buckley-Geer} E}, \bibinfo{author}{{Burke} DL},
  \bibinfo{author}{{Carretero} J}, \bibinfo{author}{{Costanzi} M},
  \bibinfo{author}{{da Costa} LN}, \bibinfo{author}{{Pereira} MES},
  \bibinfo{author}{{Desai} S}, \bibinfo{author}{{Dietrich} JP},
  \bibinfo{author}{{Doel} P}, \bibinfo{author}{{Evrard} AE},
  \bibinfo{author}{{Ferrero} I}, \bibinfo{author}{{Flaugher} B},
  \bibinfo{author}{{Frieman} J}, \bibinfo{author}{{Garc{\'\i}a-Bellido} J},
  \bibinfo{author}{{Gaztanaga} E}, \bibinfo{author}{{Gerdes} DW},
  \bibinfo{author}{{Giannantonio} T}, \bibinfo{author}{{Gschwend} J},
  \bibinfo{author}{{Gutierrez} G}, \bibinfo{author}{{Hinton} SR},
  \bibinfo{author}{{Hollowood} DL}, \bibinfo{author}{{Honscheid} K},
  \bibinfo{author}{{Hoyle} B}, \bibinfo{author}{{James} DJ},
  \bibinfo{author}{{Kuehn} K}, \bibinfo{author}{{Lahav} O},
  \bibinfo{author}{{Lima} M}, \bibinfo{author}{{Lin} H},
  \bibinfo{author}{{Maia} MAG}, \bibinfo{author}{{Marshall} JL},
  \bibinfo{author}{{Martini} P}, \bibinfo{author}{{Melchior} P},
  \bibinfo{author}{{Menanteau} F}, \bibinfo{author}{{Miquel} R},
  \bibinfo{author}{{Mohr} JJ}, \bibinfo{author}{{Morgan} R},
  \bibinfo{author}{{Palmese} A}, \bibinfo{author}{{Paz-Chinch{\'o}n} F},
  \bibinfo{author}{{Petravick} D}, \bibinfo{author}{{Pieres} A},
  \bibinfo{author}{{Plazas Malag{\'o}n} AA},
  \bibinfo{author}{{Rodriguez-Monroy} M}, \bibinfo{author}{{Sanchez} E},
  \bibinfo{author}{{Scarpine} V}, \bibinfo{author}{{Schubnell} M},
  \bibinfo{author}{{Serrano} S}, \bibinfo{author}{{Smith} M},
  \bibinfo{author}{{Soares-Santos} M}, \bibinfo{author}{{Suchyta} E},
  \bibinfo{author}{{Swanson} MEC}, \bibinfo{author}{{Tarle} G},
  \bibinfo{author}{{Thomas} D}, \bibinfo{author}{{To} C} and
  \bibinfo{author}{{DES Collaboration}} (\bibinfo{year}{2022}),
  \bibinfo{month}{Apr.}
\bibinfo{title}{{Dark Energy Survey Year 3 results: Exploiting small-scale
  information with lensing shear ratios}}.
\bibinfo{journal}{{\em \prd}} \bibinfo{volume}{105} (\bibinfo{number}{8}),
  \bibinfo{eid}{083529}. \bibinfo{doi}{\doi{10.1103/PhysRevD.105.083529}}.
\eprint{2105.13542}.

\bibtype{Article}%
\bibitem[Schaye et al.(2010)]{OWLS}
\bibinfo{author}{Schaye J}, \bibinfo{author}{Vecchia CD},
  \bibinfo{author}{Booth CM}, \bibinfo{author}{Wiersma RPC},
  \bibinfo{author}{Theuns T}, \bibinfo{author}{Haas MR},
  \bibinfo{author}{Bertone S}, \bibinfo{author}{Duffy AR},
  \bibinfo{author}{McCarthy IG} and  \bibinfo{author}{van~de Voort F}
  (\bibinfo{year}{2010}), \bibinfo{month}{02}.
\bibinfo{title}{{The physics driving the cosmic star formation history}}.
\bibinfo{journal}{{\em Monthly Notices of the Royal Astronomical Society}}
  \bibinfo{volume}{402} (\bibinfo{number}{3}): \bibinfo{pages}{1536--1560}.
ISSN \bibinfo{issn}{0035-8711}.
  \bibinfo{doi}{\doi{10.1111/j.1365-2966.2009.16029.x}}.
\eprint{https://academic.oup.com/mnras/article-pdf/402/3/1536/3114627/mnras0402-1536.pdf},
  \bibinfo{url}{\url{https://doi.org/10.1111/j.1365-2966.2009.16029.x}}.

\bibtype{Article}%
\bibitem[{Schneider} et al.(2010)]{Schneider2010}
\bibinfo{author}{{Schneider} P}, \bibinfo{author}{{Eifler} T} and
  \bibinfo{author}{{Krause} E} (\bibinfo{year}{2010}), \bibinfo{month}{Sep.}
\bibinfo{title}{{COSEBIs: Extracting the full E-/B-mode information from cosmic
  shear correlation functions}}.
\bibinfo{journal}{{\em \aap}} \bibinfo{volume}{520}, \bibinfo{eid}{A116}.
  \bibinfo{doi}{\doi{10.1051/0004-6361/201014235}}.
\eprint{1002.2136}.

\bibtype{Article}%
\bibitem[{Secco} et al.(2022)]{Secco2022}
\bibinfo{author}{{Secco} LF}, \bibinfo{author}{{Samuroff} S},
  \bibinfo{author}{{Krause} E}, \bibinfo{author}{{Jain} B},
  \bibinfo{author}{{Blazek} J}, \bibinfo{author}{{Raveri} M},
  \bibinfo{author}{{Campos} A}, \bibinfo{author}{{Amon} A},
  \bibinfo{author}{{Chen} A}, \bibinfo{author}{{Doux} C},
  \bibinfo{author}{{Choi} A}, \bibinfo{author}{{Gruen} D},
  \bibinfo{author}{{Bernstein} GM}, \bibinfo{author}{{Chang} C},
  \bibinfo{author}{{DeRose} J}, \bibinfo{author}{{Myles} J},
  \bibinfo{author}{{Fert{\'e}} A}, \bibinfo{author}{{Lemos} P},
  \bibinfo{author}{{Huterer} D}, \bibinfo{author}{{Prat} J},
  \bibinfo{author}{{Troxel} MA}, \bibinfo{author}{{MacCrann} N},
  \bibinfo{author}{{Liddle} AR}, \bibinfo{author}{{Kacprzak} T},
  \bibinfo{author}{{Fang} X}, \bibinfo{author}{{S{\'a}nchez} C},
  \bibinfo{author}{{Pandey} S}, \bibinfo{author}{{Dodelson} S},
  \bibinfo{author}{{Chintalapati} P}, \bibinfo{author}{{Hoffmann} K},
  \bibinfo{author}{{Alarcon} A}, \bibinfo{author}{{Alves} O},
  \bibinfo{author}{{Andrade-Oliveira} F}, \bibinfo{author}{{Baxter} EJ},
  \bibinfo{author}{{Bechtol} K}, \bibinfo{author}{{Becker} MR},
  \bibinfo{author}{{Brandao-Souza} A}, \bibinfo{author}{{Camacho} H},
  \bibinfo{author}{{Carnero Rosell} A}, \bibinfo{author}{{Carrasco Kind} M},
  \bibinfo{author}{{Cawthon} R}, \bibinfo{author}{{Cordero} JP},
  \bibinfo{author}{{Crocce} M}, \bibinfo{author}{{Davis} C},
  \bibinfo{author}{{Di Valentino} E}, \bibinfo{author}{{Drlica-Wagner} A},
  \bibinfo{author}{{Eckert} K}, \bibinfo{author}{{Eifler} TF},
  \bibinfo{author}{{Elidaiana} M}, \bibinfo{author}{{Elsner} F},
  \bibinfo{author}{{Elvin-Poole} J}, \bibinfo{author}{{Everett} S},
  \bibinfo{author}{{Fosalba} P}, \bibinfo{author}{{Friedrich} O},
  \bibinfo{author}{{Gatti} M}, \bibinfo{author}{{Giannini} G},
  \bibinfo{author}{{Gruendl} RA}, \bibinfo{author}{{Harrison} I},
  \bibinfo{author}{{Hartley} WG}, \bibinfo{author}{{Herner} K},
  \bibinfo{author}{{Huang} H}, \bibinfo{author}{{Huff} EM},
  \bibinfo{author}{{Jarvis} M}, \bibinfo{author}{{Jeffrey} N},
  \bibinfo{author}{{Kuropatkin} N}, \bibinfo{author}{{Leget} PF},
  \bibinfo{author}{{Muir} J}, \bibinfo{author}{{Mccullough} J},
  \bibinfo{author}{{Navarro Alsina} A}, \bibinfo{author}{{Omori} Y},
  \bibinfo{author}{{Park} Y}, \bibinfo{author}{{Porredon} A},
  \bibinfo{author}{{Rollins} R}, \bibinfo{author}{{Roodman} A},
  \bibinfo{author}{{Rosenfeld} R}, \bibinfo{author}{{Ross} AJ},
  \bibinfo{author}{{Rykoff} ES}, \bibinfo{author}{{Sanchez} J},
  \bibinfo{author}{{Sevilla-Noarbe} I}, \bibinfo{author}{{Sheldon} ES},
  \bibinfo{author}{{Shin} T}, \bibinfo{author}{{Troja} A},
  \bibinfo{author}{{Tutusaus} I}, \bibinfo{author}{{Varga} TN},
  \bibinfo{author}{{Weaverdyck} N}, \bibinfo{author}{{Wechsler} RH},
  \bibinfo{author}{{Yanny} B}, \bibinfo{author}{{Yin} B},
  \bibinfo{author}{{Zhang} Y}, \bibinfo{author}{{Zuntz} J},
  \bibinfo{author}{{Abbott} TMC}, \bibinfo{author}{{Aguena} M},
  \bibinfo{author}{{Allam} S}, \bibinfo{author}{{Annis} J},
  \bibinfo{author}{{Bacon} D}, \bibinfo{author}{{Bertin} E},
  \bibinfo{author}{{Bhargava} S}, \bibinfo{author}{{Bridle} SL},
  \bibinfo{author}{{Brooks} D}, \bibinfo{author}{{Buckley-Geer} E},
  \bibinfo{author}{{Burke} DL}, \bibinfo{author}{{Carretero} J},
  \bibinfo{author}{{Costanzi} M}, \bibinfo{author}{{da Costa} LN},
  \bibinfo{author}{{De Vicente} J}, \bibinfo{author}{{Diehl} HT},
  \bibinfo{author}{{Dietrich} JP}, \bibinfo{author}{{Doel} P},
  \bibinfo{author}{{Ferrero} I}, \bibinfo{author}{{Flaugher} B},
  \bibinfo{author}{{Frieman} J}, \bibinfo{author}{{Garc{\'\i}a-Bellido} J},
  \bibinfo{author}{{Gaztanaga} E}, \bibinfo{author}{{Gerdes} DW},
  \bibinfo{author}{{Giannantonio} T}, \bibinfo{author}{{Gschwend} J},
  \bibinfo{author}{{Gutierrez} G}, \bibinfo{author}{{Hinton} SR},
  \bibinfo{author}{{Hollowood} DL}, \bibinfo{author}{{Honscheid} K},
  \bibinfo{author}{{Hoyle} B}, \bibinfo{author}{{James} DJ},
  \bibinfo{author}{{Jeltema} T}, \bibinfo{author}{{Kuehn} K},
  \bibinfo{author}{{Lahav} O}, \bibinfo{author}{{Lima} M},
  \bibinfo{author}{{Lin} H}, \bibinfo{author}{{Maia} MAG},
  \bibinfo{author}{{Marshall} JL}, \bibinfo{author}{{Martini} P},
  \bibinfo{author}{{Melchior} P}, \bibinfo{author}{{Menanteau} F},
  \bibinfo{author}{{Miquel} R}, \bibinfo{author}{{Mohr} JJ},
  \bibinfo{author}{{Morgan} R}, \bibinfo{author}{{Ogando} RLC},
  \bibinfo{author}{{Palmese} A}, \bibinfo{author}{{Paz-Chinch{\'o}n} F},
  \bibinfo{author}{{Petravick} D}, \bibinfo{author}{{Pieres} A},
  \bibinfo{author}{{Plazas Malag{\'o}n} AA},
  \bibinfo{author}{{Rodriguez-Monroy} M}, \bibinfo{author}{{Romer} AK},
  \bibinfo{author}{{Sanchez} E}, \bibinfo{author}{{Scarpine} V},
  \bibinfo{author}{{Schubnell} M}, \bibinfo{author}{{Scolnic} D},
  \bibinfo{author}{{Serrano} S}, \bibinfo{author}{{Smith} M},
  \bibinfo{author}{{Soares-Santos} M}, \bibinfo{author}{{Suchyta} E},
  \bibinfo{author}{{Swanson} MEC}, \bibinfo{author}{{Tarle} G},
  \bibinfo{author}{{Thomas} D}, \bibinfo{author}{{To} C} and
  \bibinfo{author}{{DES Collaboration}} (\bibinfo{year}{2022}),
  \bibinfo{month}{Jan.}
\bibinfo{title}{{Dark Energy Survey Year 3 results: Cosmology from cosmic shear
  and robustness to modeling uncertainty}}.
\bibinfo{journal}{{\em \prd}} \bibinfo{volume}{105} (\bibinfo{number}{2}),
  \bibinfo{eid}{023515}. \bibinfo{doi}{\doi{10.1103/PhysRevD.105.023515}}.
\eprint{2105.13544}.

\bibtype{Article}%
\bibitem[{Sellentin} and {Heavens}(2018)]{Sellentin2018}
\bibinfo{author}{{Sellentin} E} and  \bibinfo{author}{{Heavens} AF}
  (\bibinfo{year}{2018}), \bibinfo{month}{Jan.}
\bibinfo{title}{{On the insufficiency of arbitrarily precise covariance
  matrices: non-Gaussian weak-lensing likelihoods}}.
\bibinfo{journal}{{\em \mnras}} \bibinfo{volume}{473}:
  \bibinfo{pages}{2355--2363}. \bibinfo{doi}{\doi{10.1093/mnras/stx2491}}.
\eprint{1707.04488}.

\bibtype{Article}%
\bibitem[Sheldon(2014)]{Sheldon2014}
\bibinfo{author}{Sheldon ES} (\bibinfo{year}{2014}), \bibinfo{month}{jul}.
\bibinfo{title}{{An implementation of Bayesian lensing shear measurement}}.
\bibinfo{journal}{{\em Mon. Not. R. Astron. Soc. Lett.}} \bibinfo{volume}{444}
  (\bibinfo{number}{1}): \bibinfo{pages}{L25--L29}.
ISSN \bibinfo{issn}{1745-3925}. \bibinfo{doi}{\doi{10.1093/mnrasl/slu104}}.
\bibinfo{url}{\url{http://mnrasl.oxfordjournals.org/cgi/doi/10.1093/mnrasl/slu104}}.

\bibtype{Article}%
\bibitem[Sheldon and Huff(2017)]{Sheldon2017}
\bibinfo{author}{Sheldon ES} and  \bibinfo{author}{Huff EM}
  (\bibinfo{year}{2017}), \bibinfo{month}{may}.
\bibinfo{title}{{Practical Weak-lensing Shear Measurement with
  Metacalibration}}.
\bibinfo{journal}{{\em \apj}} \bibinfo{volume}{841} (\bibinfo{number}{1}):
  \bibinfo{pages}{24}.
ISSN \bibinfo{issn}{1538-4357}. \bibinfo{doi}{\doi{10.3847/1538-4357/aa704b}}.
\eprint{1702.02601}, \bibinfo{url}{\url{http://arxiv.org/abs/1702.02601
  http://stacks.iop.org/0004-637X/841/i=1/a=24?key=crossref.f8de77f6ddca16d5b7db59d2cefcbebc}}.

\bibtype{Article}%
\bibitem[{Sheldon} et al.(2023)]{Sheldon2023}
\bibinfo{author}{{Sheldon} ES}, \bibinfo{author}{{Becker} MR},
  \bibinfo{author}{{Jarvis} M}, \bibinfo{author}{{Armstrong} R} and
  \bibinfo{author}{{LSST Dark Energy Science Collaboration}}
  (\bibinfo{year}{2023}), \bibinfo{month}{May}.
\bibinfo{title}{{Metadetection Weak Lensing for the Vera C. Rubin
  Observatory}}.
\bibinfo{journal}{{\em The Open Journal of Astrophysics}} \bibinfo{volume}{6},
  \bibinfo{eid}{17}. \bibinfo{doi}{\doi{10.21105/astro.2303.03947}}.
\eprint{2303.03947}.

\bibtype{Article}%
\bibitem[{Sugiyama} et al.(2023)]{hscy3-3x2}
\bibinfo{author}{{Sugiyama} S}, \bibinfo{author}{{Miyatake} H},
  \bibinfo{author}{{More} S}, \bibinfo{author}{{Li} X},
  \bibinfo{author}{{Shirasaki} M}, \bibinfo{author}{{Takada} M},
  \bibinfo{author}{{Kobayashi} Y}, \bibinfo{author}{{Takahashi} R},
  \bibinfo{author}{{Nishimichi} T}, \bibinfo{author}{{Nishizawa} AJ},
  \bibinfo{author}{{Rau} MM}, \bibinfo{author}{{Zhang} T},
  \bibinfo{author}{{Dalal} R}, \bibinfo{author}{{Mandelbaum} R},
  \bibinfo{author}{{Strauss} MA}, \bibinfo{author}{{Hamana} T},
  \bibinfo{author}{{Oguri} M}, \bibinfo{author}{{Osato} K},
  \bibinfo{author}{{Kannawadi} A}, \bibinfo{author}{{Hsieh} BC},
  \bibinfo{author}{{Luo} W}, \bibinfo{author}{{Armstrong} R},
  \bibinfo{author}{{Bosch} J}, \bibinfo{author}{{Komiyama} Y},
  \bibinfo{author}{{Lupton} RH}, \bibinfo{author}{{Lust} NB},
  \bibinfo{author}{{Miyazaki} S}, \bibinfo{author}{{Murayama} H},
  \bibinfo{author}{{Okura} Y}, \bibinfo{author}{{Price} PA},
  \bibinfo{author}{{Tait} PJ}, \bibinfo{author}{{Tanaka} M} and
  \bibinfo{author}{{Wang} SY} (\bibinfo{year}{2023}), \bibinfo{month}{Dec.}
\bibinfo{title}{{Hyper Suprime-Cam Year 3 results: Cosmology from galaxy
  clustering and weak lensing with HSC and SDSS using the minimal bias model}}.
\bibinfo{journal}{{\em \prd}} \bibinfo{volume}{108} (\bibinfo{number}{12}),
  \bibinfo{eid}{123521}. \bibinfo{doi}{\doi{10.1103/PhysRevD.108.123521}}.
\eprint{2304.00705}.

\bibtype{Article}%
\bibitem[{The LSST Dark Energy Science Collaboration} et al.(2018)]{DESC2018}
\bibinfo{author}{{The LSST Dark Energy Science Collaboration}},
  \bibinfo{author}{{Mandelbaum} R}, \bibinfo{author}{{Eifler} T},
  \bibinfo{author}{{Hlo{\v{z}}ek} R}, \bibinfo{author}{{Collett} T},
  \bibinfo{author}{{Gawiser} E}, \bibinfo{author}{{Scolnic} D},
  \bibinfo{author}{{Alonso} D}, \bibinfo{author}{{Awan} H},
  \bibinfo{author}{{Biswas} R}, \bibinfo{author}{{Blazek} J},
  \bibinfo{author}{{Burchat} P}, \bibinfo{author}{{Chisari} NE},
  \bibinfo{author}{{Dell'Antonio} I}, \bibinfo{author}{{Digel} S},
  \bibinfo{author}{{Frieman} J}, \bibinfo{author}{{Goldstein} DA},
  \bibinfo{author}{{Hook} I}, \bibinfo{author}{{Ivezi{\'c}} {\v{Z}}},
  \bibinfo{author}{{Kahn} SM}, \bibinfo{author}{{Kamath} S},
  \bibinfo{author}{{Kirkby} D}, \bibinfo{author}{{Kitching} T},
  \bibinfo{author}{{Krause} E}, \bibinfo{author}{{Leget} PF},
  \bibinfo{author}{{Marshall} PJ}, \bibinfo{author}{{Meyers} J},
  \bibinfo{author}{{Miyatake} H}, \bibinfo{author}{{Newman} JA},
  \bibinfo{author}{{Nichol} R}, \bibinfo{author}{{Rykoff} E},
  \bibinfo{author}{{Sanchez} FJ}, \bibinfo{author}{{Slosar} A},
  \bibinfo{author}{{Sullivan} M} and  \bibinfo{author}{{Troxel} MA}
  (\bibinfo{year}{2018}), \bibinfo{month}{Sep.}
\bibinfo{title}{{The LSST Dark Energy Science Collaboration (DESC) Science
  Requirements Document}}.
\bibinfo{journal}{{\em arXiv e-prints}} ,
  \bibinfo{eid}{arXiv:1809.01669}\eprint{1809.01669}.

\bibtype{Article}%
\bibitem[{van Daalen} et al.(2011)]{vanDaalen11}
\bibinfo{author}{{van Daalen} MP}, \bibinfo{author}{{Schaye} J},
  \bibinfo{author}{{Booth} CM} and  \bibinfo{author}{{Dalla Vecchia} C}
  (\bibinfo{year}{2011}), \bibinfo{month}{Aug.}
\bibinfo{title}{{The effects of galaxy formation on the matter power spectrum:
  a challenge for precision cosmology}}.
\bibinfo{journal}{{\em \mnras}} \bibinfo{volume}{415} (\bibinfo{number}{4}):
  \bibinfo{pages}{3649--3665}.
  \bibinfo{doi}{\doi{10.1111/j.1365-2966.2011.18981.x}}.
\eprint{1104.1174}.

\bibtype{Article}%
\bibitem[{Wechsler} and {Tinker}(2018)]{Wechsler2018}
\bibinfo{author}{{Wechsler} RH} and  \bibinfo{author}{{Tinker} JL}
  (\bibinfo{year}{2018}), \bibinfo{month}{Sep.}
\bibinfo{title}{{The Connection Between Galaxies and Their Dark Matter Halos}}.
\bibinfo{journal}{{\em \araa}} \bibinfo{volume}{56}: \bibinfo{pages}{435--487}.
  \bibinfo{doi}{\doi{10.1146/annurev-astro-081817-051756}}.
\eprint{1804.03097}.

\bibtype{Article}%
\bibitem[{Yamamoto} et al.(2025)]{Yamamoto2025}
\bibinfo{author}{{Yamamoto} M}, \bibinfo{author}{{Becker} MR},
  \bibinfo{author}{{Sheldon} E}, \bibinfo{author}{{Jarvis} M},
  \bibinfo{author}{{Gruendl} RA}, \bibinfo{author}{{Menanteau} F},
  \bibinfo{author}{{Rykoff} ES}, \bibinfo{author}{{Mau} S},
  \bibinfo{author}{{Schutt} T}, \bibinfo{author}{{Gatti} M},
  \bibinfo{author}{{Troxel} MA}, \bibinfo{author}{{Amon} A},
  \bibinfo{author}{{Anbajagane} D}, \bibinfo{author}{{Bernstein} GM},
  \bibinfo{author}{{Gruen} D}, \bibinfo{author}{{Huff} EM},
  \bibinfo{author}{{Tabbutt} M}, \bibinfo{author}{{Tong} A},
  \bibinfo{author}{{Yanny} B}, \bibinfo{author}{{Abbott} TMC},
  \bibinfo{author}{{Aguena} M}, \bibinfo{author}{{Alarcon} A},
  \bibinfo{author}{{Andrade-Oliveira} F}, \bibinfo{author}{{Bechtol} K},
  \bibinfo{author}{{Blazek} J}, \bibinfo{author}{{Brooks} D},
  \bibinfo{author}{{Carnero Rosell} A}, \bibinfo{author}{{Carretero} J},
  \bibinfo{author}{{Chang} C}, \bibinfo{author}{{Choi} A},
  \bibinfo{author}{{Costanzi} M}, \bibinfo{author}{{Crocce} M},
  \bibinfo{author}{{da Costa} LN}, \bibinfo{author}{{Davis} TM},
  \bibinfo{author}{{De Vicente} J}, \bibinfo{author}{{Desai} S},
  \bibinfo{author}{{Diehl} HT}, \bibinfo{author}{{Dodelson} S},
  \bibinfo{author}{{Doel} P}, \bibinfo{author}{{Doux} C},
  \bibinfo{author}{{Drlica-Wagner} A}, \bibinfo{author}{{Fert{\'e}} A},
  \bibinfo{author}{{Flaugher} B}, \bibinfo{author}{{Frieman} J},
  \bibinfo{author}{{Garc{\'\i}a-Bellido} J}, \bibinfo{author}{{Gaztanaga} E},
  \bibinfo{author}{{Giannini} G}, \bibinfo{author}{{Gutierrez} G},
  \bibinfo{author}{{Hartley} WG}, \bibinfo{author}{{Herner} K},
  \bibinfo{author}{{Hinton} SR}, \bibinfo{author}{{Hollowood} DL},
  \bibinfo{author}{{Honscheid} K}, \bibinfo{author}{{Huterer} D},
  \bibinfo{author}{{Krause} E}, \bibinfo{author}{{Kuehn} K},
  \bibinfo{author}{{Lahav} O}, \bibinfo{author}{{Lima} M},
  \bibinfo{author}{{Marshall} JL}, \bibinfo{author}{{Mena-Fern{\'a}ndez} J},
  \bibinfo{author}{{Miquel} R}, \bibinfo{author}{{Mohr} JJ},
  \bibinfo{author}{{Muir} J}, \bibinfo{author}{{Myles} J},
  \bibinfo{author}{{Ogando} RLC}, \bibinfo{author}{{Pieres} A},
  \bibinfo{author}{{Plazas Malag{\'o}n} AA}, \bibinfo{author}{{Porredon} A},
  \bibinfo{author}{{Prat} J}, \bibinfo{author}{{Raveri} M},
  \bibinfo{author}{{Rodriguez-Monroy} M}, \bibinfo{author}{{Roodman} A},
  \bibinfo{author}{{Samuroff} S}, \bibinfo{author}{{Sanchez} E},
  \bibinfo{author}{{Sanchez Cid} D}, \bibinfo{author}{{Scarpine} V},
  \bibinfo{author}{{Sevilla-Noarbe} I}, \bibinfo{author}{{Smith} M},
  \bibinfo{author}{{Suchyta} E}, \bibinfo{author}{{Tarle} G},
  \bibinfo{author}{{Vikram} V}, \bibinfo{author}{{Weaverdyck} N},
  \bibinfo{author}{{Wiseman} P} and  \bibinfo{author}{{Zhang} Y}
  (\bibinfo{year}{2025}), \bibinfo{month}{Jan.}
\bibinfo{title}{{Dark Energy Survey Year 6 Results: Cell-based Coadds and
  Metadetection Weak Lensing Shape Catalogue}}.
\bibinfo{journal}{{\em arXiv e-prints}} ,
  \bibinfo{eid}{arXiv:2501.05665}\eprint{2501.05665}.

\end{thebibliography*}

\end{document}